\PassOptionsToPackage{pdfpagelabels=false}{hyperref}
\documentclass[fleqn,usenatbib]{mnras}
\pdfoutput=1
\usepackage[pdftex]{graphicx}
\usepackage{amsmath}
\usepackage{amssymb}
\usepackage{geometry}
\usepackage{rotating,booktabs,multirow}
\usepackage{xcolor}
\usepackage{times}
\usepackage{tabularx}
\usepackage{longtable}
\usepackage{enumerate}

\newcommand{\bea}{\begin{eqnarray}}
\newcommand{\eea}{\end{eqnarray}}

\newcommand{\Ms}{{\rm M}_\odot}
\def \MS{M_{\rm stars}}
\def \RS{R_{\rm stars,1/2}}

\definecolor{portlandorange}{rgb}{1.0, 0.35, 0.21}

\definecolor{othergreen}{rgb}{0.0, 0.7, 0.0}

\title[Flaring of stellar disks in TNG50 MW/M31-like galaxies]{Disk flaring with TNG50: diversity across Milky Way and M31 analogs}

\author[Sotillo-Ramos et al.]
{Diego Sotillo-Ramos,$^{1}$\thanks{E-mail: sotillo@mpia.de}, 
Martina Donnari$^{1}$, Annalisa Pillepich$^{1}$, Neige Frankel$^{2}$, Dylan Nelson$^{3}$,
\newauthor Volker Springel$^{4}$, and Lars Hernquist$^{5}$ \\
$^{1}$ Max-Planck-Institut f\"{u}r Astronomie, K\"{o}nigstuhl 17, 69117 Heidelberg, Germany\\
$^{2}$ Canadian Institute for Theoretical Astrophysics, University of Toronto, 60 St. George Street, Toronto, ON M5S 3H8, Canada\\
$^{3}$ Universit\"at Heidelberg, Zentrum f\"ur Astronomie, Institut f\"ur theoretische Astrophysik, Albert-Ueberle-Str. 2, 69120 Heidelberg, Germany\\
$^{4}$ Max Planck Institut f\"ur Astrophysik, Karl-Schwarzschild-Stra\ss e 1, D-85748 Garching bei M\"unchen, Germany\\
$^{5}$ Center for Astrophysics $|$ Harvard \& Smithsonian, 6P Garden St., Cambridge, MA 02138, USA
}

\begin{document}
\maketitle

\begin{abstract}
We use the sample of 198 Milky Way (MW) and Andromeda (M31) analogs from TNG50 to quantify the level of disk flaring predicted by a modern, high-resolution cosmological hydrodynamical simulation. Disk flaring refers to the increase of vertical stellar disk height with galactocentric distance. The TNG50 galaxies are selected to have stellar disky morphology, a stellar mass in the range of $M_* = 10^{10.5 - 11.2}~\Ms$, and a MW-like Mpc-scale environment at $z=0$. 
The stellar disks of such TNG50 MW/M31 analogs exhibit a wide diversity of structural properties, including a number of galaxies with disk scalelength and thin and thick disk scaleheights that are comparable to those measured or inferred for the Galaxy and Andromeda.
With one set of physical ingredients, TNG50 returns a large variety of  flaring flavours and amounts, also for mono-age stellar populations. With this paper, we hence propose a non-parametric characterization of flaring. The typical MW/M31 analogs exhibit disk scaleheights that are $1.5-2$ times larger in the outer than in the inner regions of the disk for both old and young stellar populations, but with a large galaxy-to-galaxy variation. Which stellar population flares more, and by how much, also varies from galaxy to galaxy. TNG50 de facto brackets existing observational constraints for the Galaxy and all previous numerical findings. A link between the amount of flaring and the $z=0$ global galaxy structural properties or merger history is complex. However, a connection between the scaleheights and the local stellar vertical kinematics and gravitational potential is clearly in place.

\end{abstract}


\section{Introduction}
\label{sec:intro}
Understanding the formation and evolution of our Galaxy, of Andromeda, and of other disk galaxies is one of the main quests of modern astrophysics.
Over the last decade, large spectroscopic surveys have constrained quantities such as the ages, element abundances and phase-space properties of the stars in the Milky Way, mostly in the proximity of the Sun but also at several kpc distance, throughout the disk, bulge and stellar halo. These include LAMOST \citep{2012Deng}, RAVE \citep[][and references therein]{2020RAVE}, SEGUE/SDSS \citep{2017SEGUE}, APOGEE \citep{2017Majewski}, GALAH \citep{2017Martell}, H3 Survey \citep{2019Conroy} and
finally GAIA \citep{2016Gaia}, with the delivery of positions and proper motions for more than 1.4 billion stars in the third data release \citep{2022Gaia}. Similarly, albeit from a distance of about 750 kpc, photometric and spectroscopic surveys like PHAT \citep{2012Dalcanton} and SPLASH \citep{2009Gilbert} have mapped large portions of the disk of Andromeda and its disk-halo interface.

\subsection{The stellar disk of the Galaxy}
A remarkable feature uncovered over the past few years about the stellar disk of our Galaxy is the existence of two different stellar populations in the solar neighbourhood: on the one hand, alpha-rich and metal-poor stars seem to associate well with the \textit{geometrical} or \textit{morphological} ``thick'' disk, with scaleheight of $\sim$ 600-1400 pc and with old and kinematically-hotter stars; on the other hand, metal-rich stars with lower, i.e. solar [$\alpha$/Fe] abundances are thought to populate the \textit{geometrical} or \textit{morphological} ``thin'' disk, with scaleheight of $\sim$ 150-350 pc and characterised by young and kinematically-colder stellar populations \citep[e.g.][the latter for a compilation of several measurements]{1983Gilmore, 2008Juric, 2012Adibekyan, 2013Haywood, 2016Bland}.

Far away from the solar neighbourhood, the vertical distribution of the chemical composition, positions and kinematics of disk stars in our Galaxy are more uncertain and, possibly, more complicated. Assuming that stellar chemistry is a good proxy for stellar ages \citep{1980Twarog}, the concordance picture posits that the thick disk of the Milky Way (and possibly of other spiral galaxies) is mainly composed of old stars, whereas young stars dominate the thin disk. This is consistent with the so-called ``inside-out'' and ``upside-down'' scenario, whereby at early times stars were born in a radially-compact but vertically-thick disk and, later on, a thin and more extended disk developed \citep[e.g.][for observational inferences]{2014Robin, 2016Bovy} and \citep[][for numerical modeling]{2013Bird, 2013Stinson, 2014MinchevAA, 2020Buck, 2021Agertz, 2021Nelson, 2021Bird, 2022Yu}.

However, observations of our Galaxy have suggested that two simultaneous facts are in place at larger galactocentric distances than 8 kpc: on the one hand, the scaleheight of the Galactic stellar disk increases at larger galactocentric distances, a phenomenon called \textit{flaring} (see below; first observed in the HI disk); on the other hand, the Milky Way's geometric thick disk, here denoted as stars at large heights over the disk plane ($\gtrsim$ 2 kpc), also contains young stars  \citep{2016Ness,2017Xiang,2018Xiang,2019Feuillet}. 

\subsection{Disk flaring in the Galaxy}
The flaring of the stellar disk of our Galaxy has been studied quantitatively, namely by inferring from observations the changes in stellar disk height with galactocentric distance. 
Studies have found that the stellar disk of the Galaxy is flared in the outskirts \citep{1998Evans, 2000Alard}, but this phenomenon is unlikely to be present in the inner disk \citep{2018Mateu}. Flaring may be more appreciable when the disk stars are dissected into mono-age and/or mono-abundance populations \citep[e.g.][]{2013Stinson, 2017Minchev}, populations that are assumed to be roughly equivalent. There is strong empirical evidence also for the flaring of the Milky Way's  (low-alpha) disk \citep{2019Ness}. For example, \cite{2016Bovy} binned APOGEE stars in mono-abundance populations and quantified the changes in stellar disk height between 4  and 15 kpc from the Galactic Center.
They found that the high-[$\alpha$/Fe] population -- mainly associated with old stars -- does not show any evidence of flaring, whereas low-[$\alpha$/Fe] stars -- associated with young populations -- present clear evidence of a flaring, with scaleheights exponentially increasing as a function of galactocentric distance.
\cite{2017Mackereth} binned APOGEE RGB stars between 3 and 15 kpc from the Galactic Center in mono-age \& mono-[Fe/H] populations and reached similar albeit not identical conclusions as above (and ones shared by \citealt{2017Minchev}): all mono-age populations flare although to different levels -- the radial profiles of the scaleheight of high-[$\alpha$/Fe] stars are generally flatter, whereas the low-alpha populations flare more strongly, albeit mostly linearly.
Finally, \citet{2019Ting} studied the vertical motion of low-alpha disk stars via their vertical actions, and published an analytical function for the mean vertical action of stars at given age and radius. Their findings imply a manifest flaring of the young stellar population (3 Gyr) with scaleheights of $120$ and $500$ pc at 4 and 14 kpc, respectively (see \S\ref{sec:TNG50flaring}).

These recent results point to a consistent picture for the Galaxy, at least, and only when mono-age or mono-abundance stellar populations are analyzed separately: young stars in the Galaxy exhibit some level of flaring, at least at radii $\gtrsim10-11$ kpc \citep[see otherwise][]{2018Mateu}.
However, as of today, some confusion and uncertainties remain as to how different levels of flaring may map into the distributions of stellar ages in the height vs. radial distance plane and as to how the flaring of different mono-age stellar populations translates into the flaring of the morphological thin and thick disks. Finally, it remains unclear whether the phenomenology in the Milky Way is representative of most spiral galaxies or not.

\subsection{Thin and thick Galactic disks}
Whether the morphological or geometrical thin and thick disks of the Galaxy are two distinct components, or just the manifestation of a single variable structure, is also a matter of debate. The former is the classical view, as described for example in \citealt{2008Juric}, whereas recent analyses \citep[e.g.][]{2013Bovy,2013rix} argue for the latter, with the vertical structure of the Galactic disk being a continuum of stellar populations. The claim is that, even if the vertical stellar mass density profile is well described by a double exponential fit or similar, this does not necessarily imply a physically-originated decomposition. It is because of these arguments that it is now customary to characterize the vertical stellar disk structure in terms of mono-age or mono-abundance stellar populations, which should promise clearer physical insight. 

\subsection{Flaring of Andromeda and other spiral galaxies}
Also the stellar disk of Andromeda seems to be well described by a double vertical component, with thin and thick disks separating in both kinematics and metallicity \citep{2011Collins}, but with scaleheights approximately two to three times larger than those observed in the Milky Way: about 0.9-1.3 and 2.2-3.4 kpc,
respectively. On the other hand, the level of disk flaring in Andromeda remains unclear and, de facto, unaccessible. On the other hand, disk flaring has been suggested by observations in a number of edge-on spiral galaxies \citep{1997deGrijs, 2002Narayan, 2016Kasparova, 2019Rich, 2019Sarkar}.

\subsection{Disk flaring with theoretical and numerical models}
\label{sec:theoretical}
The formation of thick disks and disk flaring are two closely linked phenomena. Several processes have been suggested to be responsible for the thickening of the stellar disk in general or for the disk flaring in particular: radial migration, accretion of stars from satellites, heating of a thinner pre-existing disk through mergers, and in-situ star formation from gas-rich mergers.

The thickening of the stellar disk with galactic radius has been suggested to be a natural consequence of radial orbit migration by \citet{2009Sales, 2009SchonrichBin, 2011Loebman, 2013Roskar}. 
For example, \citealt{2012Minchev} suggest that radial migration leads to the thickening in the outer disk  while having the opposite effect in most part of the disk, leading to a significant effect on flaring. On the other hand, other studies support that radial migration does not contribute to the thickening of stellar disks \citep{2014Martig, 2014VeraCiro, 2016VeraCiro, 2014MinchevAA, 2016Grand}
and thus that the effect of radial migration on disk flaring cannot be large. Flaring could also be a consequence of heating caused by external triggers, i.e. not because of secular processes but rather events such as the infall of satellites or the interaction with other flying-by galaxies \citep[e.g.][with N-body only models of disks bombarded by cosmologically-consistent subhaloes]{2009Kazantzidis}.

Cosmological hydrodynamical simulations of well-resolved Milky Way (MW)-like galaxies, which have become increasingly realistic over the last decade \citep[see e.g.][for Eris, LATTE, the Auriga sample, and VINTERGATAN, respectively]{2011guedes,2016Wetzel, 2017Grand, 2021Agertz}, have also allowed to address the question of disk flaring. 

\cite{2015Minchev}, analysing two simulated galactic disks formed in a cosmological context -- one from \citealt{2012Martig} and one from \citealt{2013Aumer} --, demonstrated that a non-flaring thick disk can actually be in place even if 
several mono-age populations with different levels of flaring are superposed -- a statistical phenomenon commonly known as \textit{Simpson's paradox}: a trend can appear in several groups of data but disappear or reverse when the groups are combined.
By using a cosmological zoom-in simulation from the FIRE project of a MW-mass galaxy ($\MS \simeq 6\times10^{10}\Ms$ at $z=0$), \cite{2017Ma} found that the scaleheight of mono-age stellar populations 
shows an outward and somewhat linear flaring, being higher at larger galactocentric distances. However, differently from \cite{2015Minchev}, in the FIRE galaxy the scaleheights of both the thin and thick disks are found to be flared, with nearly the same slope of the mono-age populations. 
Also in all the 30 Auriga MW-analogs \citep{2017Grand}, an exponential flaring is a common feature: the flaring is in place for both young stars ($<$ 3 Gyr) and the whole stellar populations, even though by different amounts. Indeed, by fitting the flaring with an exponential trend, \cite{2017Grand} found that, in the majority of the Auriga galaxies, young stars show a higher degree of flaring with respect to the global ones.

With one of the APOSTLE cosmological hydrodynamical simulations, \cite{2018Navarro} demonstrated that the stellar-disk flaring reflects the flaring of the gaseous disk -- as stars inherit the properties of the gas at their birth -- and argued that the age and metallicity gradients are settled at birth and are not the result of radial migration or disk instabilities.

The flaring of mono-age populations is non negligible in all the five NIHAO-UHD MW-like galaxies \citep{2020Buck} and in the MW-mass disk galaxy VINTERGATAN \citep{2021Agertz}. However, some of them flare linearly, others flare with an exponential radial trend of the heights. Moreover, the increasing of the scaleheight in the NIHAO-UHD sample is found to be much stronger for the old stellar populations, unlike the case of the Galaxy, and mild-to-no flaring is appreciable when all stellar populations are combined, similarly to the cases by \cite{2015Minchev}. 

Finally, more recently, \cite{2021delacruz} expanded the work of \cite{2015Minchev} by showing the vertical structures of 27 MW-like galaxies with $\MS \simeq 10^{10}-2\times10^{11}\Ms$ at $z=0$: they found that in 44 per cent of their galaxies, the morphological thick disk does not flare and this typically occurs in galaxies with $\MS < 5\times10^{10}\Ms$, with a thin disk ($<$ 1kpc) and a rather quiescent merger history. On the other hand, the remaining 15 galaxies show a flared thick disk and they are more massive, have a thicker disk and have undergone a major merger with respect to their non-flaring counterparts.

Despite the many recent results on the topic put forward by the simulation community, the scientific and general interpretation of the findings above, and their applicability to the cases of the Galaxy or Andromeda, are impeded by a number of limitations. 
Firstly, most of the analyses based on state-of-the-art cosmological models remain qualitative and refer to one or just a few galaxies formed within a specific galaxy-formation model: namely, they are often reduced to the plotting of the stellar scaleheights (and/or vertical stellar velocity dispersion) as a function of radius for stars in e.g. different age bins, and are associated to only one or just a few specific realizations of galaxies that span a limited range (if any) of mass, merger history and stellar disk structure. %
Secondly, when the study of more than a handful of objects is possible, the quantification of the flaring is not consistently derived across the analyses, making the comparison of the predicted outcome problematic.

\subsection{TNG50 and the scope of this paper}
In this paper, we use the most recent and highest-resolution simulation of the IllustrisTNG project \citep{2018Pillepich-b, 2018Nelson, 2018Marinacci, 2018Naiman, 2018Springel}, TNG50 \citep{2019Pillepich,2019Nelson-b}, and quantify the stellar disk flaring of 198 MW- and M31-like galaxies, thereby tripling the number of cosmologically-simulated galaxies analyzed to this end. This is possible thanks to the mass and spatial resolution of the simulation, which returns galaxies with disks as thin as $100-200$ pc \citep{2019Pillepich, 2022SotilloRamos}, and to the encompassed volume, with realistic galaxy properties and galaxy populations across a wide range of masses, types, and environments i.e. not only for the case of disk, star-forming galaxies that form in $10^{12}\Ms$ haloes.

We hereby focus on the vertical distribution of the stellar mass in disks at $z=0$ and on its connection to the vertical stellar velocity dispersion. We assess the flaring both for the morphological thin and thick disks (i.e. when single-component vertical fits are not appropriate to obtain scaleheights) and especially by separately studying mono-age stellar populations. We again postpone to future work the study of how the latter connect to mono-abundance populations in the context of the IllustrisTNG model and enrichment, but we give particular emphasis to whether and how often --i.e. across the selected galaxy sample --, young disk stars flare more or less than old disk stars, and we explore the relationship between the degree of the flaring and $z=0$ galaxy and disk properties.

In Section \ref{sec:methods}, we hence summarize the salient aspects of the TNG50 simulation, describe the adopted selection of MW/M31-like galaxies, and define the ways we characterize the simulated stellar disks. In Section \ref{sec:properties} we show the range of stellar-disk structures encompassed by the TNG50 MW/M31 analogs, including their scaleheights as a function of radius and the cases of warped and disturbed stellar disks.
We quantify the vertical disk structure and flaring predicted by TNG50 for MW/M31-like galaxies in \S\ref{sec:TNG50flaring}. There we also argue for, and propose, a non-parametric and more-generally applicable and comparable method to quantify the amount of the disk flaring and compare the flaring of stars of different ages and to the inferences for our Galaxy. We connect stellar disk heights to the underlying stellar kinematics and potential in \S\ref{sec:kinematics}. In Section ~\ref{sec:discussion}, we quantitatively compare the TNG50 results to those from previous simulations, by casting them all under the same general and non-parametric flaring quantification; we discuss our results, limitations, and the possible origin of the diversity predicted by TNG50, and connect to observations of the distributions of the stellar ages as a function of galactocentric radius and height. Summary and conclusions are given in Section \ref{sec:conclusions}.

\section{Methods}
\label{sec:methods}

\subsection{The TNG50 simulation}
\label{sec:TNG50}

The TNG50 simulation is, among the flagship runs of the IllustrisTNG project \citep{2019Nelson-b}, the smallest in volume but best in resolution: it evolves a cubic box of $\sim$ 50 comoving Mpc a side, sampled by 2160$^3$ dark-matter particles and 2160$^3$ initial gas cells \citep{2019Nelson-b,2019Pillepich}, with a resulting gas-cell and stellar-particle mass resolution of about $8.5\times10^4 ~ \Ms$ and a dark-matter mass resolution of about $4.5\times10^5 ~ \Ms$.  

TNG50 uses the code \textsc{Arepo} \citep{2010springel} and includes the IllustrisTNG galaxy-formation model introduced and described in the method papers by \citealt{2018Weinberger, 2018Pillepich-a}: in practice, it solved for the coupled equations of gravity and magneto-hydrodynamics in an expanding Universe, in addition to prescribing the cooling and heating of the cosmic gas, star formation, stellar evolution and enrichment, as well as phenomena such as stellar feedback and the seed, growth and feedback from supermassive black holes (SMBHs). The initial conditions of TNG50 have been initialized at redshift $z=127$ and assume a cosmology compatible with the Planck 2015 results \citep{2016planck}. 

As in previous large-scale and zoom-in cosmological simulations of MW-mass galaxies, also in TNG50 stellar particles do not represent individual stars but rather simple, mono-age  stellar populations of thousands of stars characterized by an initial stellar mass function \citep[][for TNG50]{2003Chabrier}. On the other hand, a few modeling elements set apart TNG50 from the great majority of cosmological simulations that have been used so far to study the vertical structure and flaring of galactic disks: chiefly, the inclusion of magnetic fields and the effects of SMBH feedback \citep[both also in Auriga,][]{2017Grand}. Importantly, TNG50 is a relatively large uniform-volume simulation and so, differently than in many of the aforementioned zoom-in simulations, it returns a large number of massive galaxies ($\simeq$ 800 at $z=0$ above $10^{10}\,\Ms$) and hence, among them, also many MW and M31-mass objects spanning a wide range of merger histories, i.e. {\it without} any a-priori choice about the number and time of their past major mergers \citep{2022SotilloRamos}.

TNG50 is suitable for studying disk flaring thanks to its mass and spatial resolution \citep[see][for more details]{2019Nelson-b, 2019Pillepich, 2021Pillepich}. The smallest gas cell in TNG50 at $z=0$ measures 9 pc across, whereas the average gas cells within the star-forming regions of massive galaxies at $z=0$ are typically of the order $50-200$ pc: this means that processes such as star formation and feedback are implemented below such spatial scales. The gravitational potential, on the other hand, is softened on different scales for different types of resolution elements: the softening length of the stellar and DM particles read 288 pc, the smallest softening length of the gas cells is 72 pc. These are sufficient to capture half-light disk heights of $200-400$ pc for the typical massive star-forming galaxy at low redshift, but also thinner ones (i.e. thinner than the softening length for stellar particles; see \citet[][]{2019Pillepich} and next Sections). \citet[][]{2019Pillepich} presents also a study of the resolution effects on galaxy sizes and heights in TNG50: stellar disk thickness (as stellar half-mass height) can be considered to be converged in TNG50 to better than 20--40 per cent.

Finally, the choice and functioning of the IllustrisTNG model as implemented in TNG50 have been validated against observations not only of the Galaxy or Andromeda, but of large galaxy populations \citep[see][for a summary]{2021Pillepich}. 

\subsection{Galaxy selection: choosing MW and M31 analogs}
\label{sec:selection}
In the following, we identify (sub)halos within the TNG50 volume by using the Friends-of-Friends (FoF) and \textsc{SubFind} algorithms \citep{1985Davis,2001springel}.
Also, we define a ``virial'' halo boundary, $R_{200c}$, as the radius within which the mean enclosed mass density is 200 times the critical density of the Universe. We refer to the total mass enclosed within this radius as the virial mass, $M_{200c}$, of the host halo.
Additionally, all galaxies residing within one virial radius of the host center are dubbed as ``satellite'' or ``galaxy'', whereas the 
galaxy settled at the deepest potential within a FoF is named ``central'', and it is typically, but not always, the most massive one.
The galaxy stellar mass adopted in this work ($\MS$) is the sum of all stellar particles within a fixed aperture of 30 physical kpc, unless otherwise stated.

With these definitions in mind, we select MW/M31-like galaxies from TNG50 at $z=0$ by means of the following three criteria, all in turn based on {\it observable} rather than {\it halo-based} properties. Extended motivations and characterizations for this selection are given in \textcolor{blue}{Pillepich et al. in prep.} and the resulting sample has been already used in \citet[][]{2021Engler,2022Engler,2021Pillepich,2022SotilloRamos, 2023Chen, 2023Ramesh}). 

Namely, at $z=0$ we select galaxies:

\begin{itemize}
    \item with $\MS$ ($<$30 kpc) = $10^{10.5-11.2} \Ms$;
    \item with a disky stellar shape (see Section \ref{diskyness} for more details)
    \item in isolation: no other galaxies with $\MS >10^{10.5} \Ms$ within 500 kpc and host halo mass $M_{200c}<10^{13}\Ms$.
\end{itemize}

This leads to a sample of 198 TNG50 MW/M31-like galaxies, and since it is not required for a galaxy to be the central of its halo, our sample also includes pairs of a few Local Group-like systems.
Additionally, we note that, differently from the majority of zoom-in simulations of MW-mass haloes \citep[see for example][]{2011guedes,2016RocaFabrega,2021Agertz,2021Renaud}, our sample is not a-priori \textit{biased} in its history, namely, we have not imposed a recent quiescent merger history for the MW/M31-like galaxies to be part of our sample (see \citet{2022SotilloRamos} for an in-depth analysis of the merger history of all TNG50 MW/M31 analogs).

Throughout the text and in some selected Figures, we at times label \textit {MW-mass} and \textit {M31-mass} those galaxies within the MW/M31-like sample with stellar mass in the ranges $10^{10.5-10.9}\Ms$ and $10^{10.9-11.2}\Ms$, respectively. 
MW-mass (M31-mass) galaxies are sampled, on average, by {$\simeq$ $5.5 \times10^5$ ($1.3 \times10^6$)} gravitationally-bound star particles. 

\subsubsection{Stellar morphology selection}
\label{diskyness}
As stated above, each galaxy in the TNG50 MW/M31 sample has been selected to be \textit{disky}, i.e. they either satisfy a stellar-morphology constraint based on the minor-to-major axis ratio ($c/a$) of the stellar mass distribution or are disky by visual inspection. The first criterion is satisfied if $c/a\ge0.45$ \citep[see][for more details]{2019Chua,2019Pillepich}, being $c$ and $a$ the minor and major axis of the ellipsoidal distribution of stellar mass between 1 to 2 times the stellar half-mass radius ($\RS$).
Additionally, in the TNG50 MW/M31-like sample are also present galaxies that, even if with $c/a>0.45$,  clearly appear disky and with well-defined spiral arms by visual inspection, based on 3-band images in face-on and edge-on projections. Of 198 TNG50 MW/M31-like galaxies, 25 have been included via the visual-inspection step. See \textcolor{blue}{Pillepich et al. in prep} for more details.

\subsection{Measurement of stellar disk properties}

\subsubsection{Definition of disk stars}
Throughout this paper, we quantify the structures of the simulated galactic stellar disks based on mass, i.e. based on the spatial location of stellar particles and the  stellar mass densities they sample. 

We call disk stars all the gravitationally-bound (according to \textsc{SubFind}) stellar particles that are in circular orbit upon inspection, i.e. with circularity $\epsilon = L_z/L_{z,\rm{circ}}>0.7$. Here $L_z$ is the z-component of the angular momentum of a given star particle and $L_{z,\rm{circ}}$ is the angular momentum of a star located at the same radius but following a perfectly circular orbit.
The z direction for each galaxy (its ``up vector'') is chosen to be the direction of the total angular momentum of all stars within $2\times \RS$. The galactic plane is hence the plane perpendicular to this up vector. The center of a galaxy is chosen as the location of its most gravitationally-bound element, typically the location of its SMBH.

\subsubsection{Stellar disk lengths}
\label{sec:fit_procedure}
We measure the disk length, for any given galaxy, selecting all stellar particles with circular orbits (i.e. disk stars, $\epsilon > $0.7) between one and four times the half-mass radius (i.e. excluding the bulge region). We fit an exponential profile to the radial stellar surface density distribution in face-on projection, in bins of 2 kpc:
\begin{equation}
\Sigma(R) =\Sigma_{\rm d} \, \rm exp \left(-\frac{R}{R_{\rm d}}\right),
\end{equation}
where $\Sigma_d$ is the stellar mass surface density of the disk at $R=0$ and $R_{\rm{d}}$ is the disk scalelength, a characteristic scale that is commonly used as a proxy for the extension or the size of the stellar disk.

We perform 100 fits for each galaxy, by starting with random initial values around the values taken at the limits of the cylindrical shell. The fitting routine (python curve\_fit) uses a non-linear least squares method to fit our defined function to the data. The best measure of the scalelength of a given galaxy is then the mode of the distribution; an error can be obtained as the interquartile range.

\subsubsection{Stellar disk heights}
\label{sec:fit_procedure_heights}
Given the up-vector for each galaxy, we analyze its vertical stellar disk structure, by rotating its stars in edge-on projections and by extracting and fitting the vertical stellar mass density distribution of the disk stars. We determine the latter at various different radii, centered at integer multiples, from 1 to 5 in steps of 0.5, of the the scalelength of the galaxy, $R_{\rm{d}}$, and dividing the galactic disk into radial annuli (cylindrical shells) of width one integer. The latter are also centered at multiple physical radii, in cylindrical shells of 2 kpc.

We use either a single or a double parametric formula to fit the vertical mass profiles at fixed gactocentric distance. In the literature, there are a variety of formulas to describe these profiles: the most common are exponential, hyperbolic secant and squared hyperbolic secant. All three can be seen as special cases of the general formula \citep[see, e.g.,][]{1988vanderKruit}:
\begin{equation}
\label{eq:general}
f(z) \propto \rm{sech}^{2/n}(nz/2h_z),
\end{equation}
for the cases $n\rightarrow \infty, n=2$ and $n=1$, respectively. All three tend to the exponential profile as $z$ increases, being the shape of the profile for low values of $z$ the main difference among the three.

The single squared hyperbolic secant profile reads 
\begin{equation}
\label{eq:single}
\rho(z) =\rho_0 ~\rm{sech^2}(z/2h_z) ~,
\end{equation}
where $\rho_0$ is the normalization and $h_z$ is the disk scaleheight. The factor 2 in the denominator allows that the scaleheights of the exponential, and of the linear and squared hyperbolic secant cases, are comparable in magnitude. As shown in the upcoming sections, this provides a good description of the vertical mass distribution of mono-age stellar populations, for all selected galaxies. So the fit of Eq.~\ref{eq:single} is the one we adopt to quantify the stellar disk scaleheight of mono-age stellar populations. 

However, when all disk stars in a galaxy are considered, a two-component vertical formula returns a better fit for the majority of TNG50 MW/M31-like galaxies: 
\begin{equation}
\label{eq:double}
\rho(z) =\rho_{\rm{thin}} ~\rm sech^2(z/2h_{thin})+\rho_{thick} ~\rm sech^2(z/2h_{thick}) ~.
\end{equation}

This gives us the scaleheights of both a ``thin'' ($h_{z1}$) and a ``thick'' ($h_{z2}$) disk component \citep{1983Gilmore,2006Yoachim,2011Comeron741,2012Comeron,2017Ma,2020Buck,2021Agertz,2018Navarro}.
We stress here that the division into a thin and thick disk of our TNG50 MW/M31-like galaxies is purely geometrical, i.e. morphological, and does not necessarily imply a meaningful physical decomposition into two separate structures \citep[see][]{2013Bovy}.

We choose to proceed with the single and double sech$^2$ formula because it has been extensively used: e.g. by \citet{2006Yoachim, 2014Bizyaev} with observed edge-on spiral galaxies and \citet{2008VillalobosHelmi, 2013Stinson, 2017Ma, 2021Park} with simulated galaxies. The main justification is the physical motivation: it represents the vertical density variation of a self-gravitating iso-thermal population \citep{1942Spitzer, 1981vanderKruit}. However \citet{1997deGrijs, 1999Hammersley} claim that it never reproduces well the densities of the Galactic midplane.
In fact, other works have preferred the exponential formula: it has been used as fit of the vertical stellar density profile of the Galaxy \citep{1983Pritchet, 2002Siegel, 2008Juric, 2016Bovy, 2017Mackereth} and of other galaxies \citep{2011Comeron729}; it has also been used for simulations in the works by \citet{2016RocaFabrega, 2020Buck, 2021Agertz}.
\citet{1988vanderKruit} proposed that an intermediate solution, i.e. a sech profile, would reproduce better the stellar vertical densities in the MW. It is adopted in the works by \citet[][]{1997deGrijs, 2000Matthews}. 

Similarly for what is done for the scalelengths and in \citealt{2022SotilloRamos}, we perform 100 fits for each galaxy, galactocentric distance, age bins, etc. The initial guess fit values are in the ranges 20 to 1000 pc and 800 to 7000 pc for the thin and thick disks when using the double function, and 100 to 7000 for the single function. We choose the mode of the distribution as the best measure of the scaleheight and quote errors as one standard deviations of the estimated parameters, provided by the fitting function.


\section{The structural and age properties of the stellar disks in TNG50 MW/M31-like galaxies}
\label{sec:properties}

Before quantifying the stellar disk flaring according to TNG50, we first comment on the structural properties of the stellar disks of the 198 TNG50 MW/M31-like galaxies -- additional global and structural properties can be found in \textcolor{blue}{Pillepich et al. in prep.} and references therein. An extensive analysis of the merger history of each MW/M31-like galaxy, and on how stellar disks can survive major mergers, is instead given in \citealt{2022SotilloRamos}.

\begin{figure*}
\centering
\includegraphics[width=0.33\textwidth]{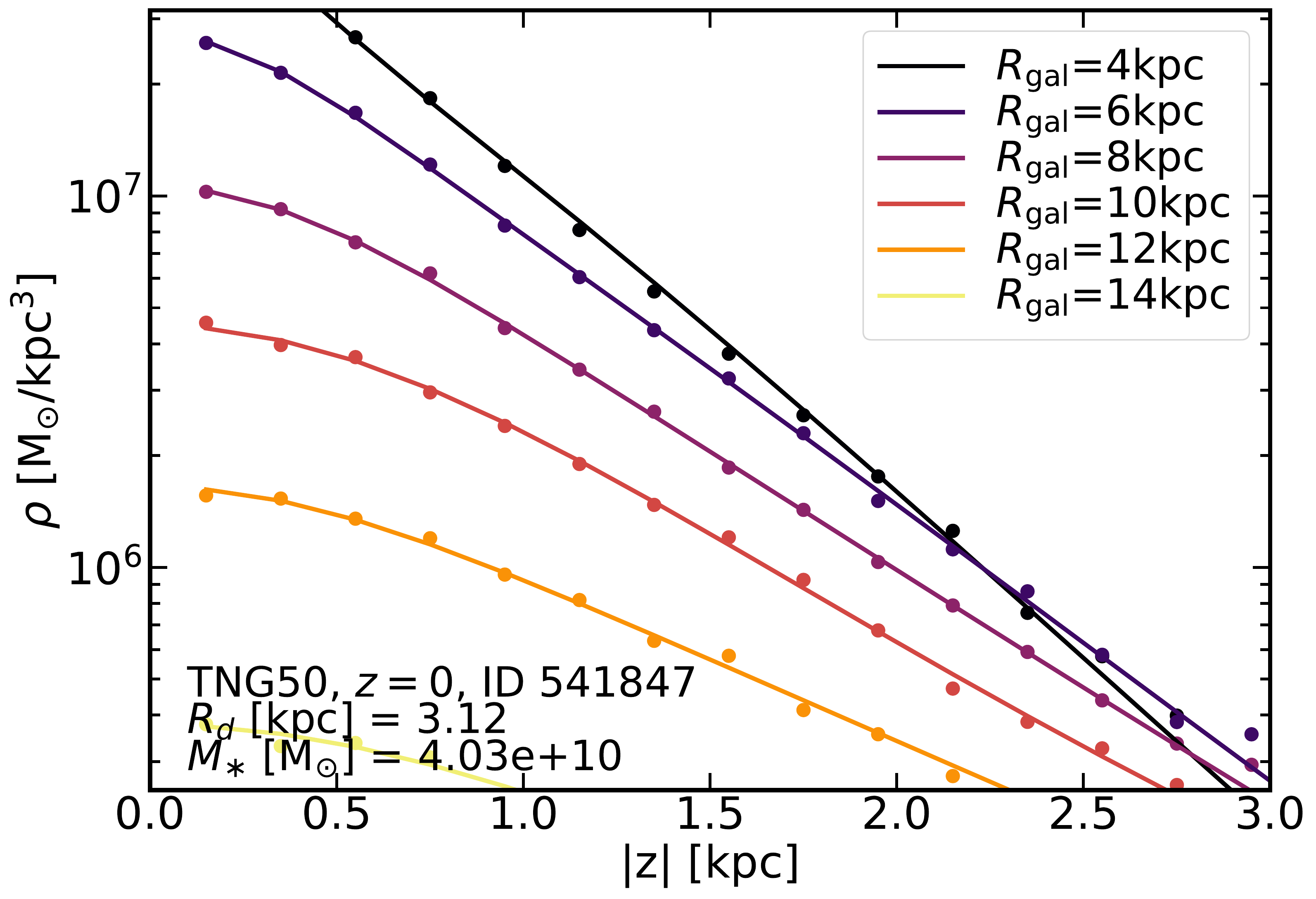}
\includegraphics[width=0.33\textwidth]{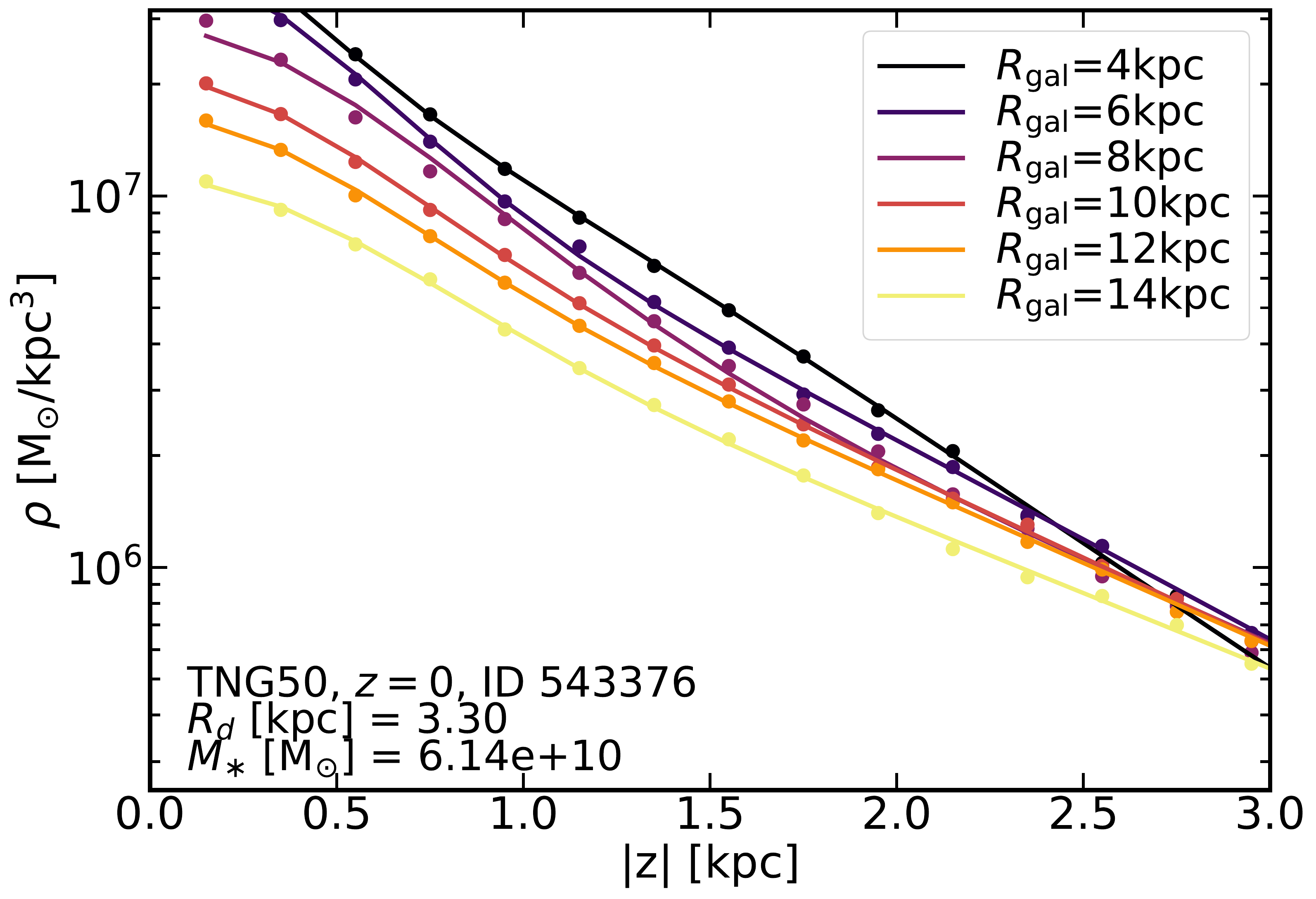}
\includegraphics[width=0.33\textwidth]{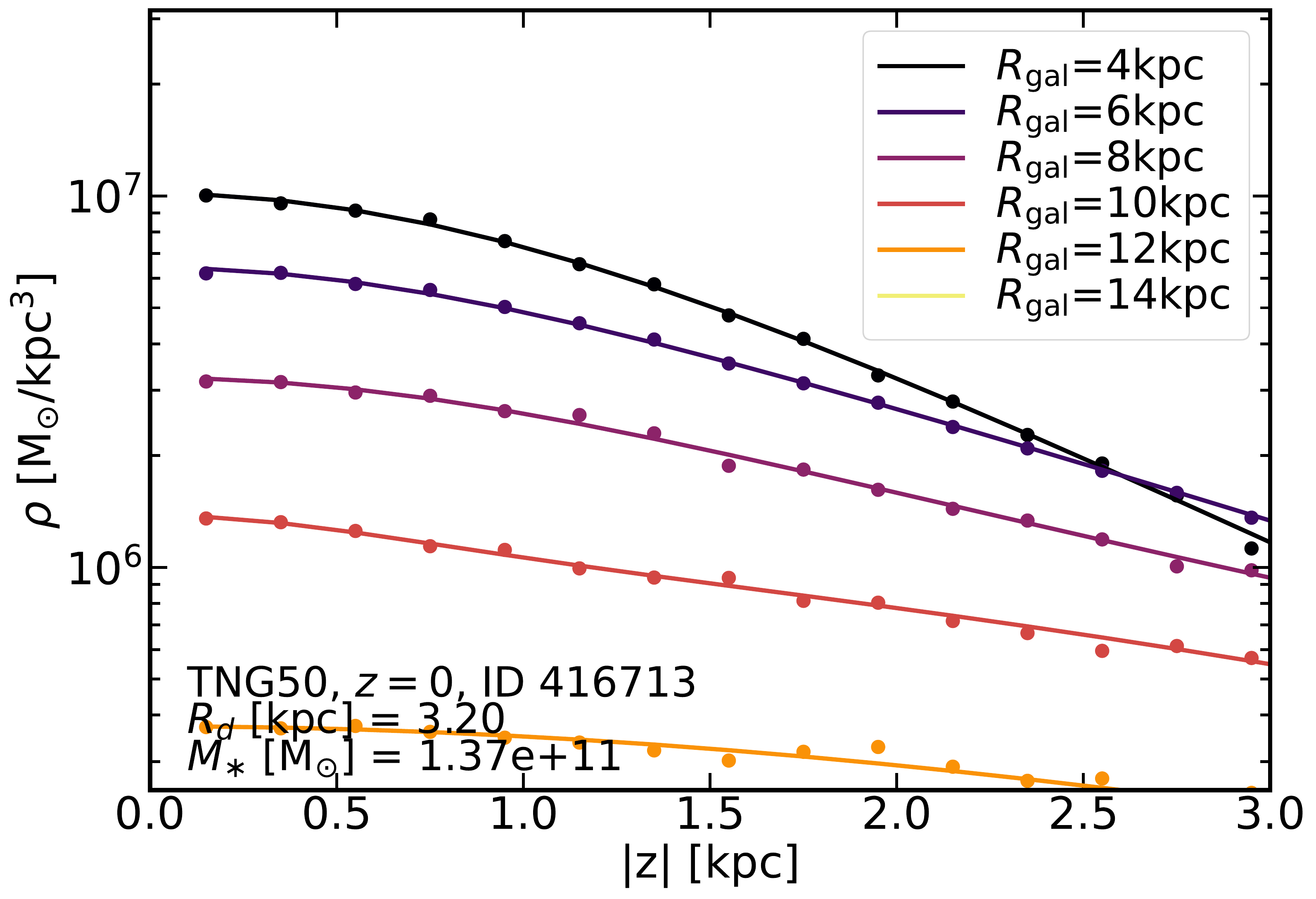}
\includegraphics[width=0.33\textwidth]{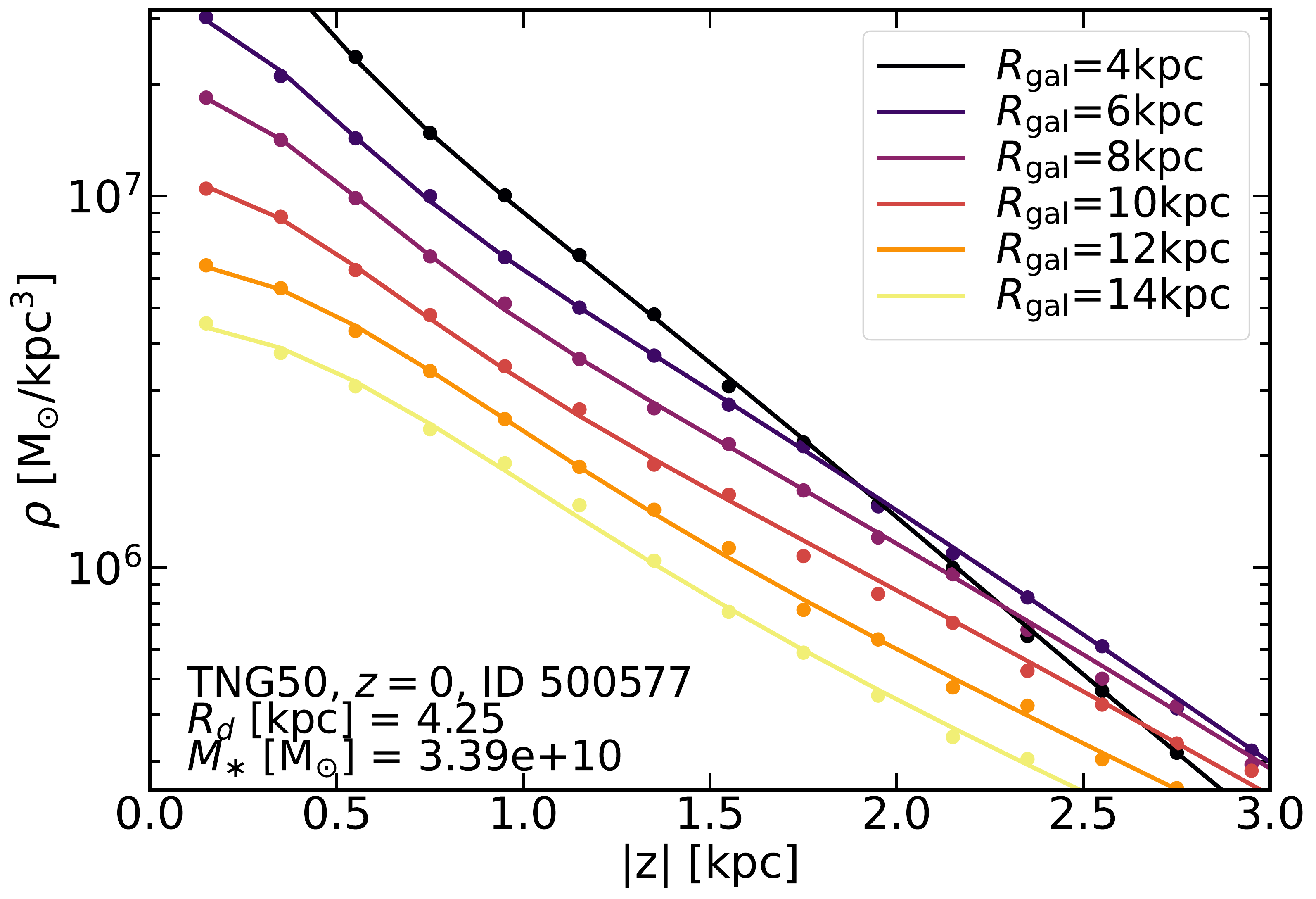}
\includegraphics[width=0.33\textwidth]{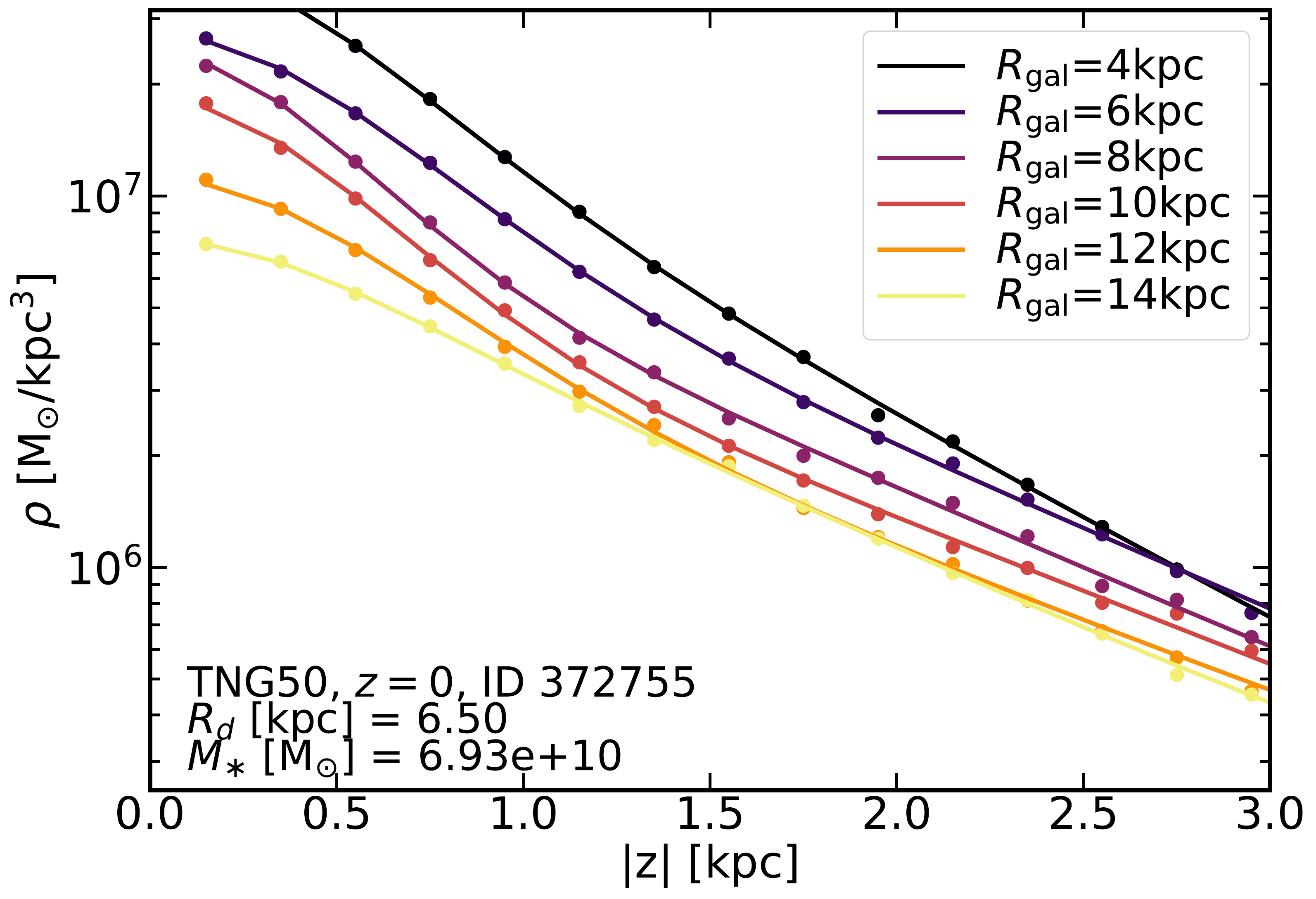}
\includegraphics[width=0.33\textwidth]{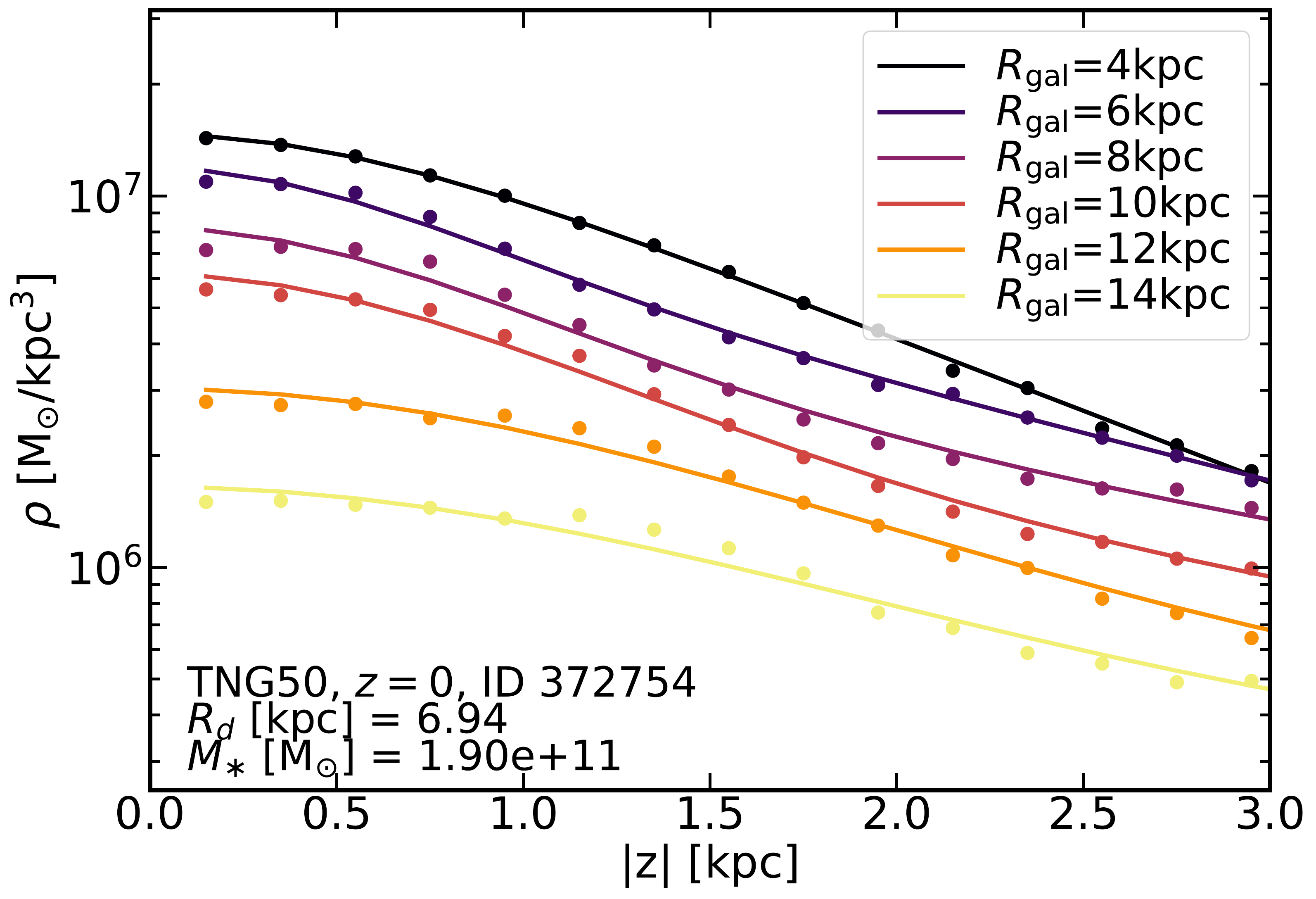}
\includegraphics[width=0.33\textwidth]{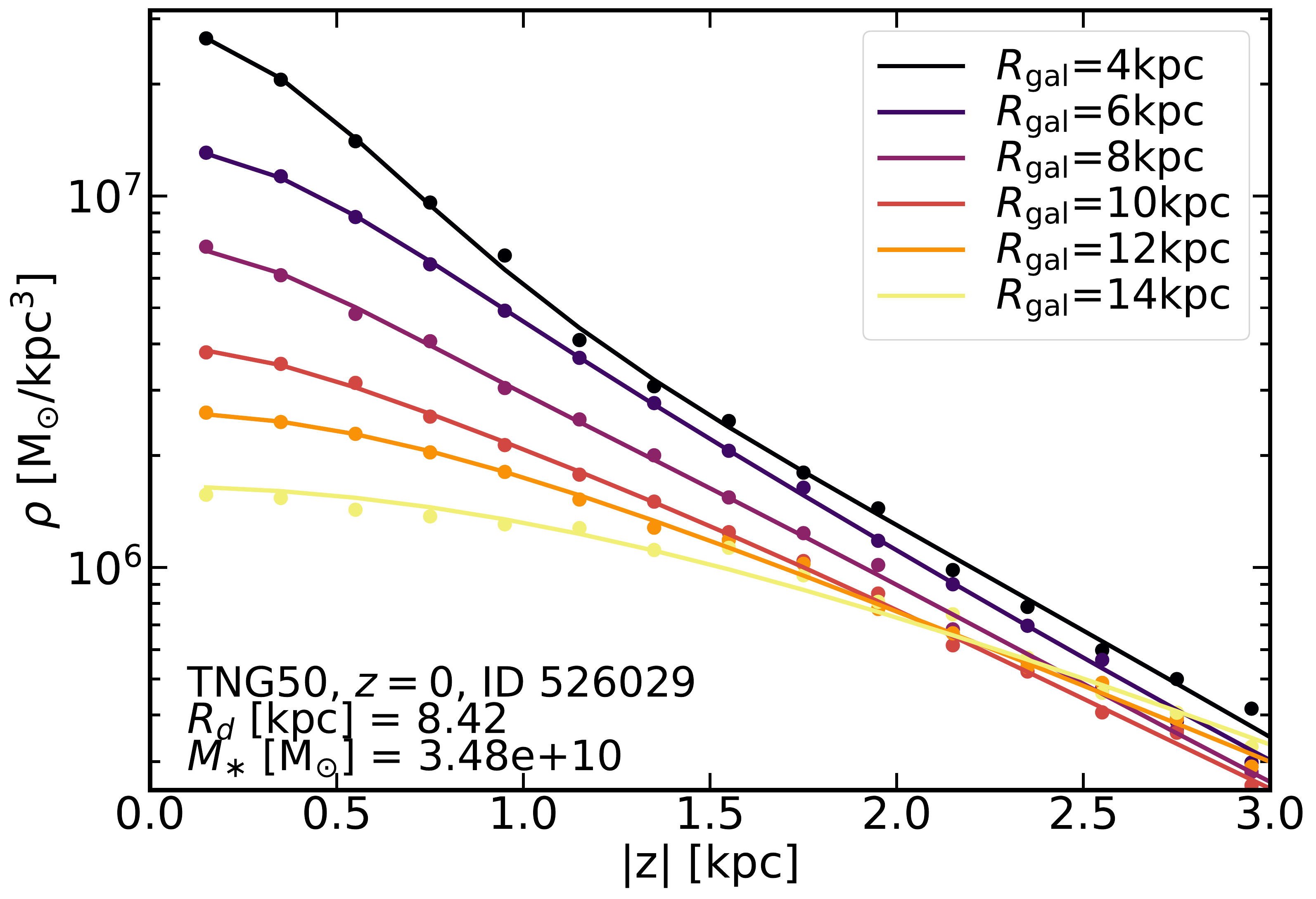}
\includegraphics[width=0.33\textwidth]{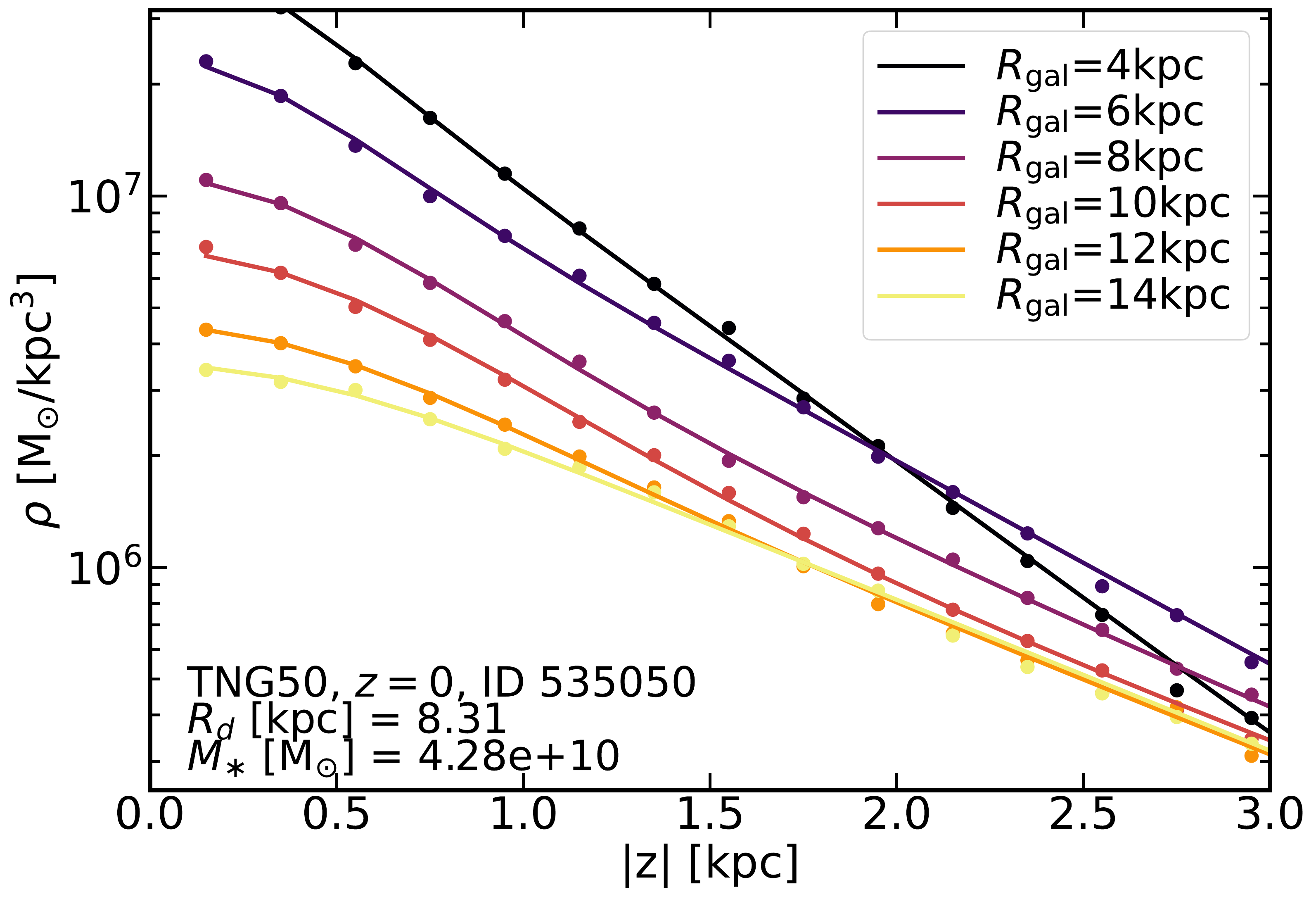}
\includegraphics[width=0.33\textwidth]{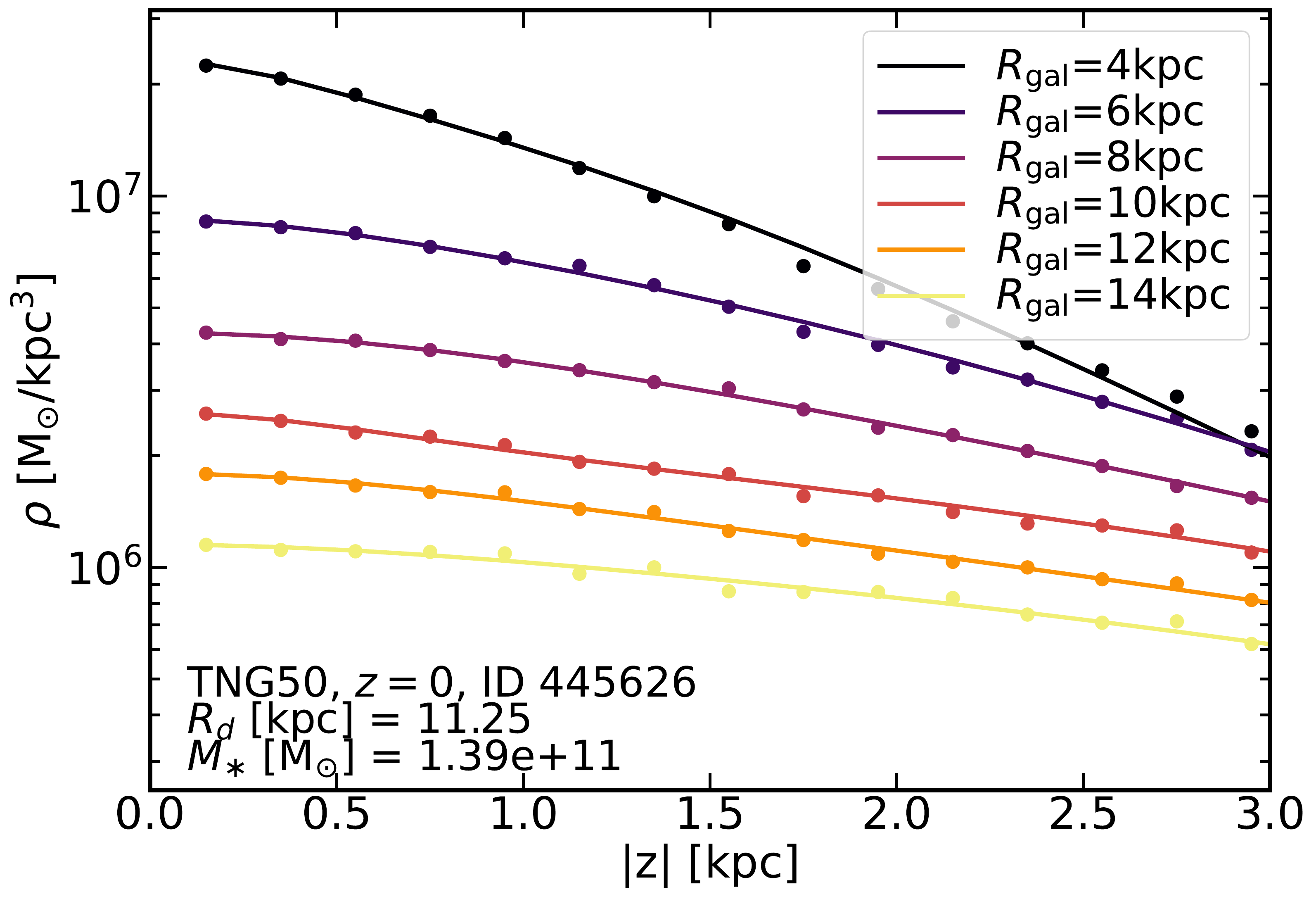}
\caption{\label{fig:vertical_profile} \textbf{Vertical surface density profiles of example TNG50 MW/M31-like galaxies.} In each panel, we show the vertical stellar surface mass density profiles at different radii for one of nine TNG50 MW/M31 analogs. Solid curves represent the two-component fit, as described in the text (see Eq.~\ref{eq:double}). These galaxies are selected to be representative of the whole sample, below the 25$^{th}$ percentile, in between 25$^{th}$ and 75$^{th}$ percentile and above the 75$^{th}$ percentile, of both galaxy stellar mass (left to right columns) and scalelength (top to bottom rows).}
\end{figure*}

\subsection{Diversity of stellar disk lengths and heights}
\label{sec:StellarSizes}

An in-depth analysis of the global structural properties of the stellar disks of the 198 TNG50 MW/M31-like galaxies can be found in \textcolor{blue}{Pillepich et al. in prep.} We refer the reader to that work for details, whereas here we report the most relevant facts. 

TNG50 predicts a wide range of stellar disk sizes, also at fixed stellar mass. Within the TNG50 MW/M31 sample, the stellar disk scalelengths vary between $\sim 1.5$ and $\sim$17 kpc, denoting a remarkable variety of disk extents in such a narrow range of stellar mass (\textcolor{blue}{Pillepich et al. in prep.}, their Fig.11, top). These sizes are consistent with previous zoom-in simulations of $\sim10^{12}\,\Ms$ haloes, e.g. Auriga \citep[][]{2017Grand}, Eris \citep[][]{2011guedes}, NIHAO-UHD \citep[][]{2020Buck} and VINTERGATAN \citep[][]{2021Agertz}. Also, TNG50 disk sizes are compatible with those measured for local disky and spiral galaxies \citep[based on stellar light rather than stellar mass][]{2009Gadotti,2016Lelli} and for the Galaxy \citep{2007Hammer,2008Juric,2013Bovy} and Andromeda \citep{2005Worthey,2006Barmby,2007Hammer}. We note that a number of TNG50 MW/M31-like galaxies fall within the observed values for the scalelength and stellar mass of the Galaxy and Andromeda, whereas the rest have more or less extended stellar disks for their mass: compared to the total TNG50 sample of MW/M31 analogs, the Milky Way has a rather compact stellar disk given its mass, as it settles at the lower end of the TNG50 distribution, while for Andromeda the value is rather average.

The scaleheights of TNG50 galaxies, evaluated at galactocentric distances of a few times the disk length, can be as small as $\simeq$ 200 pc (lowest 10th percentiles). Yet, stellar disks of TNG50 MW/M31-like galaxies (as selected in Section~\ref{sec:selection}) can be as thick as a few kpc (\textcolor{blue}{Pillepich et al. in prep.}, their Fig.11, bottom panels). As is the case for the disk extent, TNG50 MW/M31-like galaxies have typically thicker thin disks than the Galaxy but not necessarily than Andromeda. TNG50 disk heights are consistent with those of zoom-in simulations \citep[e.g.][]{2011guedes,2017Grand,2017Ma,2020Buck}. There is even a number of TNG50 MW/M31-like galaxies that exhibit thin and thick disks with similar heights as the observational estimates of both the Galaxy and Andromeda.

We in particular highlight and take note of six galaxies (Subhalo IDs 516101, 535774, 538905, 550149, 552581, 536365), which we refer to as \textit{MW-analogs}
and whose stellar disk properties are within the observational estimates for the Galaxy. These MW-analogs are chosen among the TNG50 MW/M31-like galaxies that have thin and thick disk heights consistent with those of the Galaxy (approximately in the range $175-360$ pc and $625-1450$ pc, respectively), measured at either 7-9 kpc or $2.7-4.7\times R_{\rm{d}}$; and with disk scalelength and stellar mass in the ranges $1.7-2.9$ kpc and $10^{10.5-10.9}\,\Ms$, encompassing available literature contraints. There is also one galaxy (Subhalo ID 432106) that could be considered a \textit{M31-analog}, based on its stellar mass, disk scalelengths and thickness (\textcolor{blue}{Pillepich et al. in prep.}).
%

\subsection{Vertical stellar mass profiles}
\label{sec:mass_profile}

In Fig.~\ref{fig:vertical_profile} we show the vertical surface mass density profiles at different radii of disk stars for nine TNG50 MW/M31 analogs. These are selected to be representative of the whole sample, namely below, in between and above the 25$^{th}$ and 75$^{th}$ percentiles, in either galaxy stellar mass (from left to right) or in stellar disk scalelength (top to bottom). The scalelength and stellar mass are labeled in each panel. Dots represent the measured values of the stellar density, solid curves represent the resulting function (as per Eq.~\ref{eq:double}), whereas different colors denote different radii.

Amid differences in the details, all galaxies in Fig.~\ref{fig:vertical_profile} (and all the others not shown here) exhibit vertical stellar mass disk profiles that, in general, are steeper towards higher altitudes (or shallower closer to the galactic midplane, i.e. better described with a $\rho(z)\propto {\rm sech}^2$ profile than with a pure exponential) and whose normalization decreases with galactocentric radius. More massive galaxies (right-hand columns) typically exhibit shallower profiles as a function of physical vertical distance from the mid plane, consistently with the mild trend with mass of the scaleheights at fixed galactocentric distance present in the whole sample.

\begin{figure}
\centering
\includegraphics[width=0.45\textwidth]{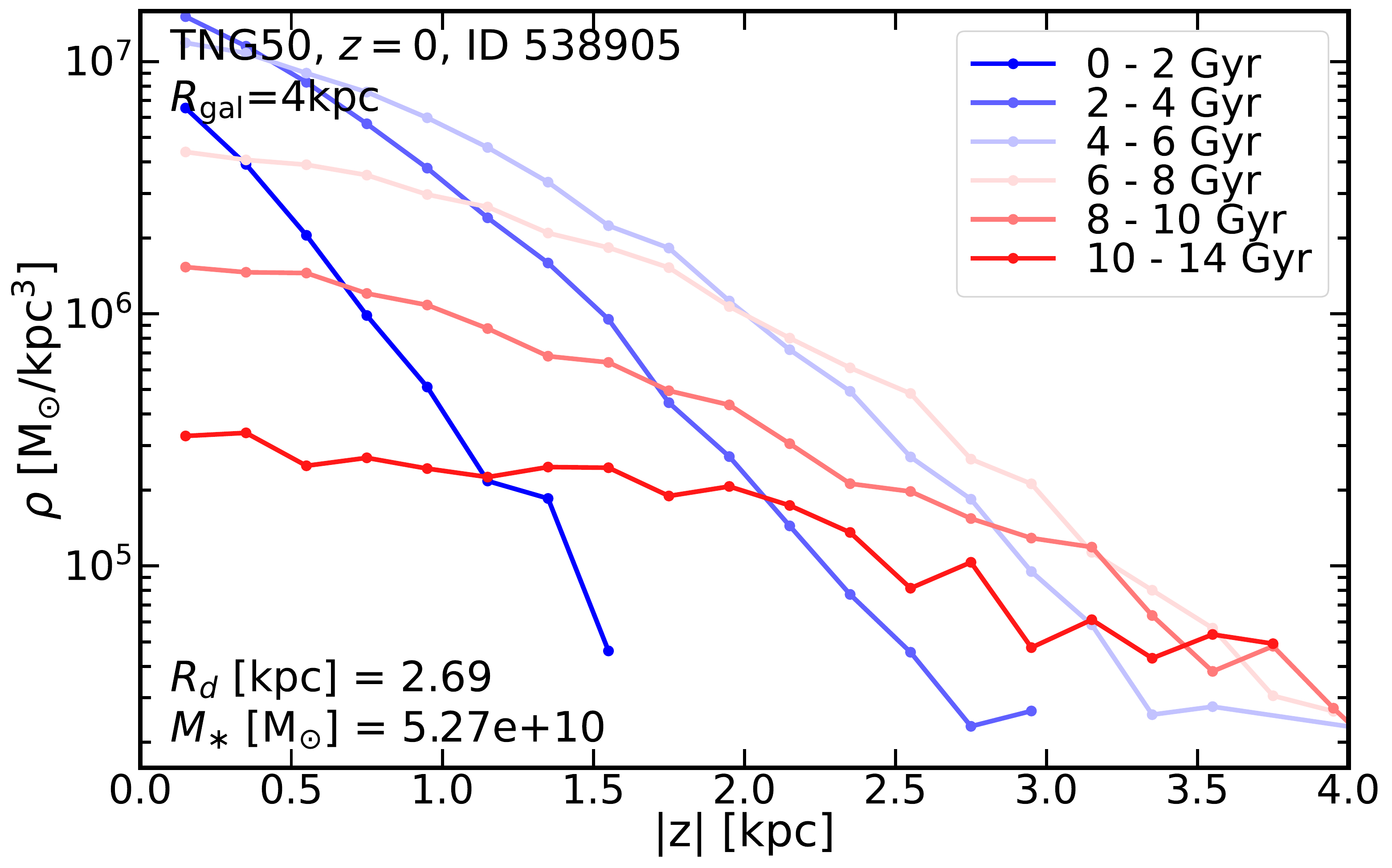}
\includegraphics[width=0.45\textwidth]{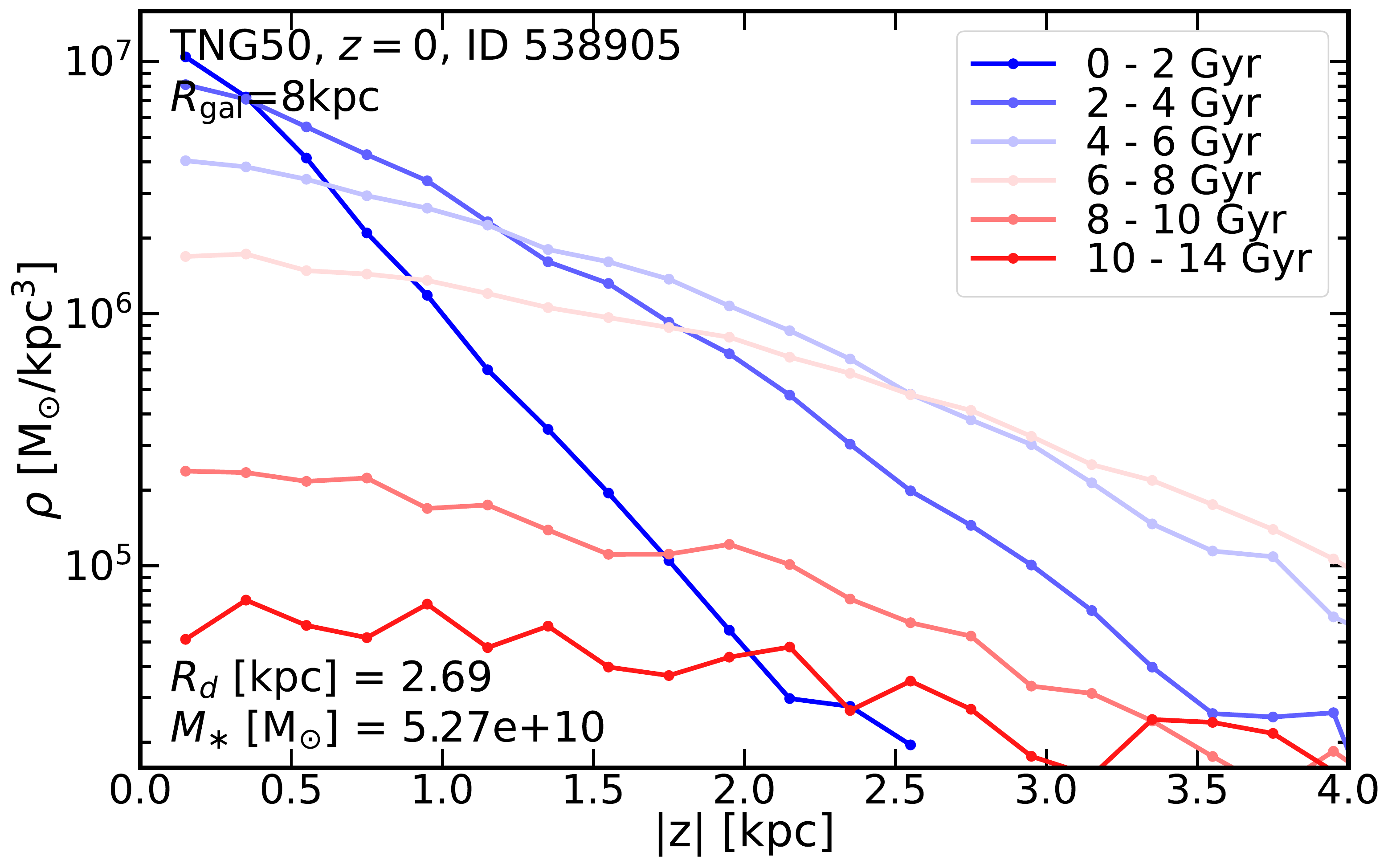}
\includegraphics[width=0.45\textwidth]{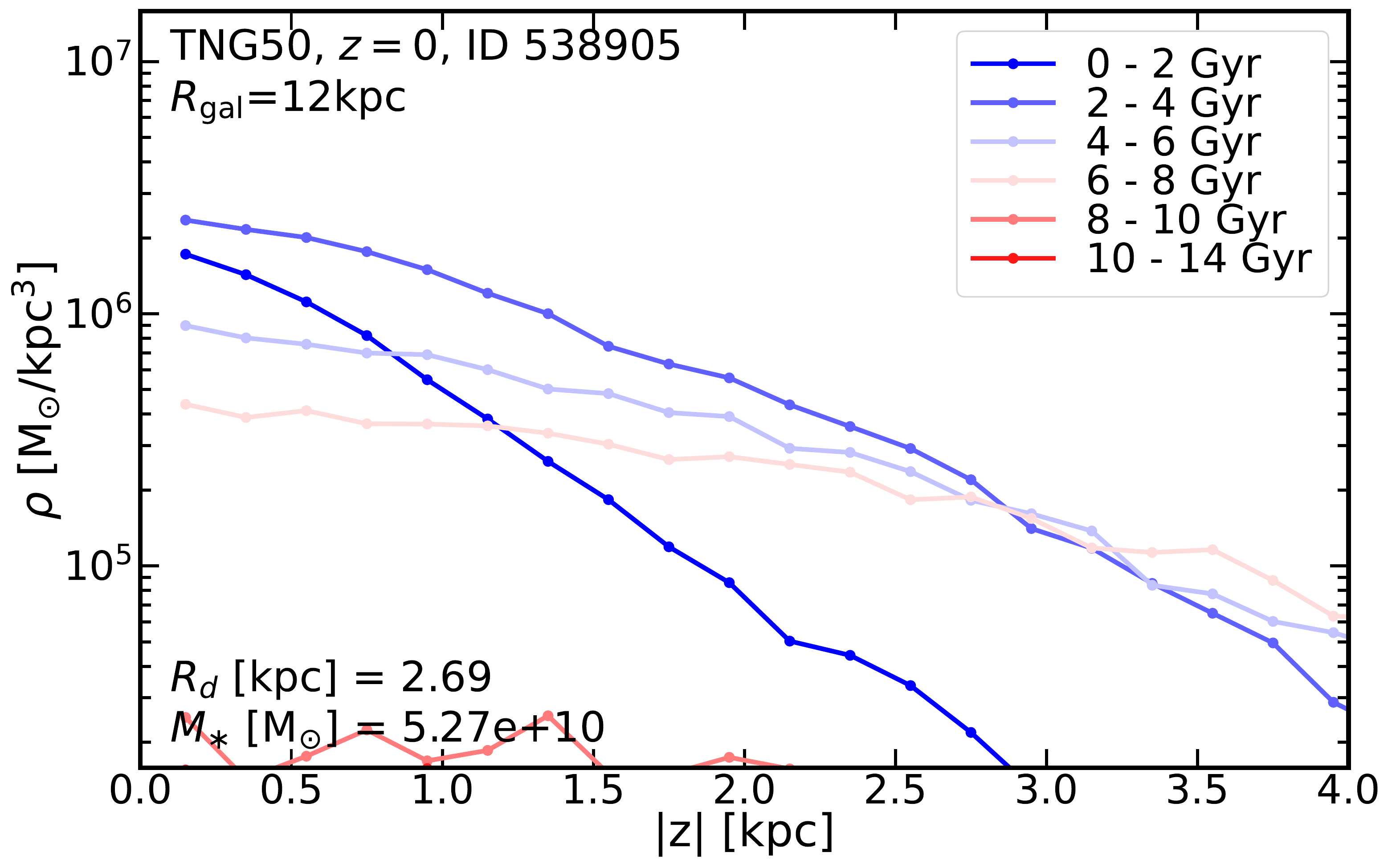}
\caption{\label{fig:vertical_profile_ages} \textbf{Vertical surface density profiles of mono-age stellar populations in an example TNG50 MW/M31-like galaxy}. We show the vertical stellar surface mass density profile of mono-age stellar populations at three different radii: 3-5 kpc (top), 7-9 kpc (middle), and 11-13 kpc (bottom). Different colors in each panel denote different stellar ages, as labeled in the legend.}
\end{figure}

By selecting disk stars according to their ages, we find that the vertical mass distribution of each stellar population can be better fitted with a single-function profile, in agreement with observations \citep[e.g.][]{2014Martig,2018Xiang} and with previous simulations \citep{2017Ma,2020Buck} -- this further justifies our choice of using a single hyperbolic secant squared functional form (Eq.~\ref{eq:single}). In Fig.~\ref{fig:vertical_profile_ages}, we show an example of the vertical surface mass density profiles of mono-age disk stars (in bins of stellar ages of $\Delta t_{\rm age}$=2~Gyr and a final bin for all stars older than 10~Gyr): this is the TNG50 object with Subhalo ID 538905, selected to have a stellar mass and a disk scalelength similar to our Galaxy. 
Here we show the vertical stellar mass profiles of disk stars at three different radii, centered at $\sim$ 4, 8, and 12 kpc, and covering a radial range of $\pm1$ kpc (from top to bottom, respectively). We see that the scaleheight at all radii is smallest for the youngest stars or, in other words, that the vertical profile is flatter, i.e. thicker or more extended, with increasing stellar age. Also, in the case of this galaxy, at fixed age, the vertical profiles are shallower at increasing radius, indicating a clear presence of a flaring. 

A qualitatively similar vertical stellar mass profile is found for our Galaxy by \citealt{2018Xiang} in the solar neighborhood with data from the LAMOST Galactic Survey, who however speculate that the sharp density profiles recovered for the old stellar populations may have suffered contamination from young main sequence stars, which in turn may increase the density near the disk mid-plane (see their Figures 19 and 20). 

\begin{figure}
\centering
\includegraphics[width=0.45\textwidth]{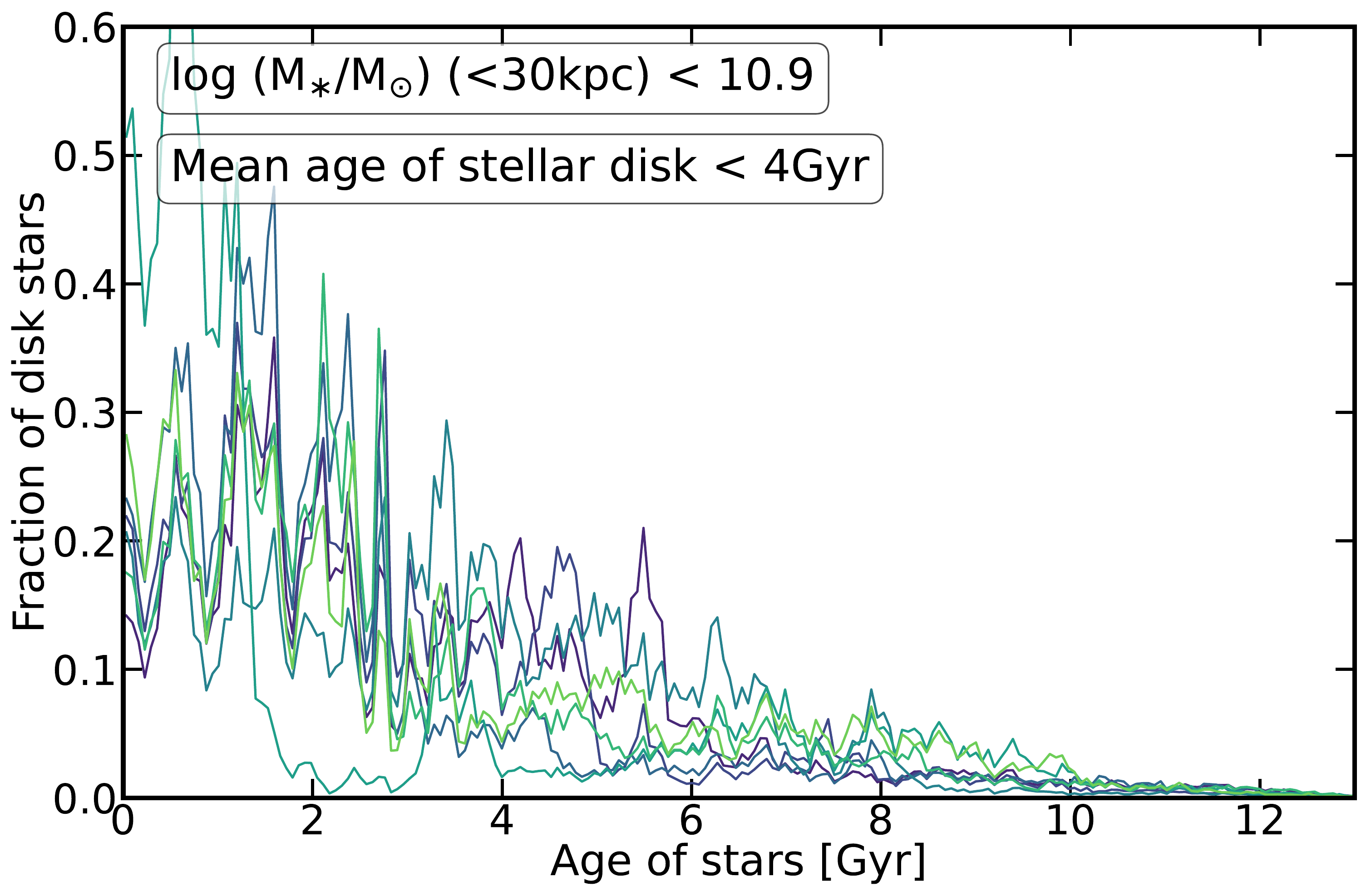}
\includegraphics[width=0.45\textwidth]{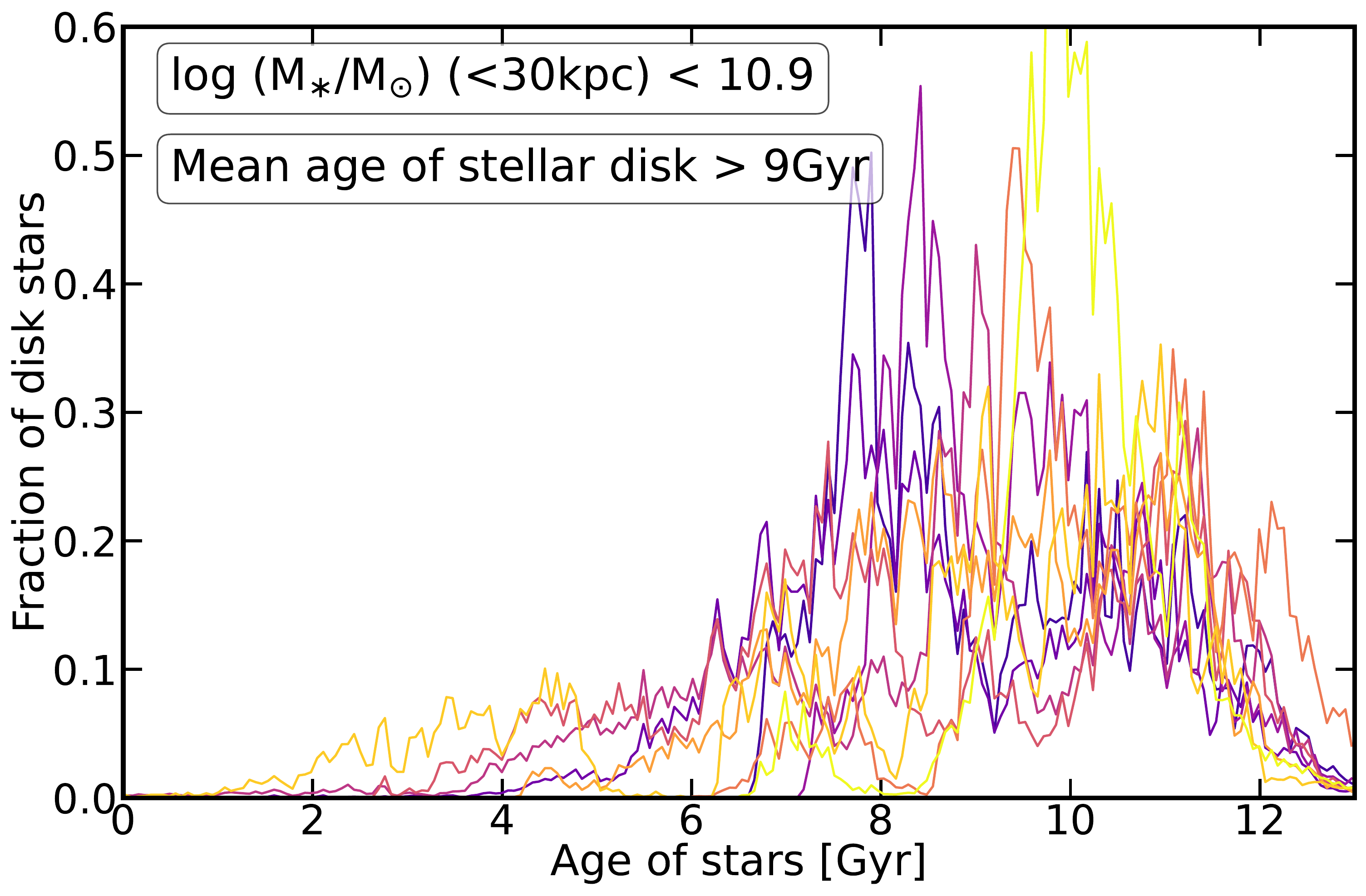}
\includegraphics[width=0.45\textwidth]{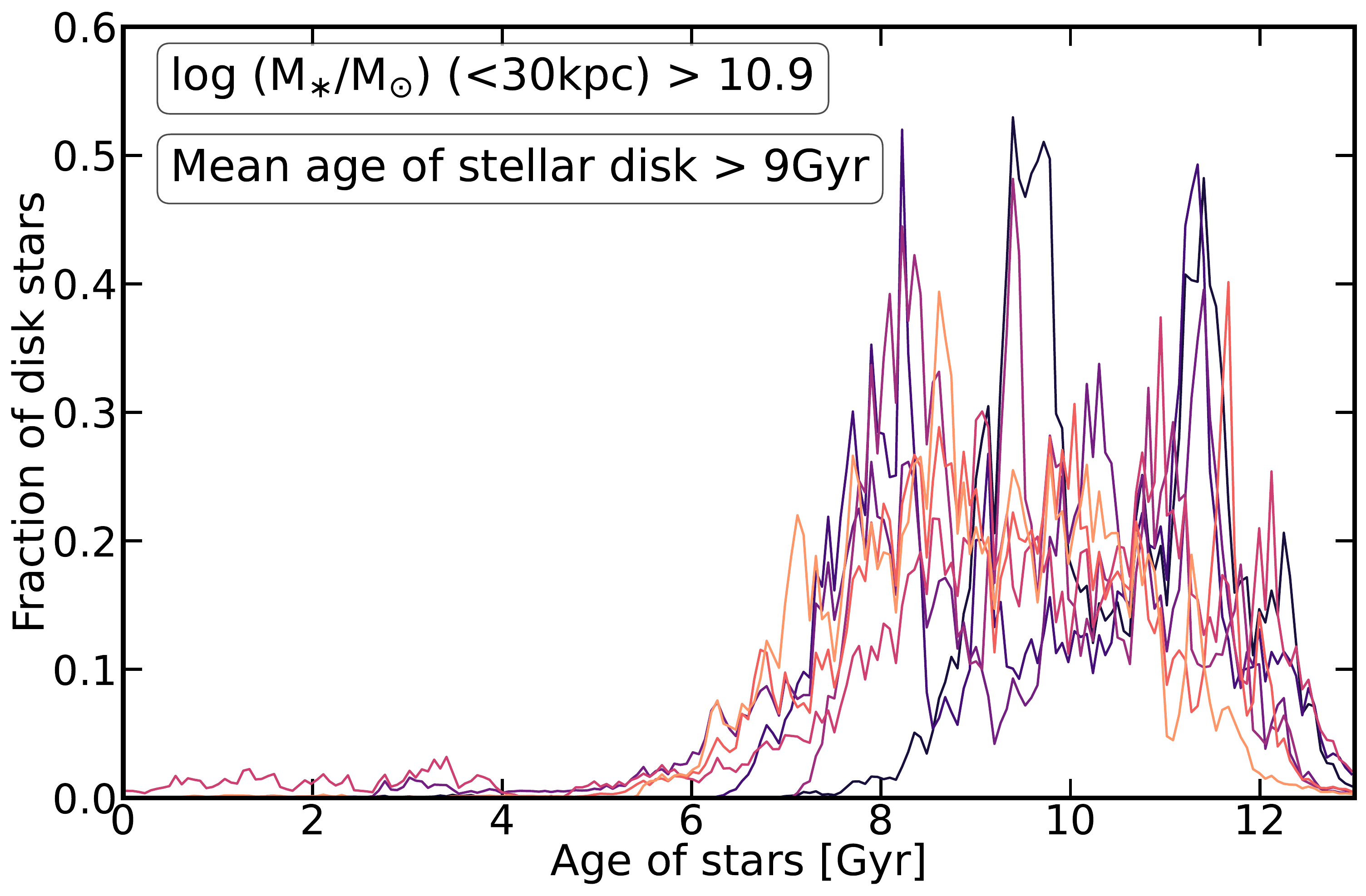}
\caption{\label{fig:age_distribution} \textbf{Age distributions of disk stars in selected TNG50 MW/M31-like galaxies.} Each line represents one galaxy. In the top, we show example galaxies with a \textit{young} disk, i.e. with a mean stellar age younger than 4 Gyr. In the middle and bottom panels, we show example galaxies with an \textit{old} disk, i.e. with a mean stellar age older than 9 Gyr. Additionally, we split this sample into two bins of stellar mass, below (middle) and above (bottom) 10$^{10.9} \Ms$. Only low-mass galaxies show a relatively young stellar disk (7 galaxies, top panel) while old stellar disks are present in both low and high-stellar mass MW/M31-like galaxies (9 and 8 galaxies in the middle and bottom panel, respectively). All the other MW/M31-like galaxies (174 of 198, not shown) have an intermediate-age stellar disk.} 
\end{figure}

\subsection{Age distributions of disk stars}
\label{sec:ages}

The great diversity in the disk structures of TNG50 MW/M31 analogs is also reflected in the ages of the stars that populate the disks. As we aim at characterizing the vertical stellar disk structure and disk flaring for mono-age stellar populations, here we give an overview of the age distributions of disk stars in TNG50 MW/M31-like galaxies. 

Within the sample, TNG50 returns both disks characterized by a sizeable young component, as well as disks that are mostly old -- consistently with the fact that not all TNG50 MW/M31-like galaxies are indeed star forming \citep[][and \textcolor{blue}{Pillepich et al. in prep}]{2021Pillepich}. 
In Fig.~\ref{fig:age_distribution}, we show the age distribution of disk stars for a selection of MW/M31-like galaxies in TNG50, divided in \textit{young} disks, i.e. with a mean stellar age younger than 4 Gyr, and \textit{old} disks, i.e. with a mean stellar age older than 9 Gyr. We further divide the MW/M31 sample into two bins of stellar mass, to separate them in MW-mass and M31-mass galaxies. 

Now, young stellar disks are found only in MW-mass galaxies (7 galaxies, top panel), whereas old stellar disks are returned in both low and high stellar-mass galaxies (9 galaxies, middle panel, and 8 galaxies, bottom panel, respectively). All the other TNG50 MW/M31-like galaxies (174 of 198, not shown) have intermediate-age stellar disks, i.e. with broad distributions across all stellar ages. 

Interestingly, most of the TNG50 MW/M31-like analogs exhibit many short periods of intense star formation. In \citealt{2022SotilloRamos} we have shown that these can be caused by merger events and close pericentric passages of gas-rich companions and in \citealt{2023Boecker} we have shown that the latter can induce bursts of star formation also in the innermost regions of the disks.

\subsection{The cases of warped and disturbed stellar disks}
\label{sec:warps}
Vertical perturbations within or in the outer regions of galactic disks may affect our statistical quantification of their structural properties. They may also lead to an incorrect impression and quantification of the flaring (see next Sections). These can appear under the form of the common \textit{S-shaped} warps, as well as asymmetries, and have been generally ascribed to e.g. the tidal distortions imparted by an external (merging) satellite in fly-by \citep{1989Ostriker,2009Kazantzidis,2013Gomez,2016Donghia,2017Gomez,2020Semczuk} or to a misaligned accretion of high angular momentum cold gas \citep[see for example][]{2010Roskar,2013Aumer}.

In this paper we do not attempt an accurate and in-depth analysis of warped stellar disks, which we leave to future works, but we at least signal those TNG50 MW/M31-like galaxies that may exhibit some warps or disturbed stellar disks based on a visual inspection of their edge-on stellar maps. Among the TNG50 MW/M31 sample, we identify 21 galaxies that have a well-defined S-shaped warp and 17 galaxies with a generally disturbed or distorted stellar disk, which we refer to as \textit{disturbed} disk. When needed, we will properly identify such galaxies in plots and discuss them in the text (e.g. Section \ref{sec:youngVSold}). A catalog with corresponding flags is released with this paper.

\section{Disk flaring with TNG50}
\label{sec:TNG50flaring}

Equipped with the vertical stellar mass distributions of stars and of mono-age stellar distributions throughout the simulated stellar disks, we can quantify how, and by how much, if at all, the disk scaleheights of all, young and old stellar populations increase with galactocentric distance. Namely, in the following, we provide results from TNG50 about the the vertical structure of stellar disks in MW/M31 analogs across galactocentric distances. The questions we would like to answer are the following: how often does flaring occur in MW/M31-like galaxies according to TNG50? And across all MW/M31-mass disk galaxies, how often do young and old stellar populations display the same or different amount of flaring?

\begin{figure*}
\centering
\includegraphics[width=0.49\textwidth]{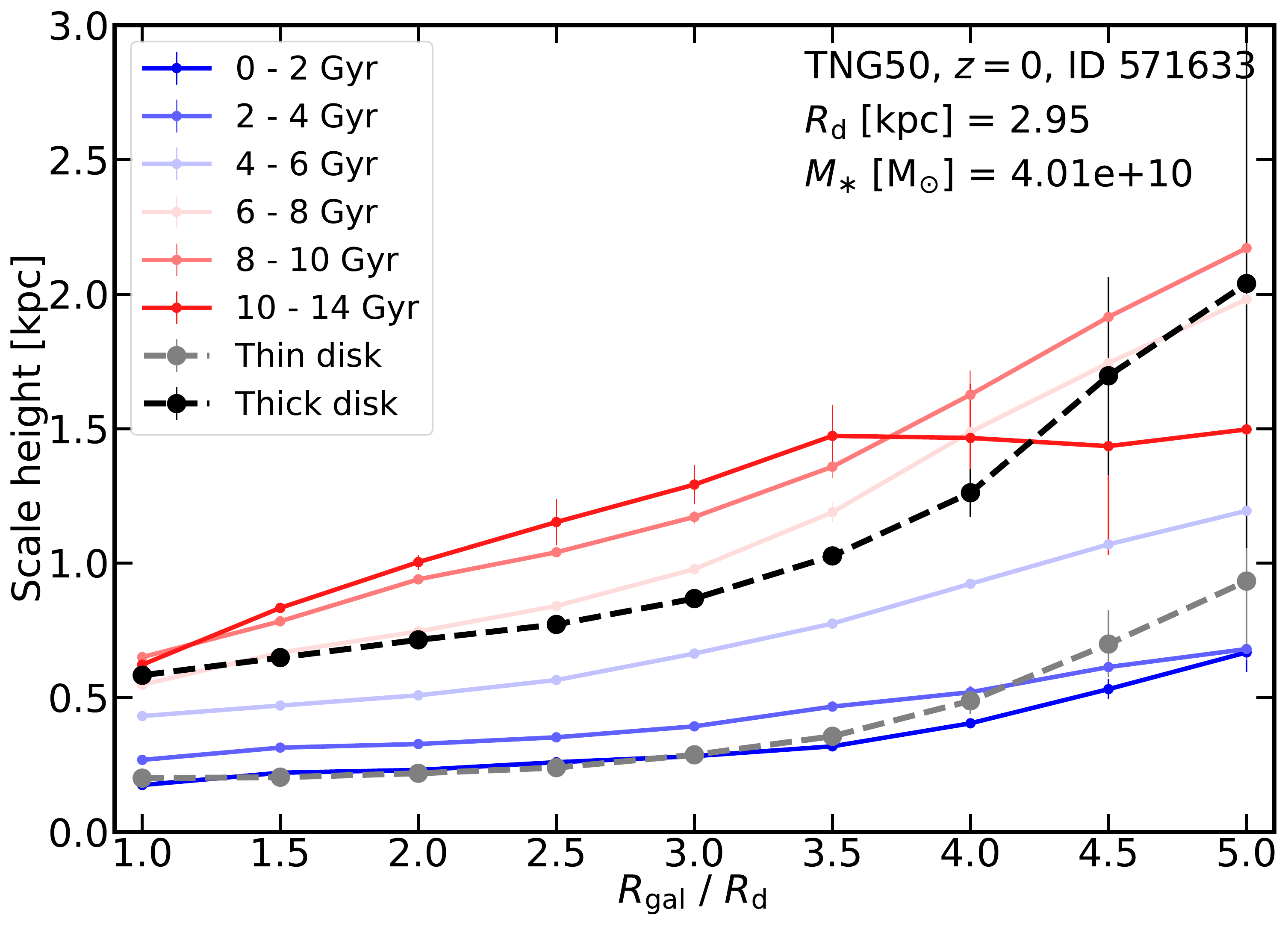}
\includegraphics[width=0.49\textwidth]{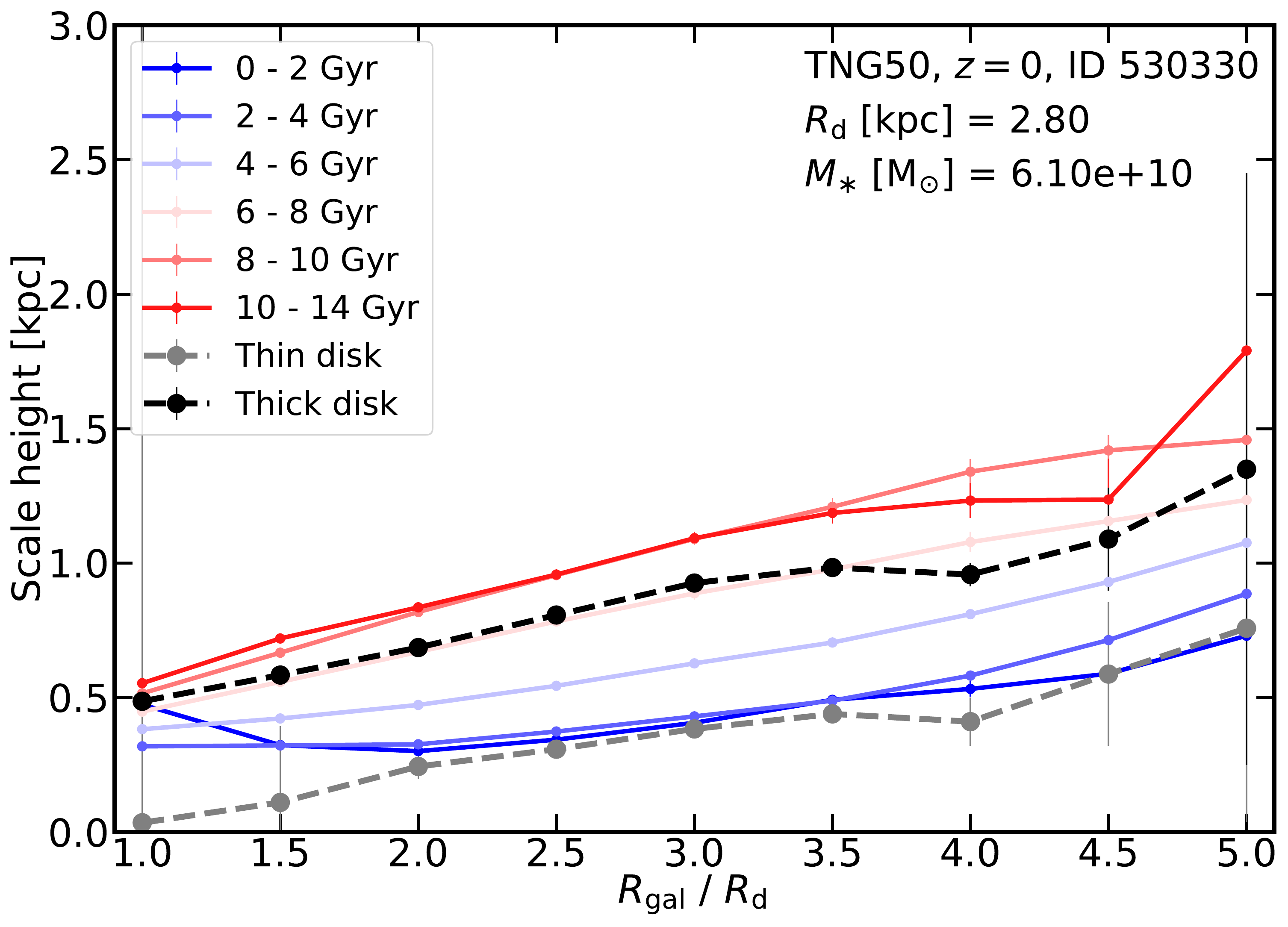}
\includegraphics[width=0.49\textwidth]{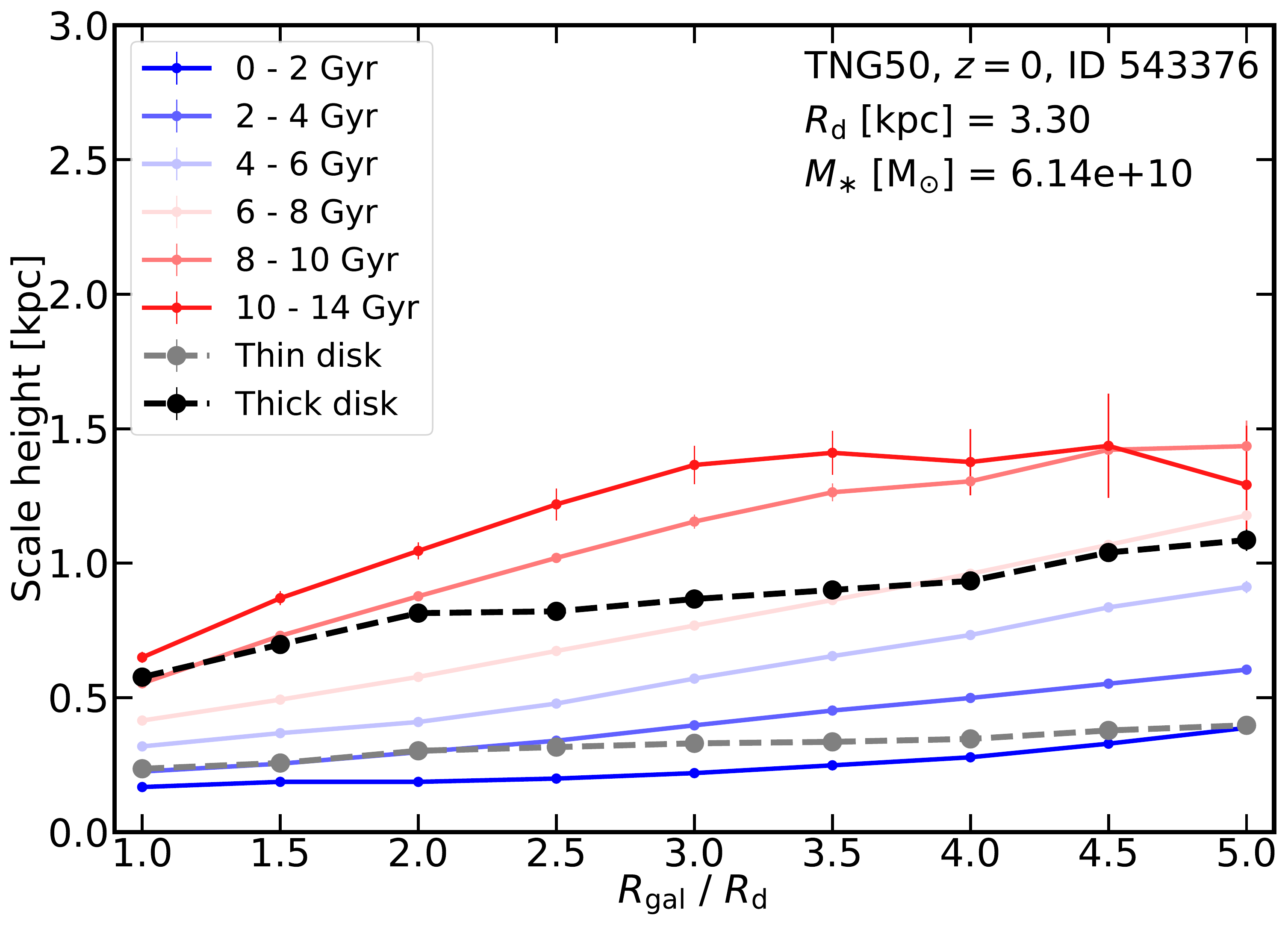}
\includegraphics[width=0.49\textwidth]{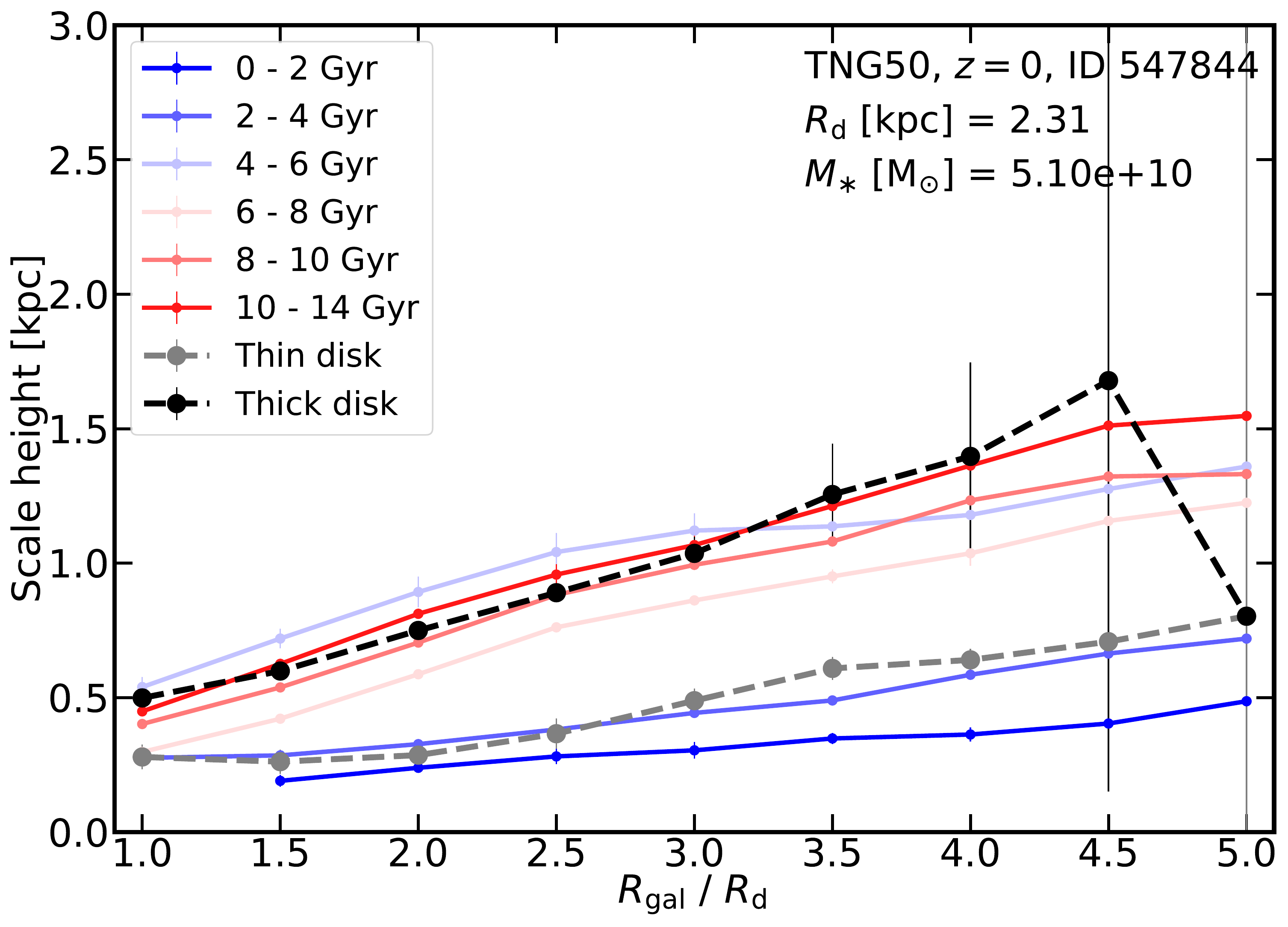}
\includegraphics[width=0.49\textwidth]{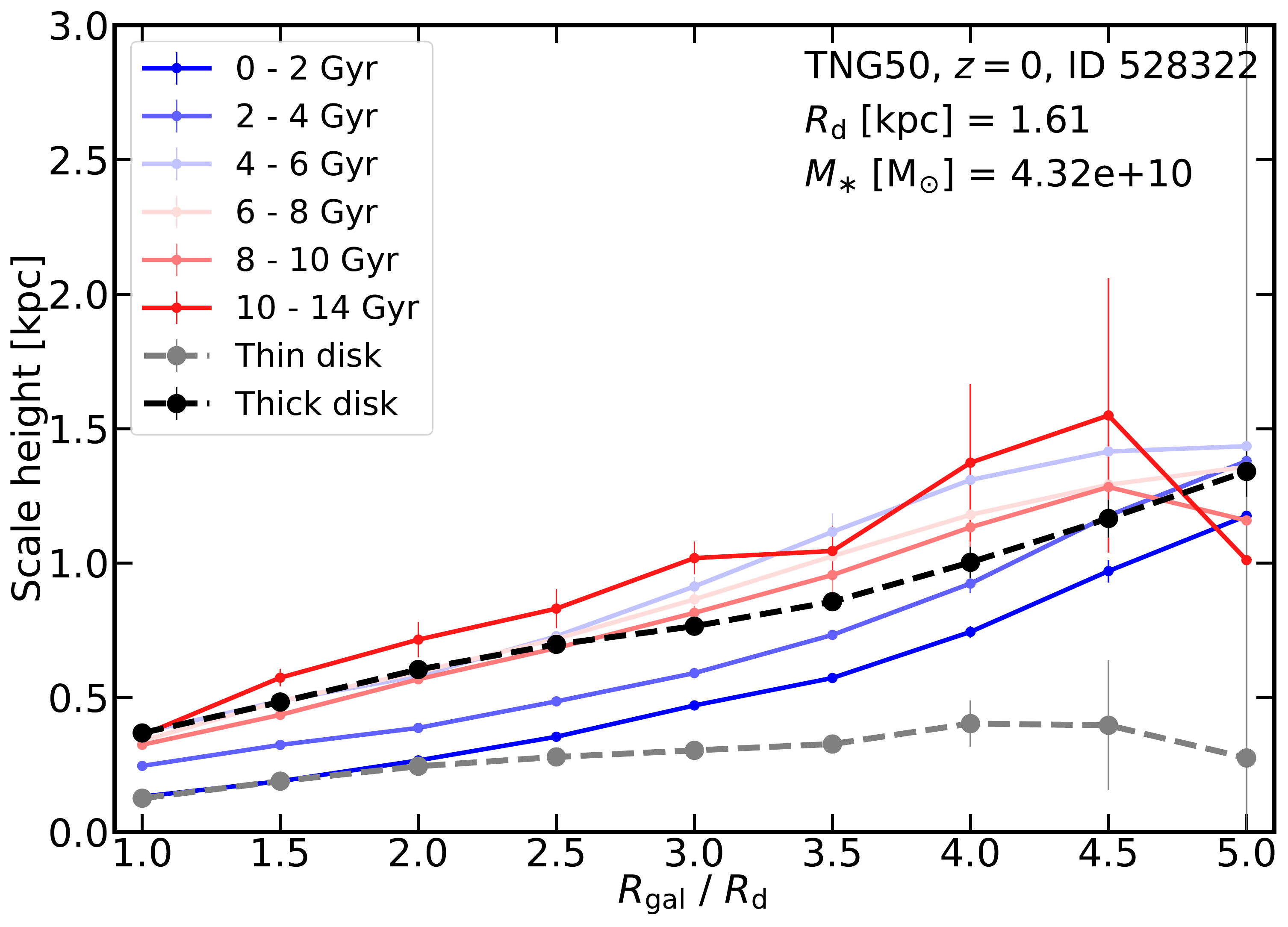}
\includegraphics[width=0.49\textwidth]{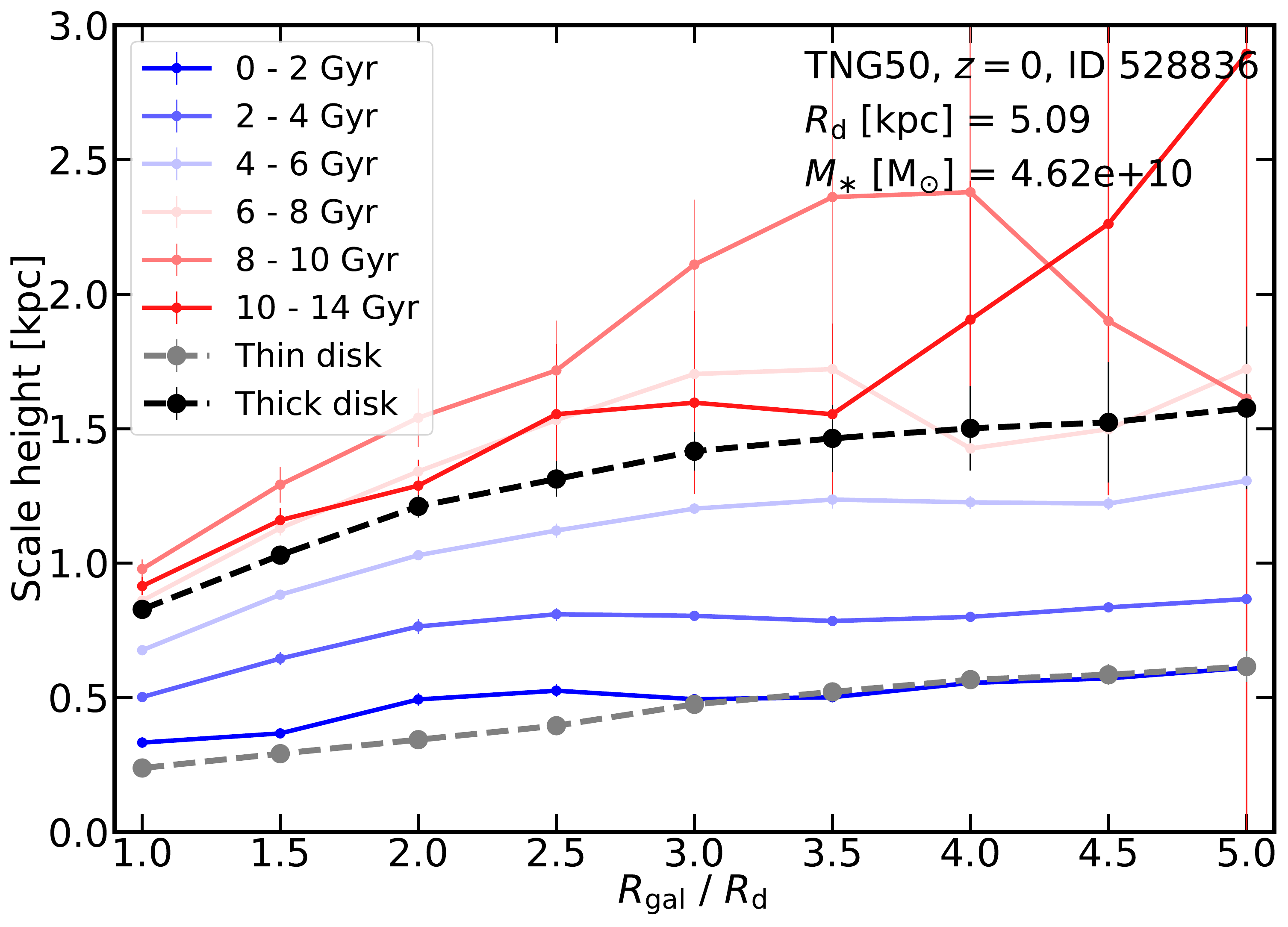}
\caption{\label{fig:flare} {\bf Visualization of disk flaring in a few example TNG50 MW/M31-like galaxies}. We show disk scaleheight as a function of galactocentric distance normalized by the scalelength for six TNG50 MW/M31 analogs, chosen to highlight the diverse ways in which the flaring is manifested across the whole galaxy sample (see the text for details). Curves of different colors represent different stellar ages, as labeled in the legend. The black (grey) curves with circles denote the geometrical thick (thin) disk, i.e. all disk stars without splitting in stellar ages.}
\end{figure*}

\subsection{Diversity in disk flaring across TNG50 MW/M31 analogs} 

Fig.~\ref{fig:flare} shows the disk scaleheights as a function of galactocentric distance of mono-age stellar populations in six MW/M31-like galaxies from the TNG50 simulation.
Curves of different colors represent different bins of stellar ages, as labeled in the legend (blue to red from young to old), with scaleheights obtained via a single-component fit (Eq.~\ref{eq:single}). The dashed, dotted curves accounts for all disk stars in the simulated galaxies, with scaleheights obtained via a double-component fit (Eq.~\ref{eq:double}): black for geometrical thick disk and grey for the geometrical thin disk. These galaxies are chosen to highlight the variety of disk flaring that we find across the TNG50 MW/M31-like sample:

\begin{enumerate}[i)]
  \item \textit {Top left panel}: both mono-age and total stellar populations exhibit a substantial flaring, with the older stars flaring, upon visual impression, a bit more than younger ones. In this galaxy, the flaring of most mono-age populations is almost exponential (in analogy with \citealt{2015Minchev,2017Grand} and with some galaxies of \citealt{2020Buck}).\\

  \item \textit{Top right panel}: both mono-age and total stellar populations exhibit a flaring, which appears in this case linear rather than exponential (in analogy with \citealt{2017Ma,2021Agertz,2021delacruz} and some galaxies of \citealt{2020Buck}). \\

  \item \textit{Middle left panel}: flaring is more evident when considering mono-age populations than when considering all disk stars at once \citep[colored curves vs. black and grey ones; consistently with the results of][]{2015Minchev}. \\

  \item \textit{Middle right panel}: in this galaxy, young and old stellar populations follow two somewhat different trends, that can be identified, respectively, with the geometrical thin and thick disks.\\

  \item \textit{Bottom left panel}: in this galaxy, all mono-age stellar populations and the thick disk are flared, but not the thin disk (or at least not out to 4.5$\times R_d$). The latter does not follow any of the mono-age populations: clearly a double functional profile was necessary to unravel it.\\
  
  \item \textit{Bottom right panel}: a linear flaring of both mono-age and total populations is manifest in the inner part of the galaxy. The flaring disappears in the outer parts, as the scaleheights remains either constant or non monotononic.
\end{enumerate}

The significance of the results uncovered here is remarkable: with a single set of physical-model ingredients, TNG50 returns all the manifestations of flaring that have been found so far in individual simulations and simulated galaxies. This demonstrates that the diversity of disk flaring outcomes can arise naturally from the diversity of the galaxy population.

It is also manifest from Fig.~\ref{fig:flare} that the change of scaleheights with galactocentric distance is not always smooth nor monotonic. This can be explained by the fact that realistic disk galaxies, even if simulated, can have complex structures, including over- and underdense regions across their disks. Namely, TNG50 galaxies are certainly not akin to idealized smooth exponential disks: this is reflected in their more complex radial and vertical stellar density distributions, which in turn may simply be not well described by parametric functions (see Sections~\ref{sec:fit_procedure} and \ref{sec:fit_procedure_heights}).  We notice that these effects are more accentuated towards the disks' outskirts, whereby, together with sparser populations of stars, the error bars on the scaleheight estimates get typically larger.

\subsection{A new non-parametric and generally-accessible quantification of disk flaring}
\label{sec:demographics}

As TNG50 returns a wide variety of disk flaring, a unique prescription to quantify it would be inadequate. We have seen, for example, that the radial trends of scaleheights can be both exponential and linear. Past works have fitted the flaring with an exponential formula only \citep{2002Lopez,2017Grand} or with a linear function only \citep{1998Evans,2000Alard,2021delacruz,2022Lian}. The TNG50 phenomenology suggests that a non-parametric quantification of the flaring is of the essence. 

We propose a quantification of the flaring that is independent of the shape of the flaring (linear, exponential or otherwise) and that can be applied in the most general manner to any data, so long as the latter allows the estimate of the disk scaleheight at two different galactocentric distances. Namely, we advocate for a quantification of the flaring based simply on the \textit{relative enhancement} of disk heights at two locations (inner and outer disk), i.e. the difference between the scaleheights at two fixed galactocentric distances divided by the height in the innermost location.
Now, as discussed in \S\ref{sec:StellarSizes}, stellar disks can span a wide diversity of extents even in a relatively narrow range of galaxy mass: $\sim 1.5-17$ kpc. This hence requires to quantify the amount of flaring upon normalizing the galactocentric distance by the scalelength of each MW/M31-like galaxies. We then propose and evaluate the amount of flaring in between (1-5)$\times R_{\rm{d}}$ \citep[see also][for a similar approach]{2014MinchevAA}:

\begin{equation}
\tau_{\rm flare} = \frac{h_{z,5 \times R_{\rm{d}}} - h_{z,1 \times R_{\rm{d}}}}{h_{z,1 \times R_{\rm{d}}}} \, ,
\label{eq:tau_flare}
\end{equation}
where $h_{z,r}$ denotes the vertical scaleheight in a narrow radial annulus at distance $r$, in our case according to \S\ref{sec:methods}. A $\tau_{\rm flare}$ equal to 1 (3) means that the disk height is 2 (4) times larger at $5 \times R_{\rm{d}}$ than at the inner radius.

\begin{figure*}
\centering
\includegraphics[width=0.99\textwidth]{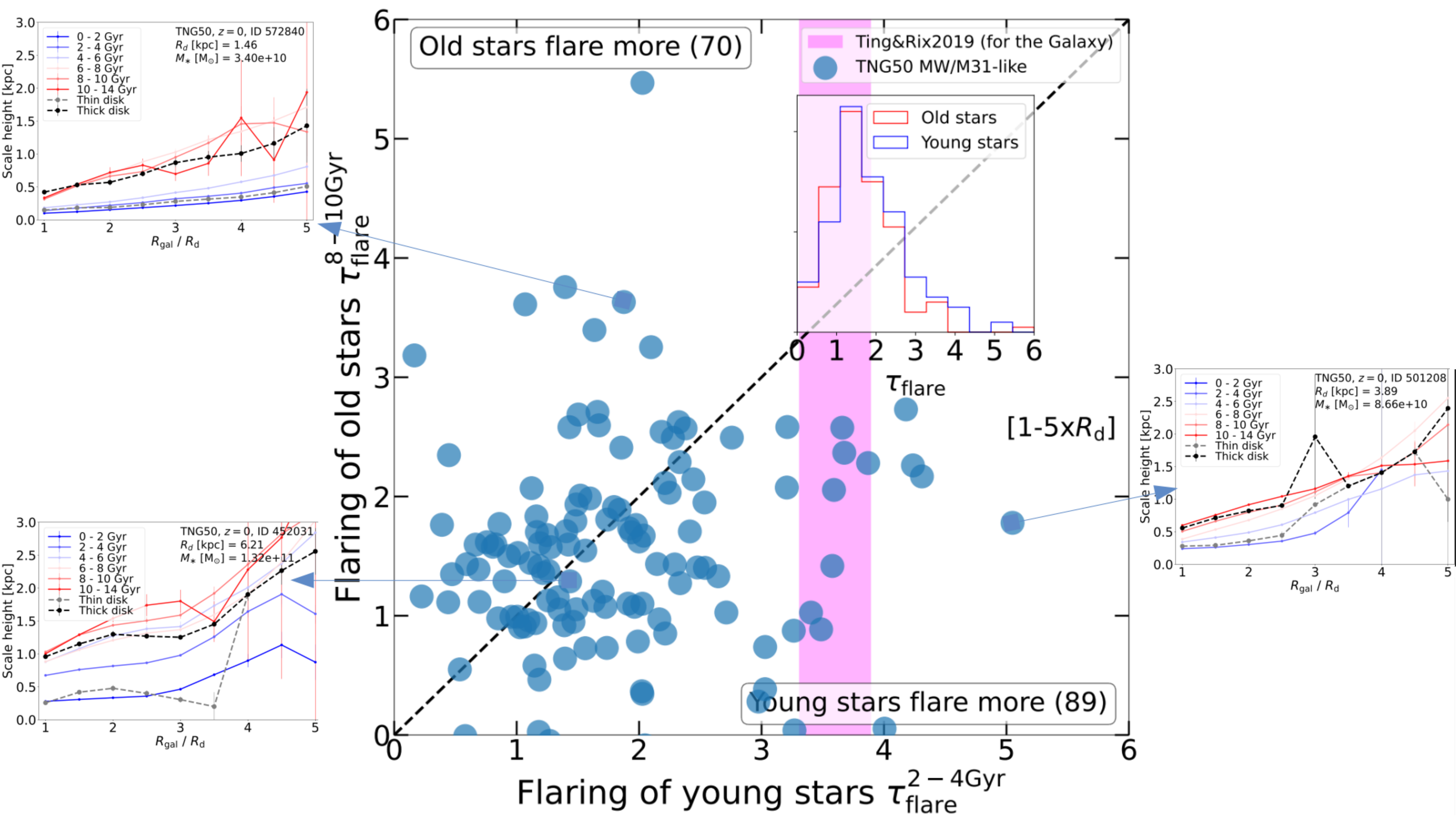}
\includegraphics[width=0.4\textwidth]{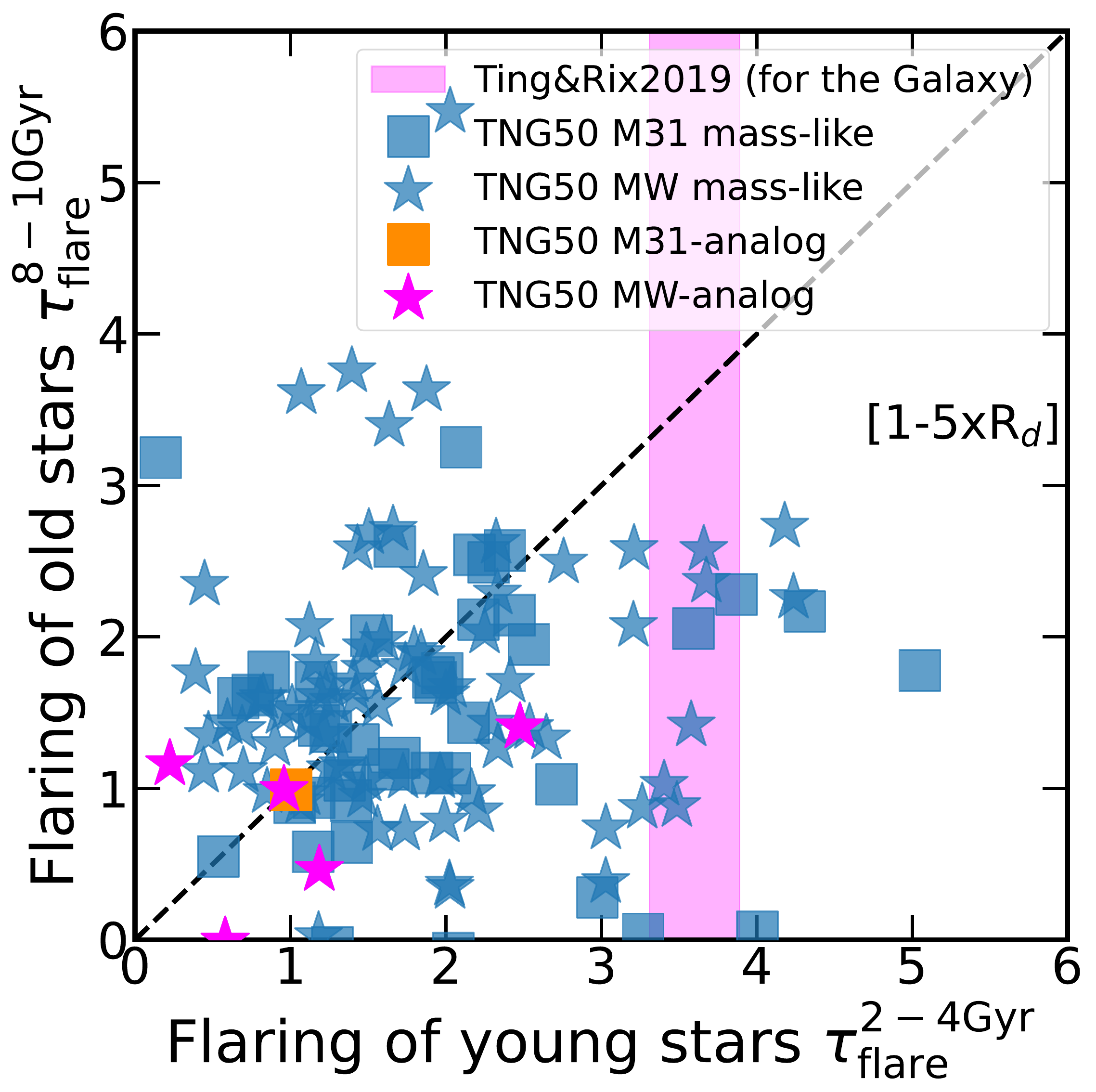}
\includegraphics[width=0.4\textwidth]{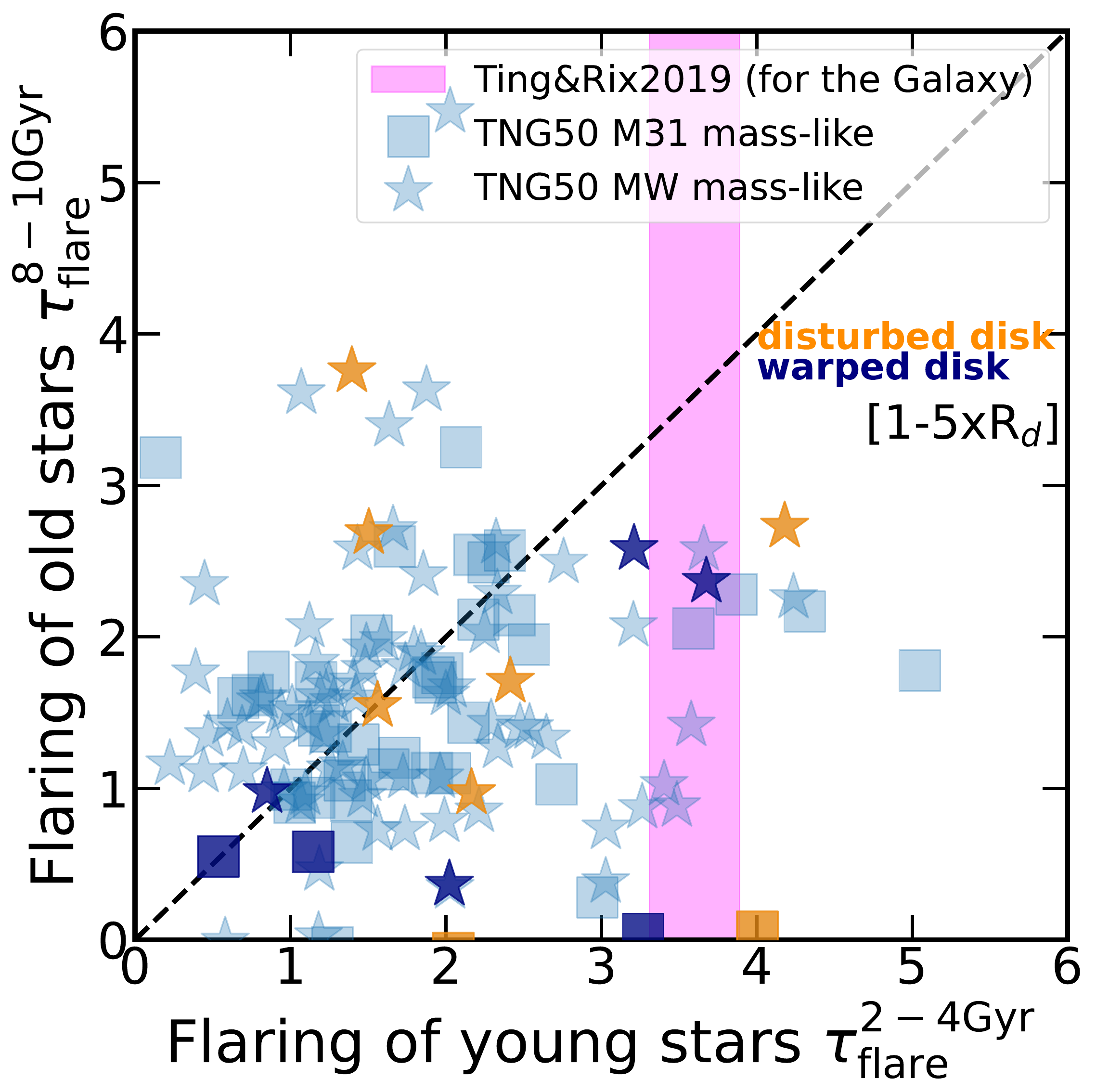}
\caption{\label{fig:young_old} {\bf Flaring of young vs. old stellar populations in TNG50 MW/M31-like galaxies.} We compare between the flaring of young (2-4 Gyr) and old (8-10 Gyr) stellar populations. The black solid line separates galaxies in which old stars flare more than the young ones (above the bisector) or vice versa (below the bisector). The vertical magenta areas indicate the flaring of young stellar population of our Milky Way, inferred from \citet{2019Ting} -- see the text for more details. We have selected three galaxies from three different regions of the plane and show in the small panels their scaleheights vs. radius. In the bottom left panel, TNG50 MW/M31-like galaxies are separated according to their stellar mass, above (squares, M31-mass like) and below (stars, MW-mass like) ${\rm log} ~ (\MS/\Ms)= 10.9$. Moreover, galaxies dubbed as MW-analogs (i.e. having stellar mass, scalelength, thick and thin scaleheight similar to the Milky Way) are depicted with pink stars. 
In the bottom right panel, we highlight the cases in which galaxies show a warped (in navy) or disturbed (in orange) stellar disk, showing that these visually-identified features do not seem to systematically bias our quantification of flaring.}
\end{figure*}

\subsection{What flares more? Young or old stars?} 
\label{sec:youngVSold}
We proceed our analysis by quantifying the degree of flaring of the stellar disk populations as per Eq.~\ref{eq:tau_flare}, separately for old and young stars, i.e. with stellar ages of $8-10$ and $2-4$ Gyr, respectively. As noted before, the vertical distributions of stars in relatively narrow age intervals are well described by a single-component formula (Eq.~\ref{eq:single}), which hence gives the corresponding scaleheight, in agreement with previous simulations \citep{2014Martig,2015Minchev,2017Ma} and observations \citep[e.g.][assuming the chemistry as a good proxy for ages]{2016Bovy}.

The main results of this paper are shown in Fig.~\ref{fig:young_old}. In the top panel, we show the comparison between the flaring of young and old stellar populations across 159 TNG50 MW/M31-like galaxies\footnote{In fact, in 39 of the 198 TNG50 MW/M31, one or both the stellar age bins are not sufficiently populated to ensure a good vertical profiling and fitting of the stellar vertical mass distribution.}.

The black dashed line separates galaxies in which old stars flare more than the young ones (above the bisector) or vice versa, young stars flare more than the old ones (below the bisector): according to TNG50, a slight majority of MW/M31-like galaxies (89) galaxies, i.e. 56 per cent) have stellar disks whereby the young stars flare more than the old ones.
However, a non-negligible fraction of TNG50 galaxies display a similar amount of flaring between young and old stars, settling on top of (or very close to) the dashed line. For the average or typical TNG50 MW/M31-like galaxy, young and old stellar populations exhibit scaleheights in the outer disk that are $\sim1.5-2$ times larger than those in the inner part of the disk (median values of 1.44 and 1.66 for young and old stars, respectively; see inset of the main panel of Fig.~\ref{fig:young_old}).

A few galaxies populate the bottom right and top left corners, where the flaring of the young stars is considerably more pronounced than of the old stars, or vice versa. Still, across the studied TNG50 MW/M31 sample, slightly more frequent are galaxies where the young, rather than the old, stars reach high levels of flaring, e.g. $\tau_{\rm flare} \gtrsim 4$, corresponding to scaleheights at large distances that are $\gtrsim 5$ times larger than in the inner disk regions. Young stars show a somewhat broader diversity of flaring (with a weak peak at $\tau_{\rm flare} = 1-4$), whereas the old stellar populations exhibit a narrower distribution concentrated around $\tau_{\rm flare} = 0-3$.

\begin{figure*}
\centering
\includegraphics[width=\textwidth]{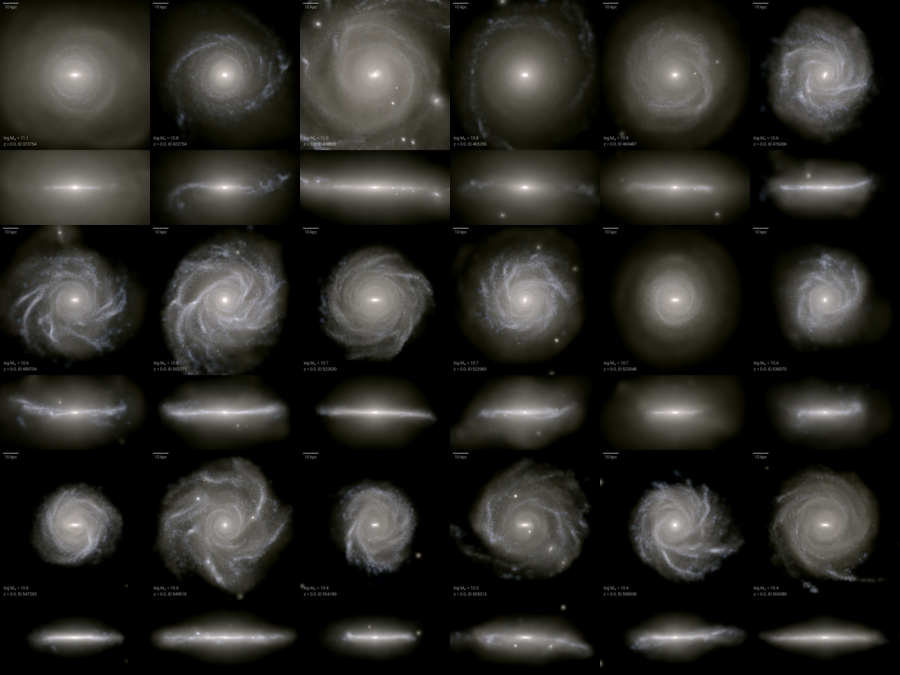}
\caption{\label{fig:images} \textbf{Stellar-light composite images of selected TNG50 MW/M31-like galaxies}. We show the face-on and edge-on projections of 18 MW/M31 analogs from TNG50 at z = 0 that exhibit substantial flaring (i.e. $\tau > 3$).}
\end{figure*}

\begin{figure*}
\centering
\includegraphics[width=0.45\textwidth]{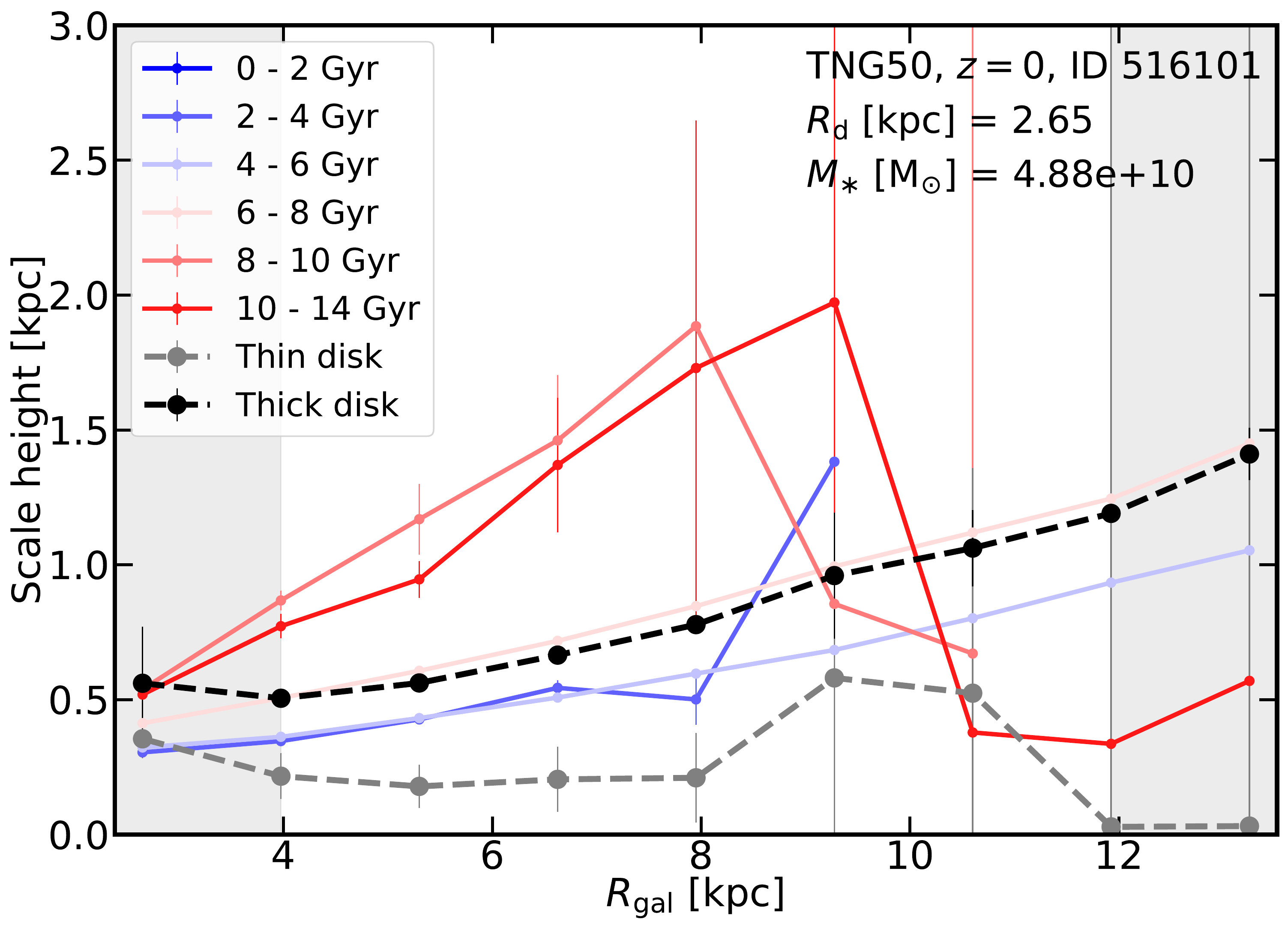}
\includegraphics[width=0.45\textwidth]{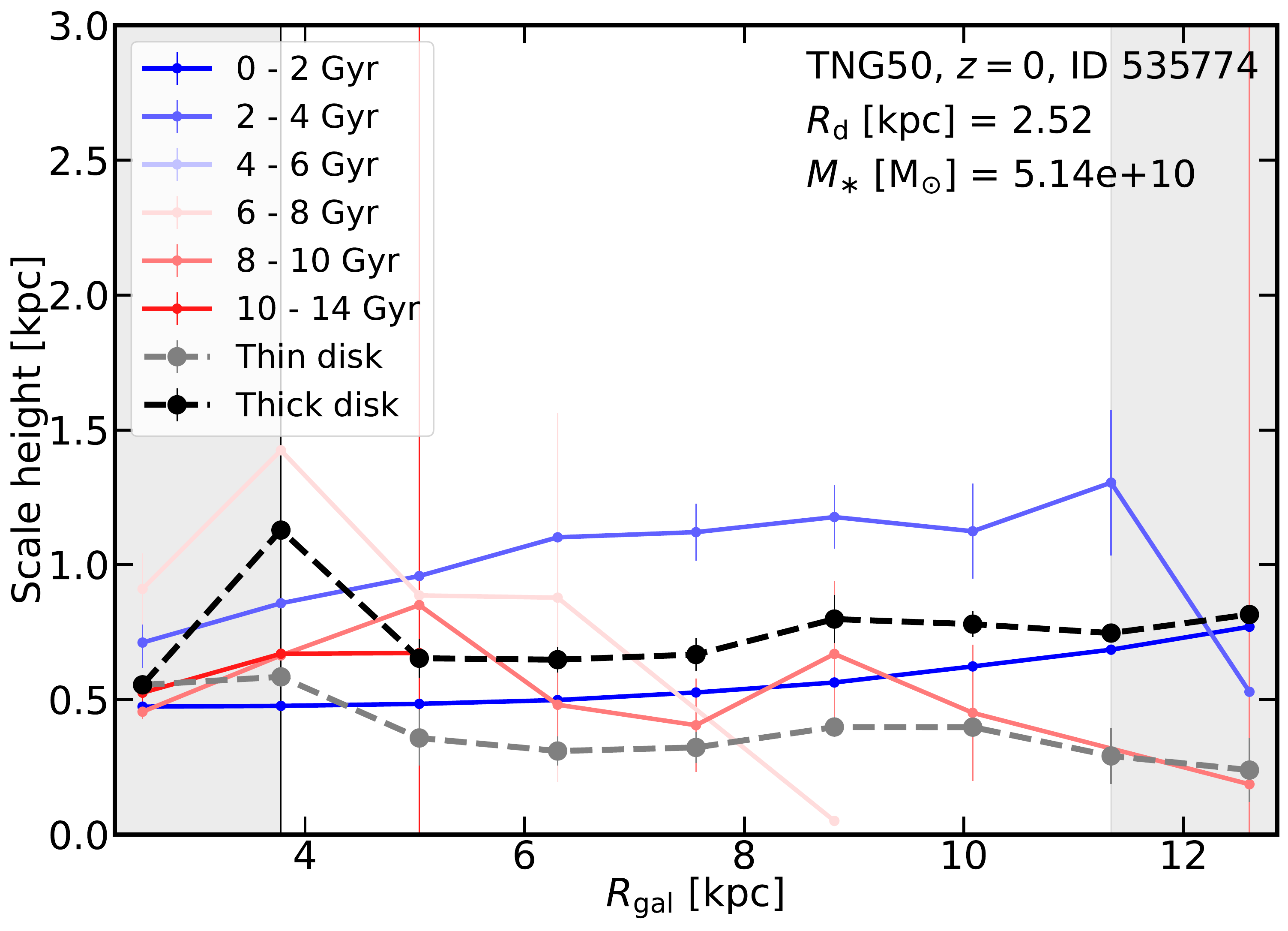}
\includegraphics[width=0.45\textwidth]{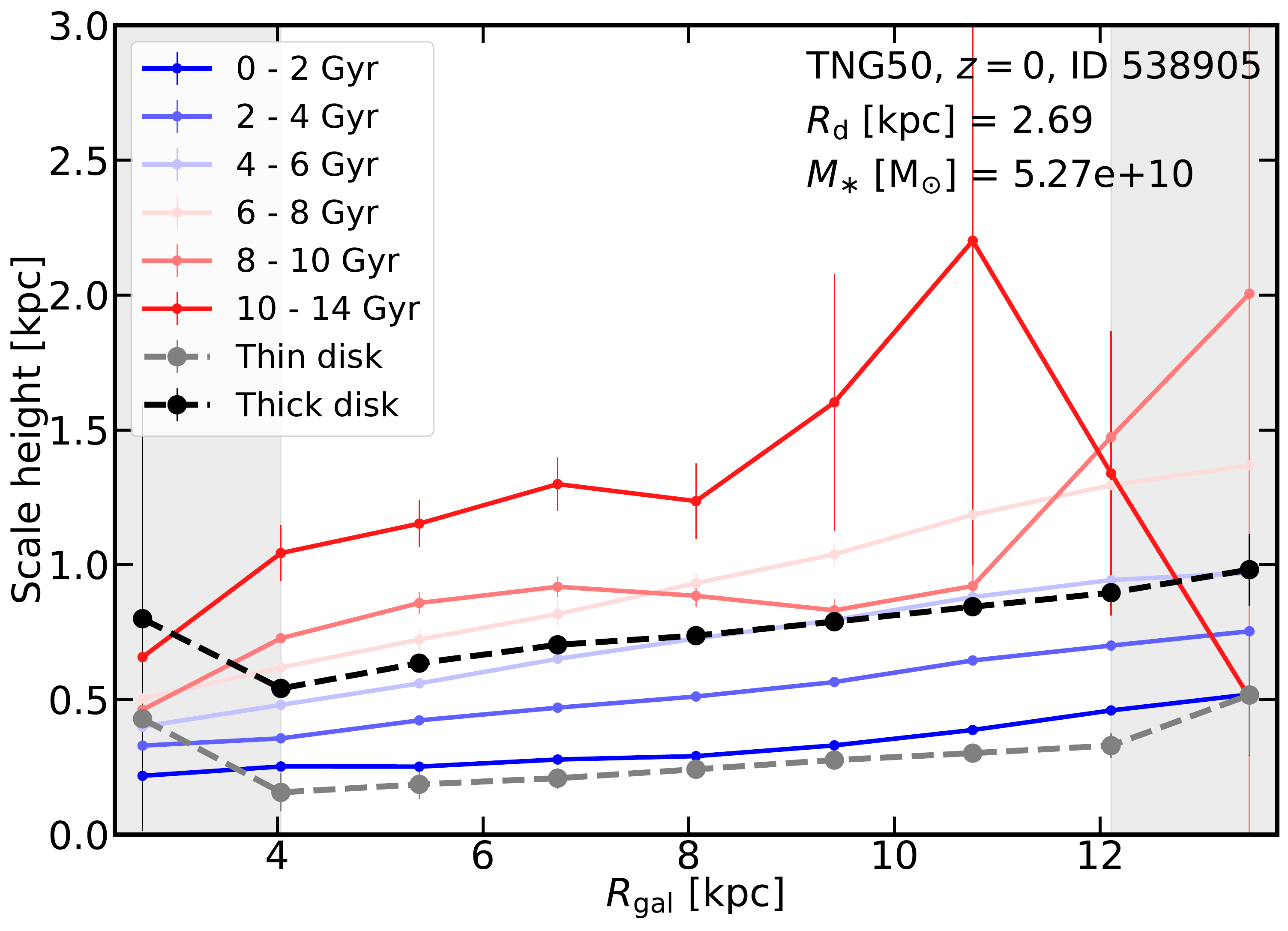}
\includegraphics[width=0.45\textwidth]{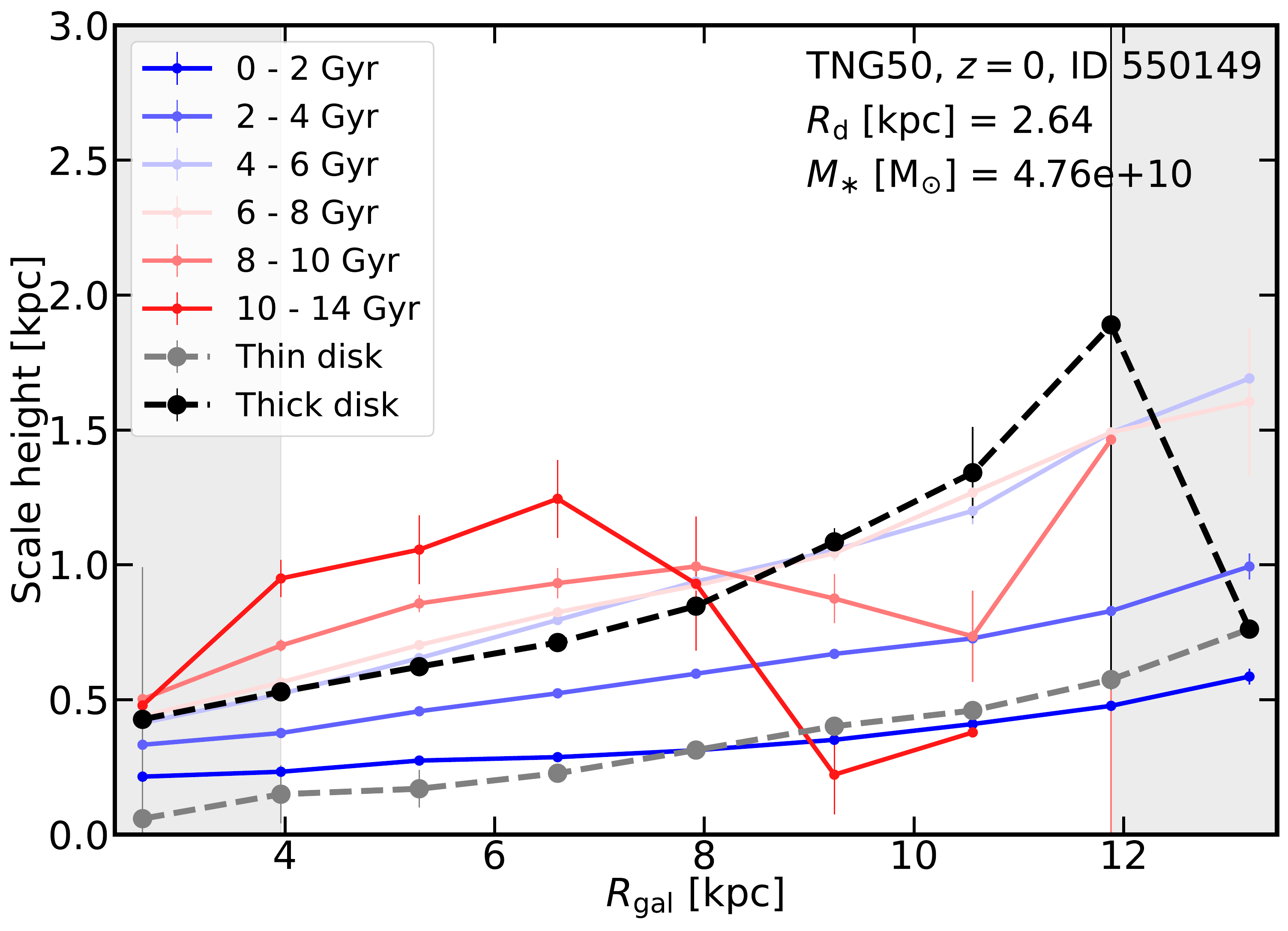}
\includegraphics[width=0.45\textwidth]{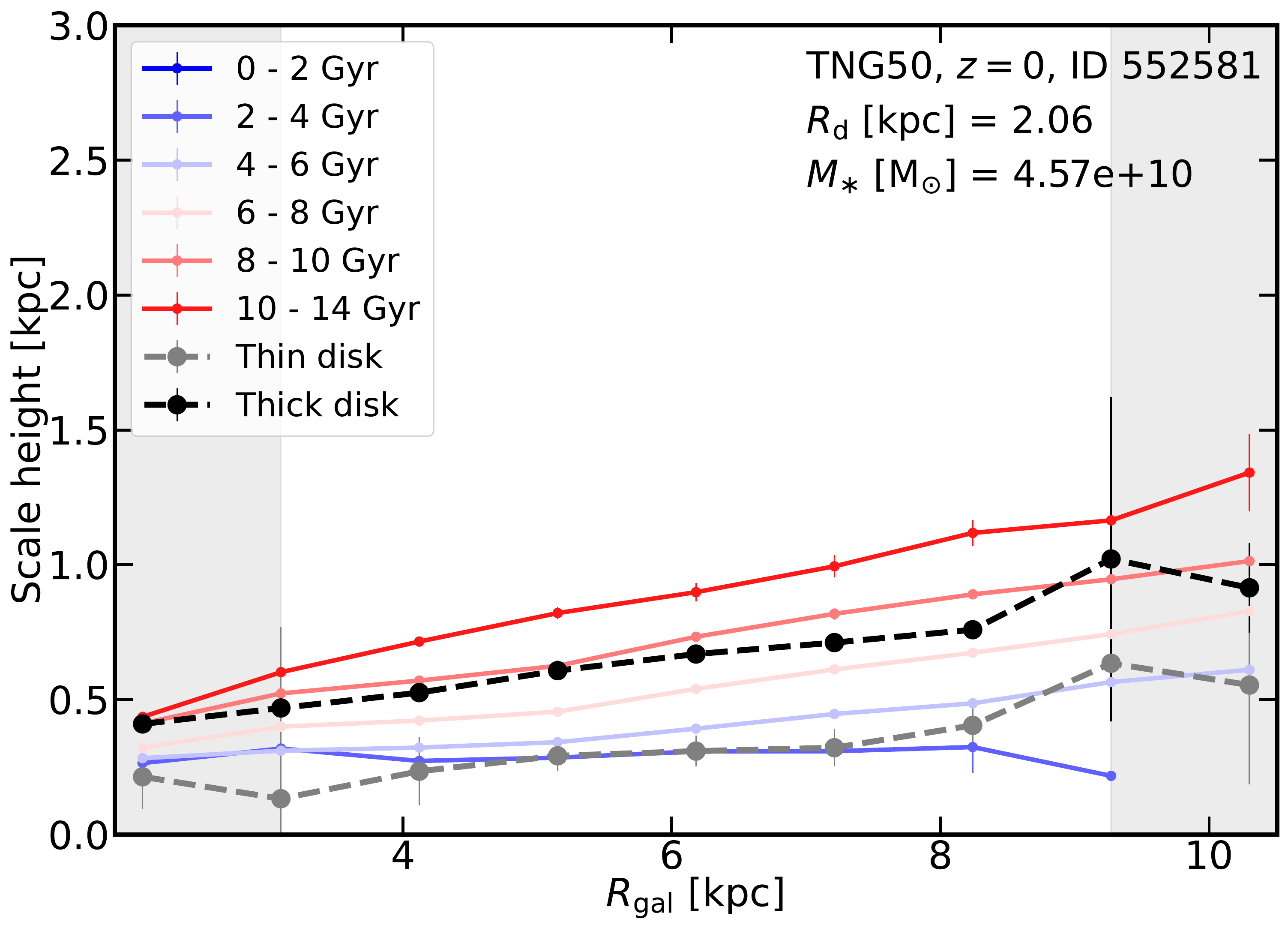}
\includegraphics[width=0.45\textwidth]{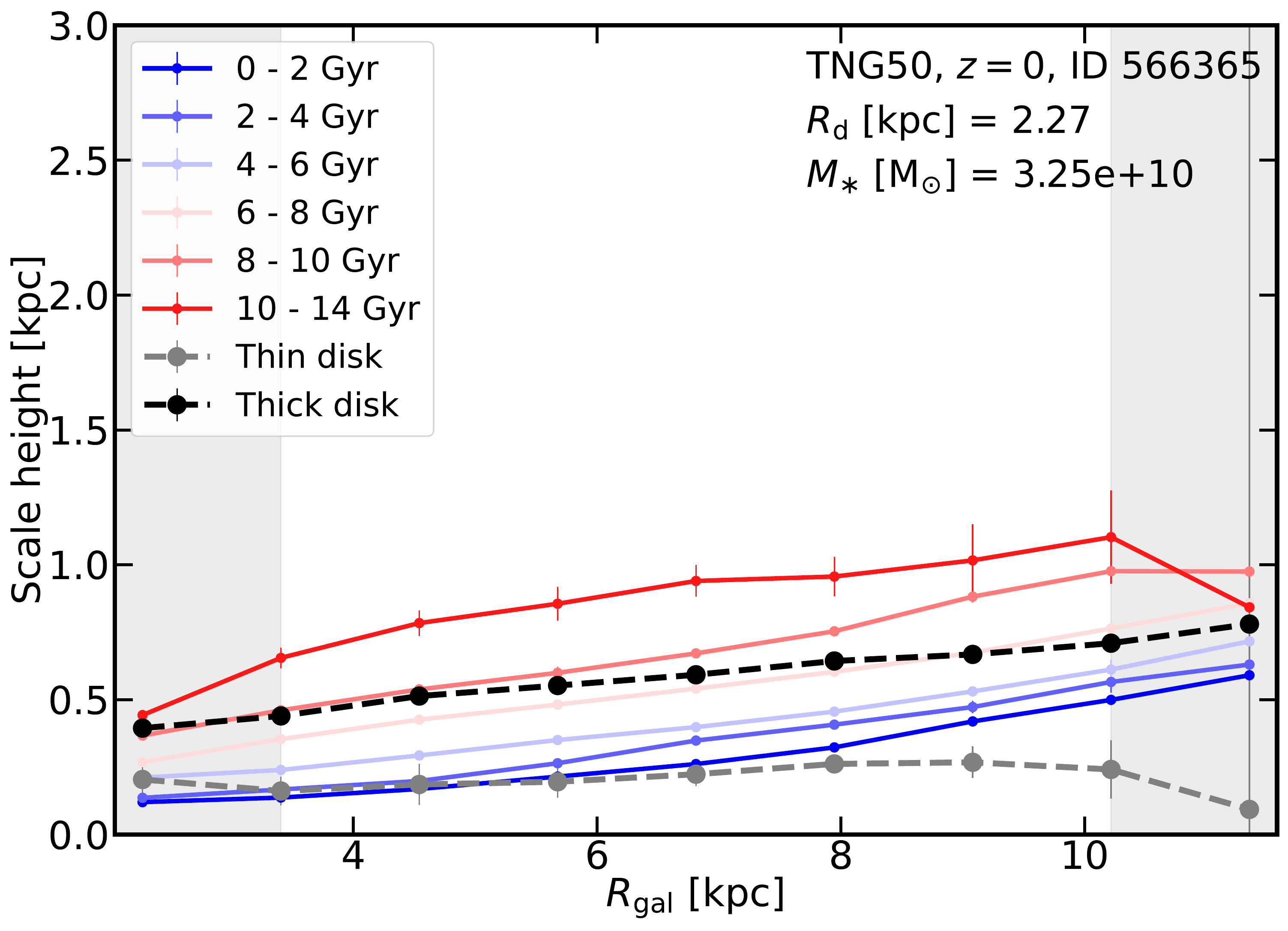}
\caption{\label{fig:flareMWanalogs} {\bf Vertical disk structure and flaring of six TNG50 MW-like galaxies.} The scaleheight as a function of the galactocentric distance (normalized by the scalelength) is shown for the six MW analogs with stellar mass and disk properties most similar to the Galaxy. Error bars represent one standard deviation errors of the parameters, provided by the fitting function.}
\end{figure*}


To visualize how the flaring quantification corresponds to diverse vertical structures, we show the change of scaleheight as a function of galactocentric distance for three galaxies, each representing different regions of the $\tau_{\rm flare}^{\rm young}- \tau_{\rm flare}^{\rm old}$ plane. A series of stellar light images of TNG50 MW/M31-like galaxies with substantial flaring are shown in Fig.~\ref{fig:images}. 

Importantly, a few galaxies from TNG50 seem to reproduce the flaring phenomenology inferred observationally for our Galaxy. The vertical magenta areas in the panels of Fig.~\ref{fig:young_old} represent the flaring (evaluated using Eq.~\ref{eq:tau_flare}) of the ``young'' stellar population of our Milky Way, with the scaleheights inferred from the vertical-action values estimated in  \cite{2019Ting}. In the iso-thermal regime, a population of stars with a mean vertical action $\langle{J_z}\rangle (R, t)$ has vertical distribution $p(z) \sim {\rm sech}^2(\frac{z}{2h_z})$ with a scaleheight $h_z(R, t) = \sqrt{\frac{\langle J_z \rangle (R,t) }{2\nu_z(R)}}$, where $\nu_z(R)$ is the local vertical frequency which we determine with Galpy \citep{2015Bovy} in the same way as \cite{2019Ting}. Therefore, assuming a vertical stellar density distribution proportional to ${\rm sech}^2$ and a Milky Way gravitational potential from \citet{2015Bovy}, we can infer a level of flaring for our Galaxy. About seven galaxies in the TNG50 sample show a flaring similar to the one inferred observationally for the Galaxy.


In particular, a good fraction of TNG50 galaxies with flaring similar to the Galaxy also have similar galaxy stellar mass. In the bottom left panel of Fig.~\ref{fig:young_old}, we show the same plot as in the top but with different symbols denoting different subsamples of the TNG50 MW/M31-like galaxies: squares indicate M31-mass objects (53 in total, $\ge 10^{10.9} \Ms$), stars indicate the 106 MW-mass galaxies. The magenta star (orange square) symbols represent the MW analogs (M31 analogs) identified in \S\ref{sec:properties}, i.e. the six (one) galaxies that have detailed stellar disk structural properties consistent with the Galaxy (Andromeda).
Within TNG50, there is no simulated galaxy with the same stellar disk structure, including extent, thickness, and flaring, of the Galaxy. However, our Galaxy represents one among many realizations of disky galaxies and, in terms of flaring of the young stars per se, according to TNG50, it appears rather common. 


Finally, the right bottom panel of Fig.~\ref{fig:young_old} is meant to convince us that the general picture depicted so far is not systematically affected or biased by cases of warps or disturbed disks (Section \ref{sec:warps}). These are highlighted in blue or orange, based on the visual inspection presented above. Although our measurements are all azimuthally-averaged, distorted disks could potentially imply an under or over estimation of the amount of flaring. There is no manifest bias toward large or small degrees of disk flaring when we focus on warped and disturbed stellar disks, namely they populate all regions of the depicted space as the rest of the population. However, the few identified warped disks seem to tend to have a young stellar population that flares more than the old one. 

\subsection{The cases of TNG50 galaxies with stellar disk properties compatible with the Galaxy's}

As already mentioned in Section \ref{sec:StellarSizes}, in the TNG50 MW/M31 sample there are six galaxies with stellar mass and disk sizes similar to the Milky Way and one galaxy with disk properties similar to Andromeda.
For the MW-analogs, and to connect more directly with the observational opportunities in our Galaxy, we show in Fig.~\ref{fig:flareMWanalogs} 
the change of the scaleheight as a function of galactocentric distance.

As already pointed out throughout this work, even for galaxies sharing the main disk properties, the flaring can be qualitatively very diverse. Indeed, for two MW-analogs, Subhalo IDs 538905 and 566365, the flaring is quite linear, whether we consider mono-age stellar populations (colored curves) or all disk stars (black and grey curves); on the other hand, the galaxy with Subhalo ID 552581 shows an exponential flaring. Additionally, the MW-analogs 516101 and 535774 show more irregular trends where mono-age population and the thin and thick disks flare quite differently.
Once we evaluate the degree of flaring in the way proposed in this work (i.e. by using Eq.~\ref{eq:tau_flare}), the relative enhancement between 1 and 5$R_{\rm{d}}$ turns out to be diverse (see bottom left panel of Fig.~\ref{fig:young_old}, magenta stars with black contours): for two of them, young stars flare more than old ones, with different level of flaring. Additionally, we have an analog where young and old stars flare equally, and an analog where old stars flare much more than young stars.

\begin{figure*}
\centering
\includegraphics[width=0.45\textwidth]{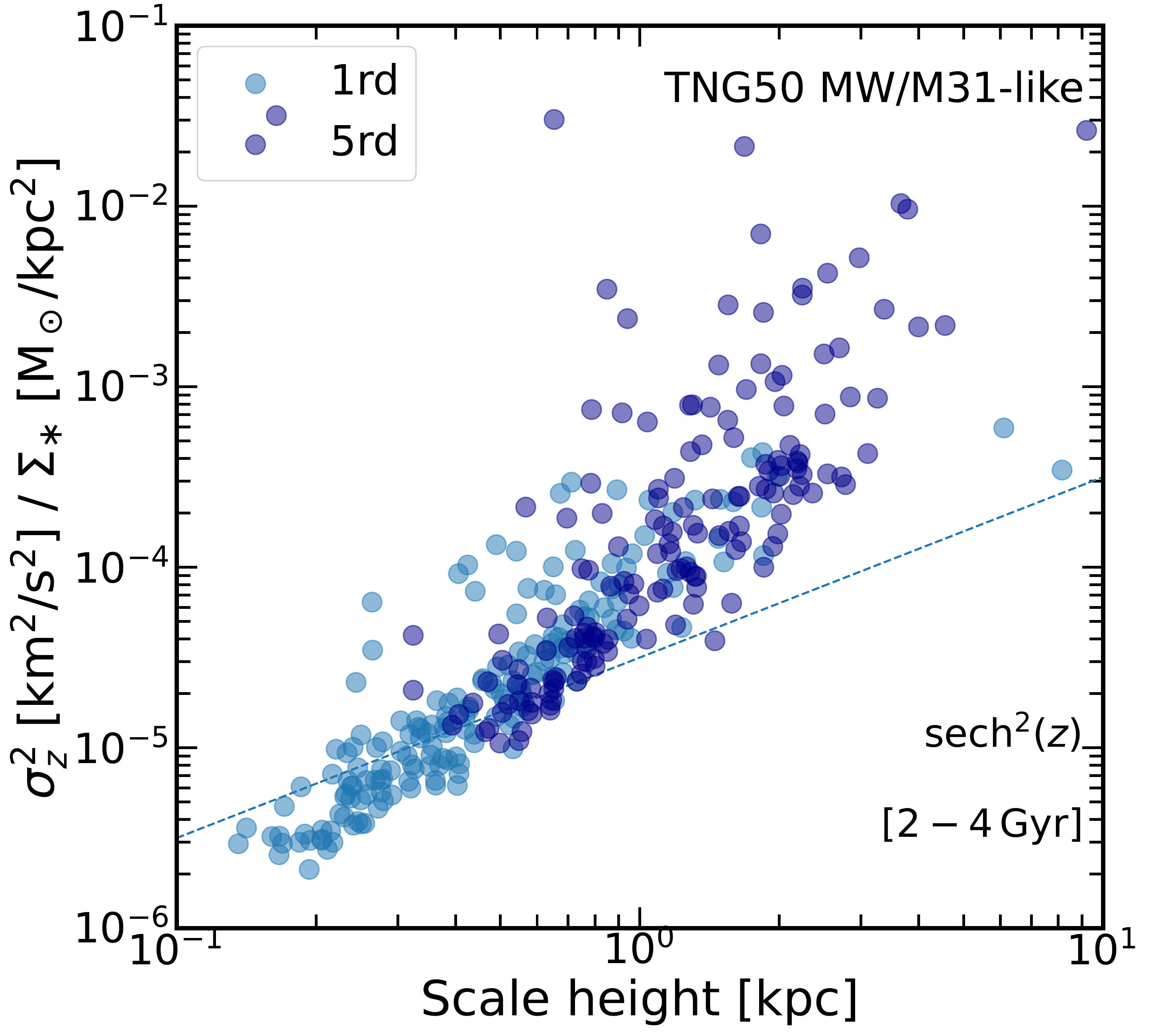}
\includegraphics[width=0.45\textwidth]{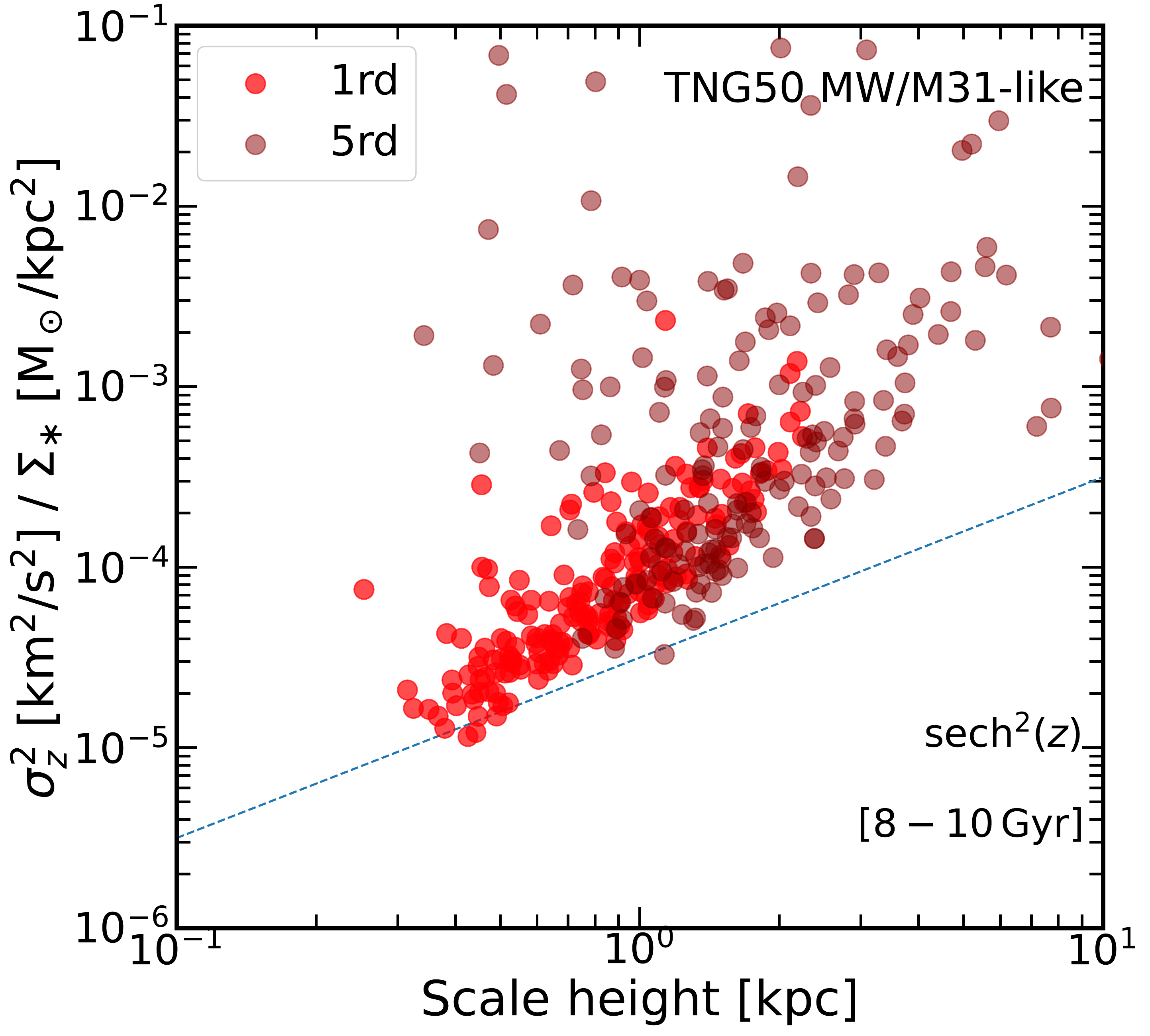}
\caption{\label{fig:young_old_sigmaVSurfDens} \textbf{Relationship between stellar heights and stellar kinematics for TNG50 MW/M31-like galaxies}. We plot the ratio between the squared local stellar vertical velocity dispersion and the stellar surface density vs. stellar disk scaleheight, for young (left) and old (right) stellar populations. We do so at different locations within the stellar disk.
The dashed line defines the linear correspondence that is  expected for an idealized self-gravitating disk.}
\end{figure*}

\begin{figure*}
\centering
\includegraphics[width=0.45\textwidth]{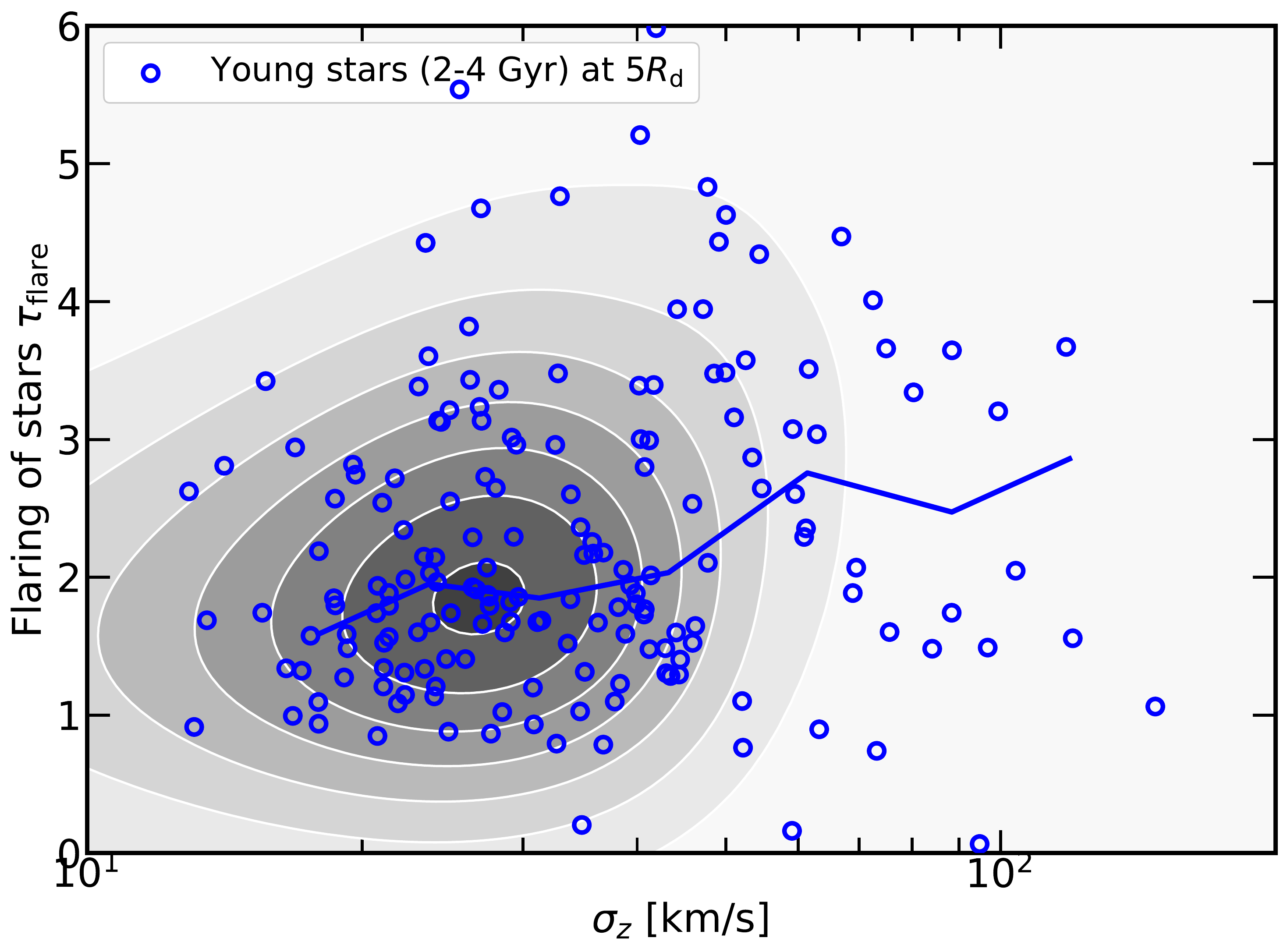}
\includegraphics[width=0.45\textwidth]{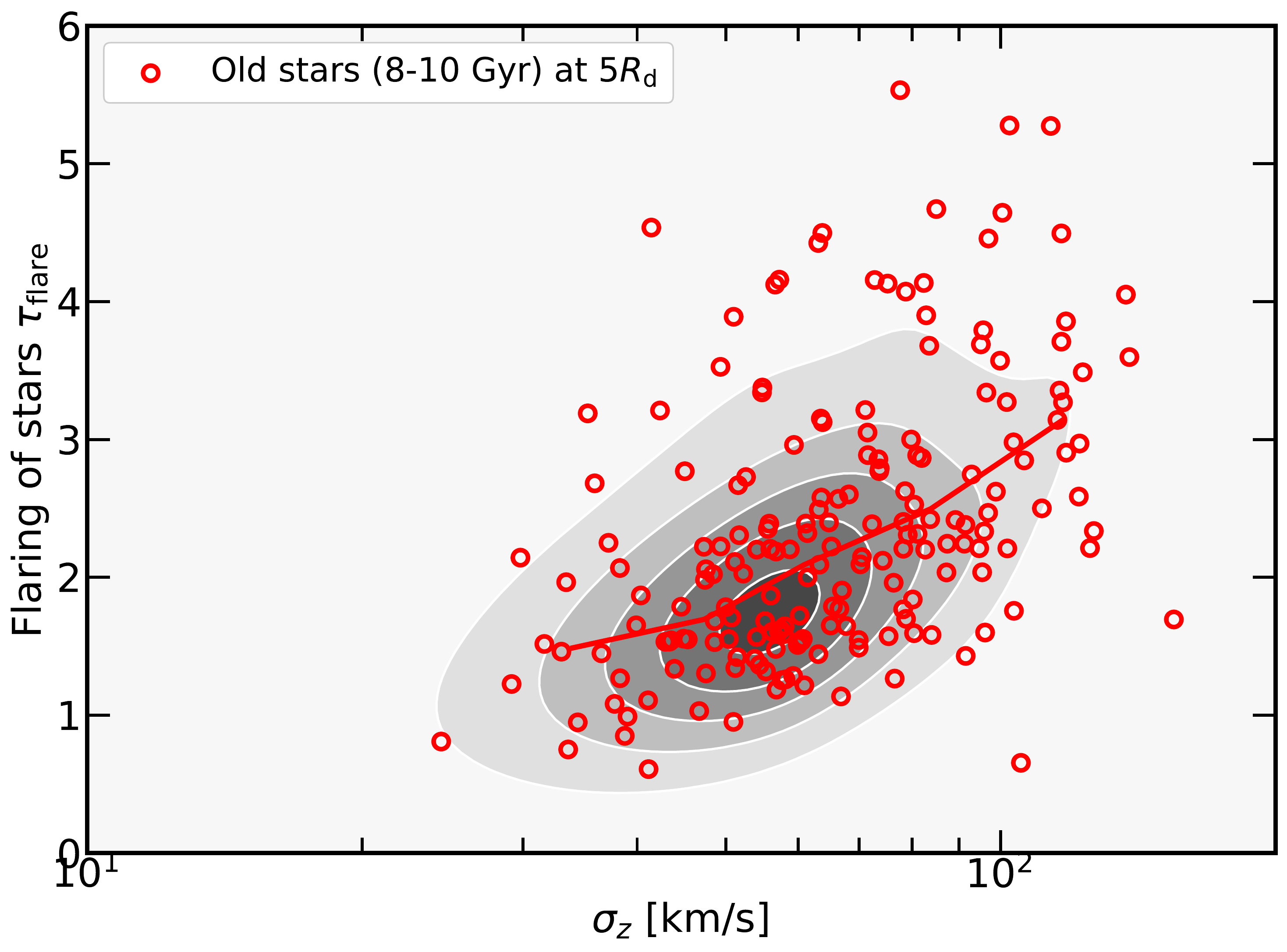}
\includegraphics[width=0.45\textwidth]{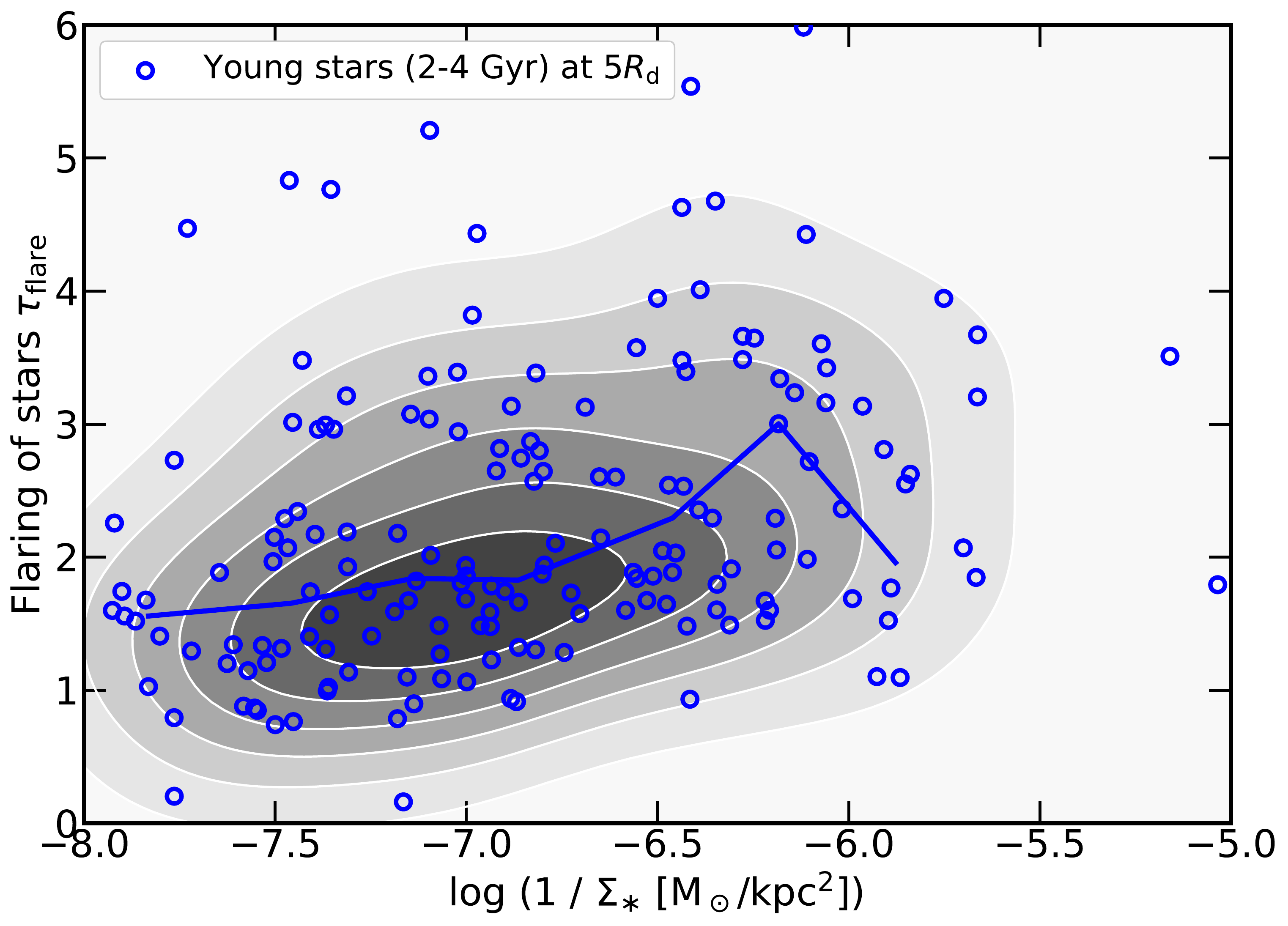}
\includegraphics[width=0.45\textwidth]{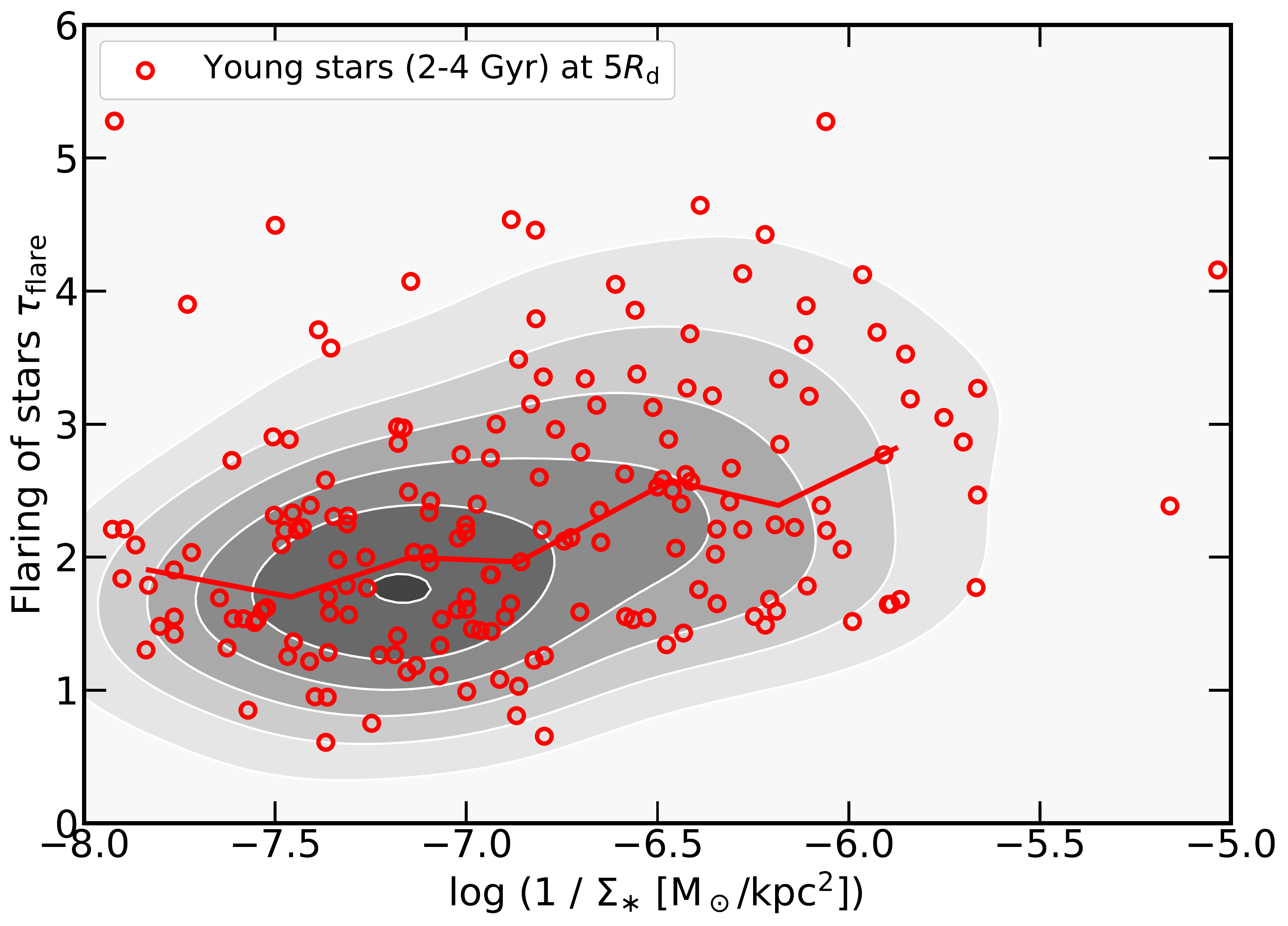}
\includegraphics[width=0.45\textwidth]{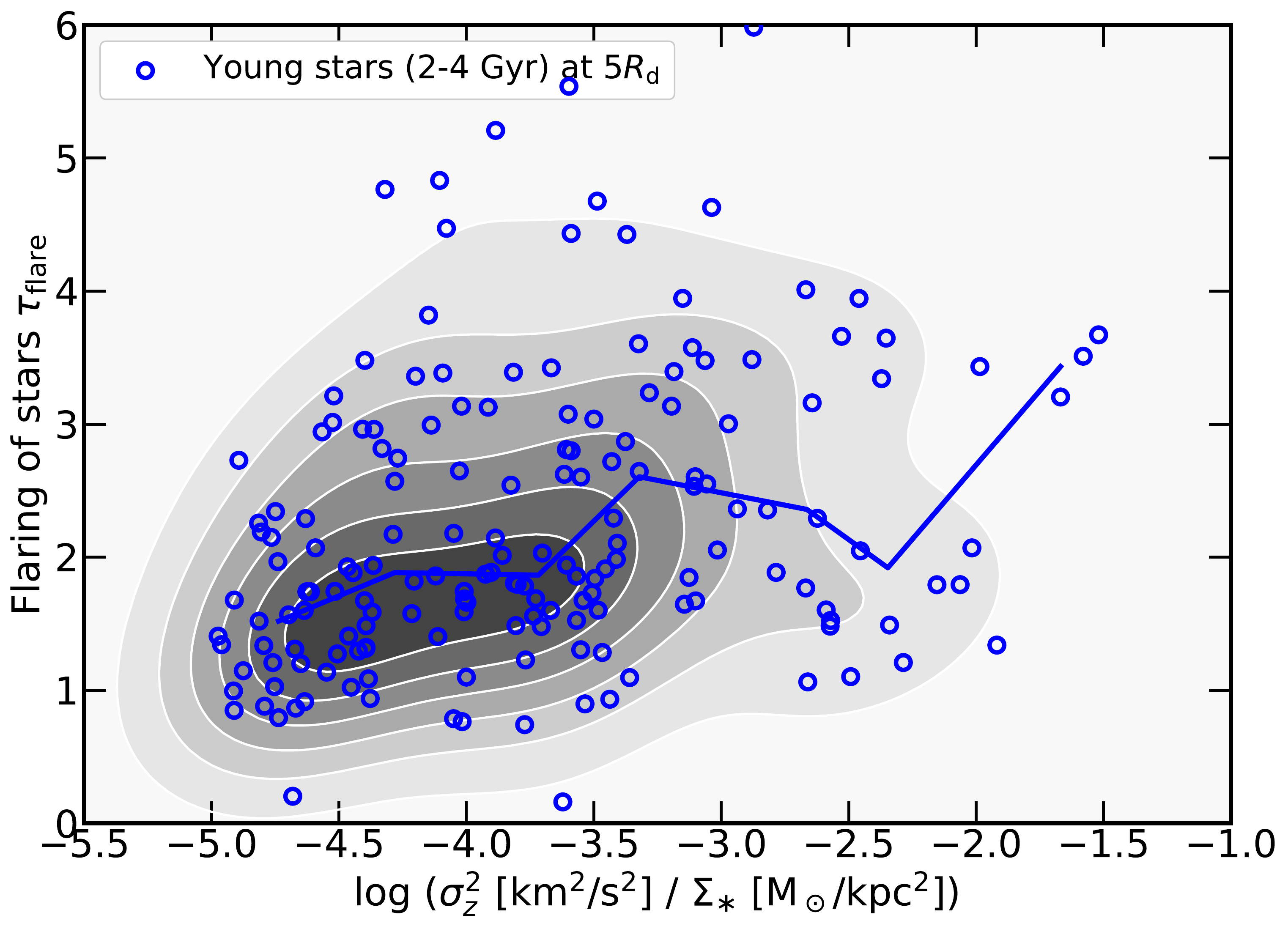}
\includegraphics[width=0.45\textwidth]{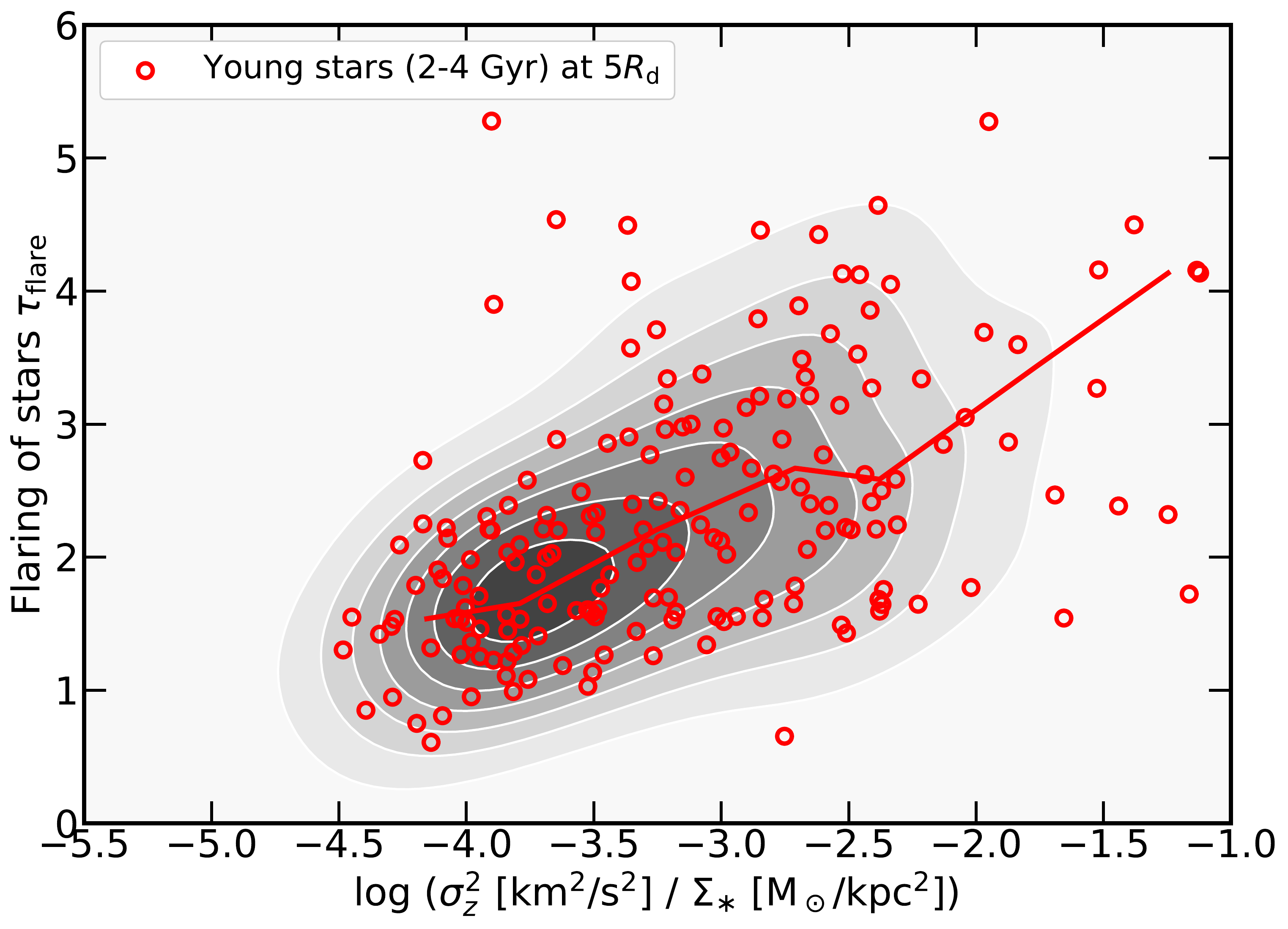}
\caption{\label{fig:tau_kinematics} \textbf{Disk flaring vs. kinematic properties for TNG50 MW/M31-like galaxies}. We plot the amount of flaring of young (left column, in blue) and old (right
column, in red) stellar populations as a function of the stellar vertical velocity dispersion at 5$R_{\rm d}$ (top) and the inverse of the stellar surface density at 5$R_{\rm d}$ (middle), and finally their ratio (bottom). In each panel, the galaxy number density is estimated with a Gaussian kernel and represented with the shaded contour areas. Solid curves are medians in bins of the quantity on the $x$-axes. Galaxies with hotter or less dense outer stellar disk flare more.}
\end{figure*}

\section{Disk flaring and kinematics in TNG50}
\label{sec:kinematics}



Our fiducial quantification and definition of disk flaring is based on a geometrical estimation of the stellar disk scaleheights. However, the structural properties of galaxies are expected to be the global manifestation of the underlying orbital configuration and interaction of all present matter components. In \citealt{2019Pillepich}, we showed that the vertical structure of both the stellar and gaseous components of TNG50 star-forming galaxies across epochs indeed is the resolved outcome of an ensemble of physical ingredients -- chiefly, the shape and depth of the overall gravitational potential, which in turn is the result of the interaction and orbital mixture of both collisional and collisionless material. Here we expand upon that analysis by focusing on MW/M31-like galaxies, and link the flaring with the kinematics of disk stars: we hence examine their vertical velocity dispersion. 

From a theoretical point of view, the scaleheight for an isothermal sheet is related, above and below a certain position, to the local vertical velocity dispersion ($\sigma^2_{\rm z}$) and the local stellar surface density ($\Sigma_{\ast}$), according to the relation \citep{1942Spitzer}: $ h_{\rm z} \propto \sigma^2_{\rm z} / \Sigma_{\ast}$.

We show in Fig.~\ref{fig:young_old_sigmaVSurfDens} this connection for the disk stars of the TNG50 MW/M31 analogs, separating between star in the inner vs. outer disk (different shades of colors) and young vs. old stellar populations (left vs. right panels, respectively). Here we adopt our fiducial choice for the heights measurement as in the previous Sections. 

A number of interesting considerations can be made. Firstly, despite the complexity of the realistic disks realized by TNG50 and even though stars are not the only matter components in the disk regions, the stellar disk heights of both young and old stellar populations relate to the underlying vertical velocity dispersion and average mass surface density: the higher the $\sigma^2_{\rm z} / \Sigma_{\ast}$ ratio, the thicker the disk.
However, and secondly, the relation may be not perfectly linear and is steeper than $ h_{\rm z} \propto \sigma^2_{\rm z} / \Sigma_{\ast}$ (dotted lines): this is probably due to the gas and dark matter contributing to the disk potential and also to the morphological complexity of a real galactic stellar disk compared to the idealized isothermal sheet.

We have quantified the same relation by using the half-mass heights of the disks instead of the best fit to a squared hyperbolic secant function, as a more robust estimate of the scaleheight against disk internal structure and inhomogeneities: we notice, although we do not show, that the scatter in the relationships of Fig.~\ref{fig:young_old_sigmaVSurfDens} at fixed galactocentric distance is considerably reduced in such a case. This suggests that the galaxy-to-galaxy scatter in Fig.~\ref{fig:young_old_sigmaVSurfDens} is not all due to physical effects and the most severe cases of galaxy outliers are those where the parametric functional forms are not a very good description of the vertical stellar mass distribution in the disk.

Fig.~\ref{fig:young_old_sigmaVSurfDens} shows that, also according to TNG50 and as argued in previous works, at fixed galactocentric distance in the disk, older stars are not only thicker but also hotter than younger ones, i.e. exhibit overall larger  velocity dispersions (left vs. right panel). The same phenomenology has been measured in the Milky way \citep{2004Nordstrom,2015Dorman} and also in other cosmological simulations \citep{2020Buck}.

We also find, although do not show, that at fixed galactocentric distance, the stellar disk heights exhibit a correlation with both the local stellar velocity dispersion and the local stellar mass surface density, individually: namely, stellar disks are thicker when they are hotter or less dense. But so, what does determine the flaring of the stellar heights?

In Fig.~\ref{fig:tau_kinematics}, we show how the level of disk-height flaring of TNG50 MW/M31-like galaxies depends on stellar velocity dispersion (top), inverse of the stellar mass surface density (middle) and the ratio of the two, i.e. $\sigma^2_{\rm z} / \Sigma_{\ast}$ (bottom), whereby these quantities are evaluated in the outer disk regions. To avoid any issue with fitting the vertical mass profiles with parametric functions, here we quantify the flaring by measuring the stellar half-mass heights (see also \S\ref{sec:discussion}). We also quantify the flaring and its relationship with stellar kinematics and potential for young and old stellar populations separately: left vs. right panels of Fig.~\ref{fig:tau_kinematics}.

From this analysis we can see that, according to TNG50, galaxies with hotter or less dense outer stellar disks flare more strongly, in both young and old stellar populations. By comparing the top to the middle panels, we can also conclude that the diversity in flaring is mostly driven by the diversity in outer-disk temperatures, i.e. in stellar kinematics, rather than by a diversity in stellar surface density.

\section{Discussion}
\label{sec:discussion}


\subsection{On other methods to quantify the disk flaring}
One of the main messages of this paper is that, not only across differently-simulated galaxies, but also within one given simulation, the nature and amount of disk flaring can be very diverse, even for disk galaxies within a narrow range of galaxy stellar mass and environments. We hence advocate for a non parametric quantification of flaring, as provided in \S\ref{sec:demographics}. 

One possible drawback of this proposition is that, at least in our fiducial implementation, it still relies on a parametric measure of the stellar disk heights. In \S\ref{sec:fit_procedure_heights}, we have described different methods that are commonly used to quantify disk heights and all results in the previous sections, unless otherwise stated, are based on our fiducial choice: profile fitting with single and double squared hyperbolic secant, for mono-age stellar populations and all-ages disk stars, respectively. 

\begin{figure*}
\centering
\includegraphics[width=0.45\textwidth]{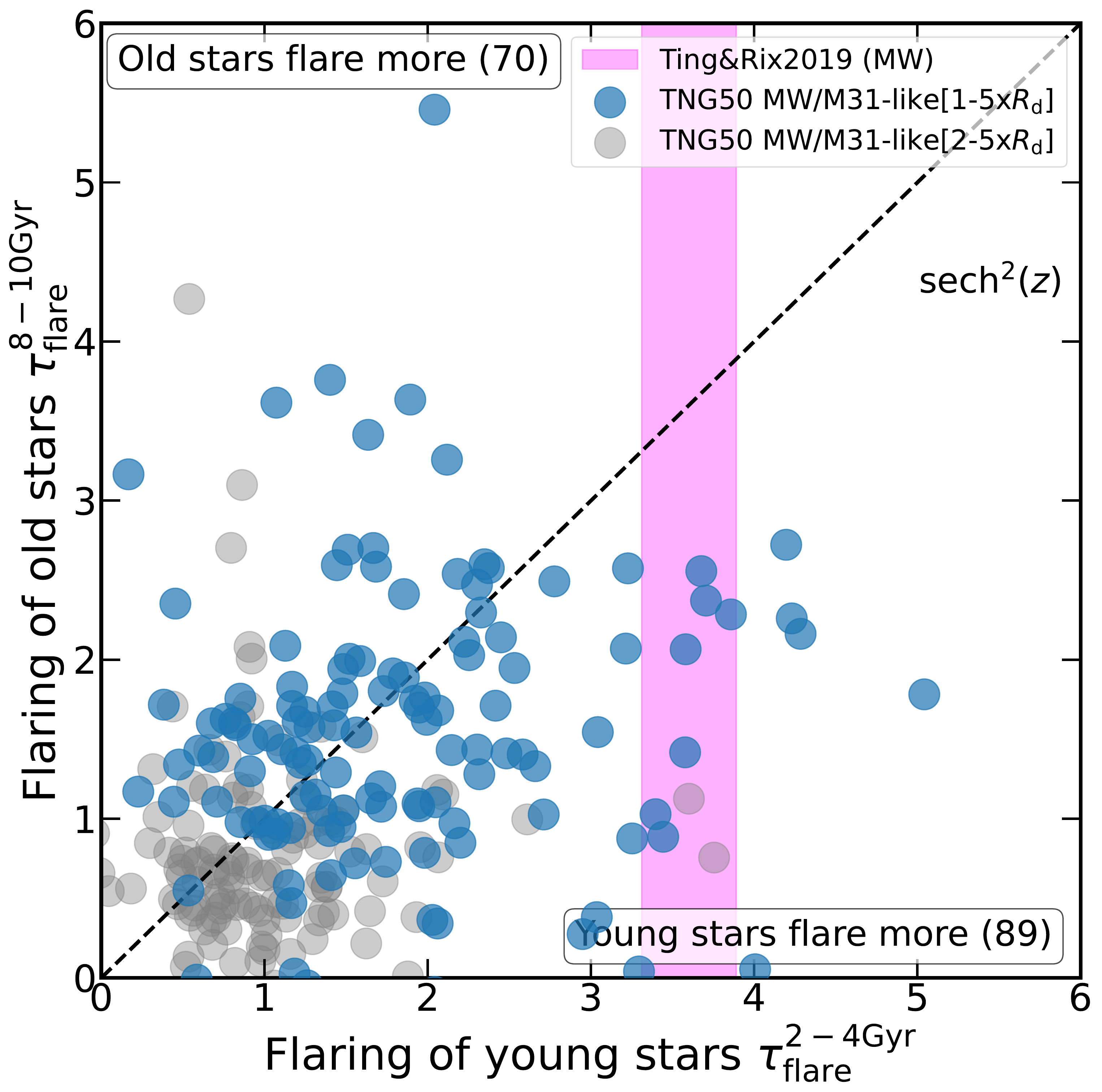}
\includegraphics[width=0.45\textwidth]{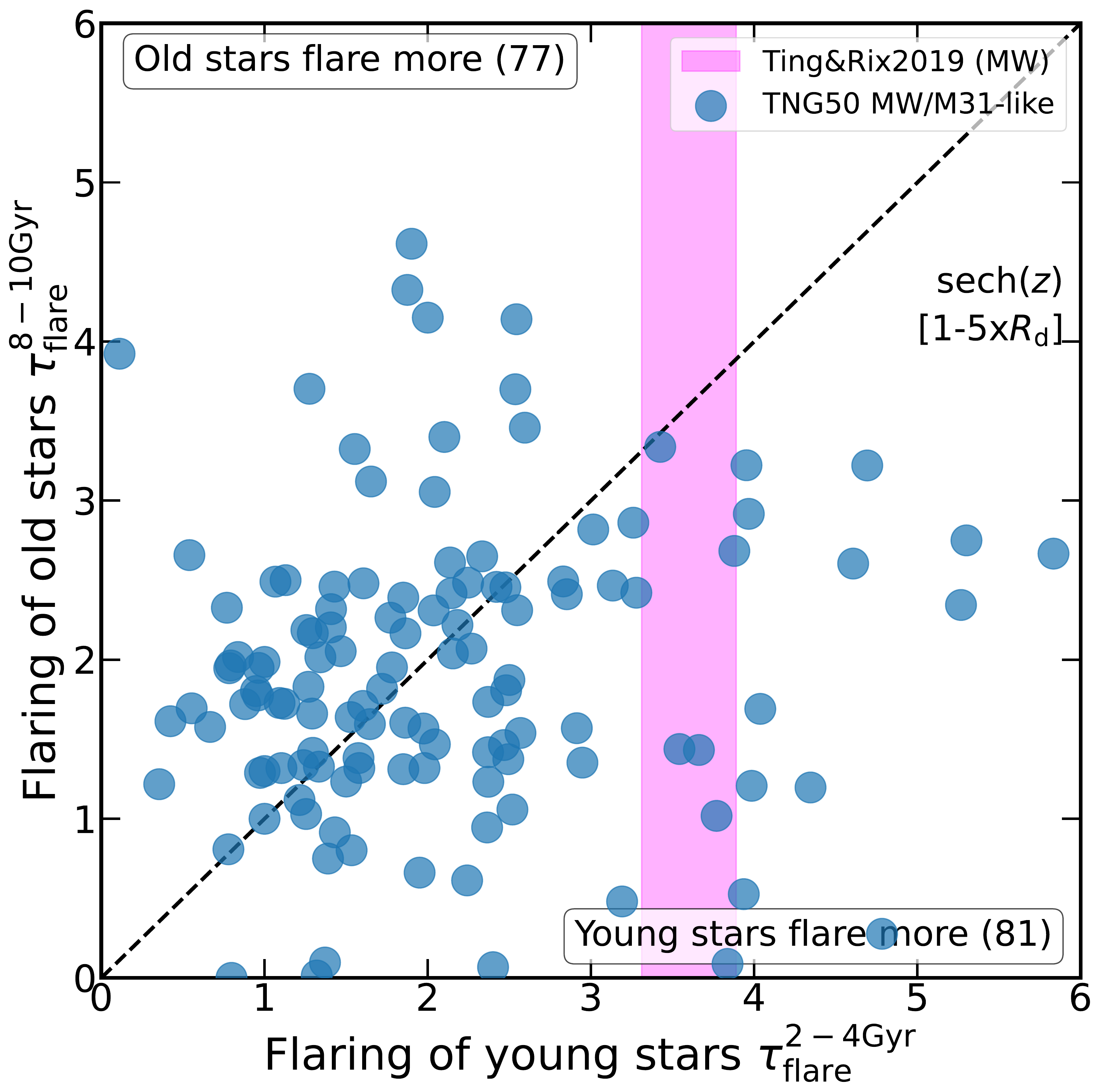}
\includegraphics[width=0.45\textwidth]{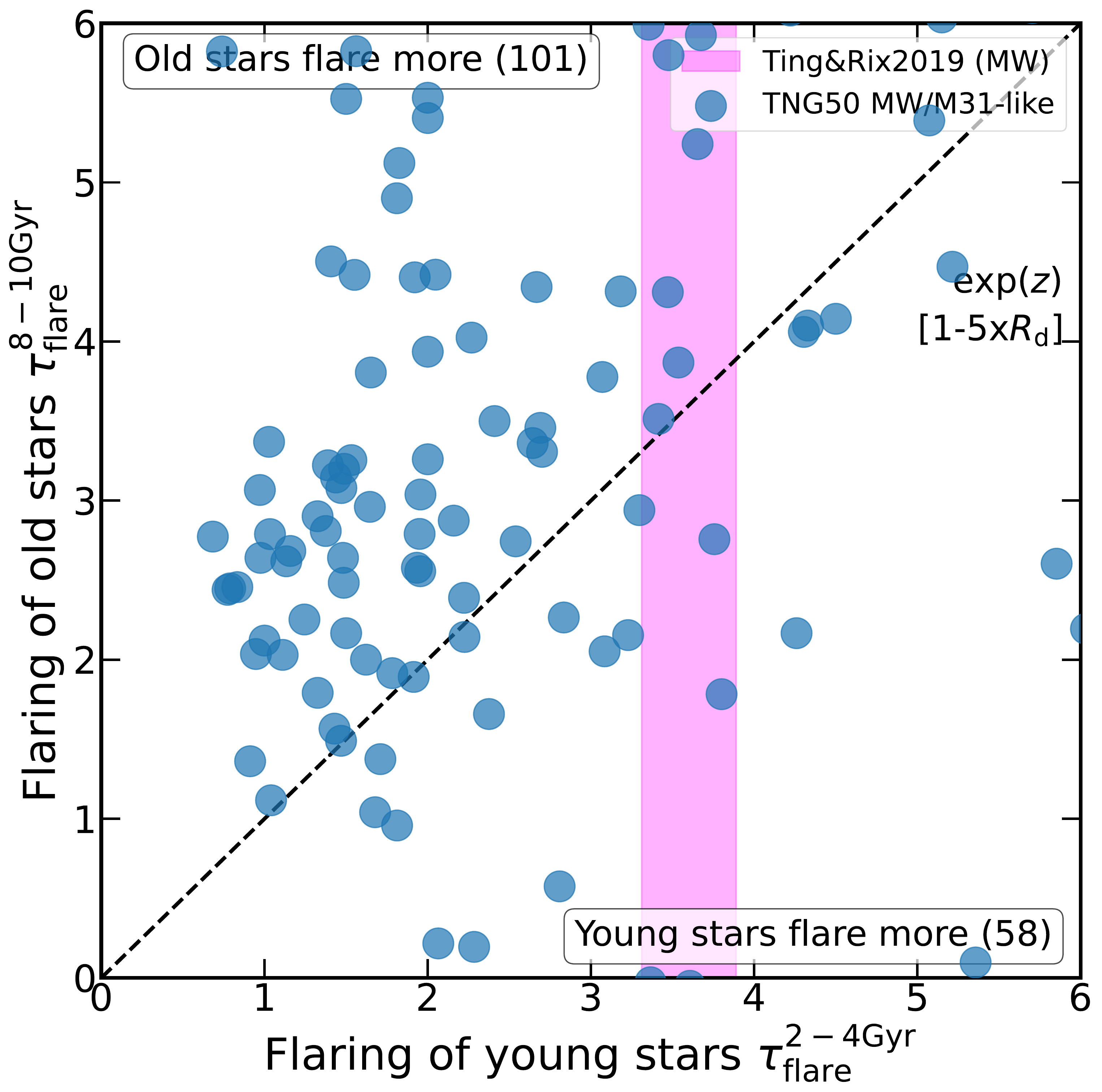}
\includegraphics[width=0.45\textwidth]{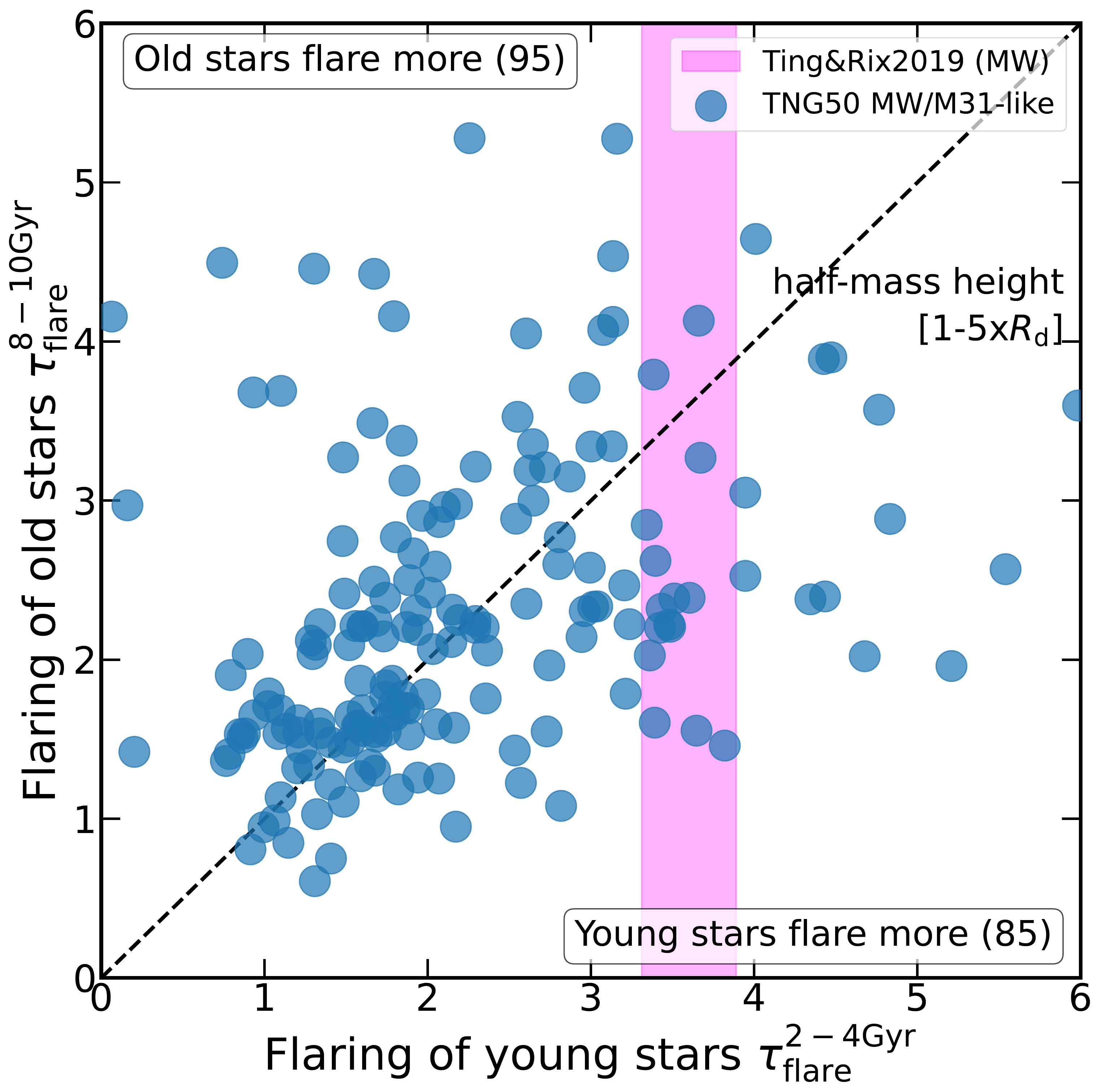}
\caption{\label{fig:young_old_4methods} {\bf Flaring of young vs. old stellar populations in TNG50 MW/M31-like galaxies for different methods to measure disk heights.} Top left: quantification of the flaring with the fiducial height measurements based on fitting a squared hyperbolic secant profile (as in all figures so far): blue data points refer to the flaring measured between 1 and 5$\times R_{\rm d}$ (same as main panel of Fig.~\ref{fig:young_old}), whereas in grey the flaring is evaluated at 2 vs. 5$\times R_{\rm d}$.} Top right:  hyperbolic secant profiles. Bottom left: exponential profiles. Bottom right: half-mass heights. Estimating disk heights with the exponential fit returns, among the others, the most different results, both quantitatively and qualitatively.
\end{figure*}

We show now in Fig.~\ref{fig:young_old_4methods} how the flaring values for young and old stellar populations depend on the method for height measurements. We show again the main panel of Fig.~\ref{fig:young_old} in the top left, unchanged, and repeat it for the additional parametric cases in the top right (hyperbolic secant) and bottom left (exponential). The bottom right panel shows the flaring measurement based on non-parametric half-mass height measurements, as in Fig.~\ref{fig:tau_kinematics}.

All four cases exhibit roughly similar flaring ranges and three of them also have similar distributions on the depicted plane (squared and single hyperbolic secant and half-mass heights), whereas for the exponential profiles the distribution appears more sparse and dissimilar.
The fraction of galaxies where the young stars flare more than the old stars (and vice-versa) also changes from method to method, although not in qualitatively significant manners, barring the case of the exponential fits. The latter is the method that returns the most different qualitative and quantitative quantification of flaring. The quantification based on the non-parametric measurements of the stellar half-mass disk heights provides confidence that, in the population, fit-based assessments are not dramatically affected by possible numerical/fitting issues or by the possibility that parametric functional forms do not describe the stellar mass distributions well. On the other hand, individual galaxies may exhibit somewhat different amounts of flaring depending on the adopted method to measure their disk heights. Comparisons can hence only be done once the same operational definitions are adopted on all sides.

Considering that some of these galaxies may have a bar at a galactocentric distance of $1R_{\rm d}$, we have repeated the measurements of the flaring parameter in the range $(2-5)\times{R_{\rm d}}$ instead of $(1-5)\times{R_{\rm d}}$: the results are shown in the top left panel of Fig.~\ref{fig:young_old_4methods}, in grey. The values are smaller (as we can expect from the flaring phenomenon) but they are distributed similarly to the original definition on both sides of the identity line. 

\subsection{A note to observers: on the ``flaring'' based on the spatial distribution of stellar ages}
\label{sec:agesInDisks}

As introduced in Section \ref{sec:intro}, it has become customary to inspect the vertical structure of the Galactic stellar disk by looking at the mean or median stellar ages as a function of galactocentric radius $R_{\rm gal}$ and as a function of vertical distance from the midplane $|z|$. This is typically done in observations of the Galaxy \citep{2016Ness,2017Xiang,2019Feuillet} as well as with simulation data \citep{2017Ma,2020Buck,2021Agertz}. 

\begin{figure*}
\centering
\includegraphics[width=\textwidth]{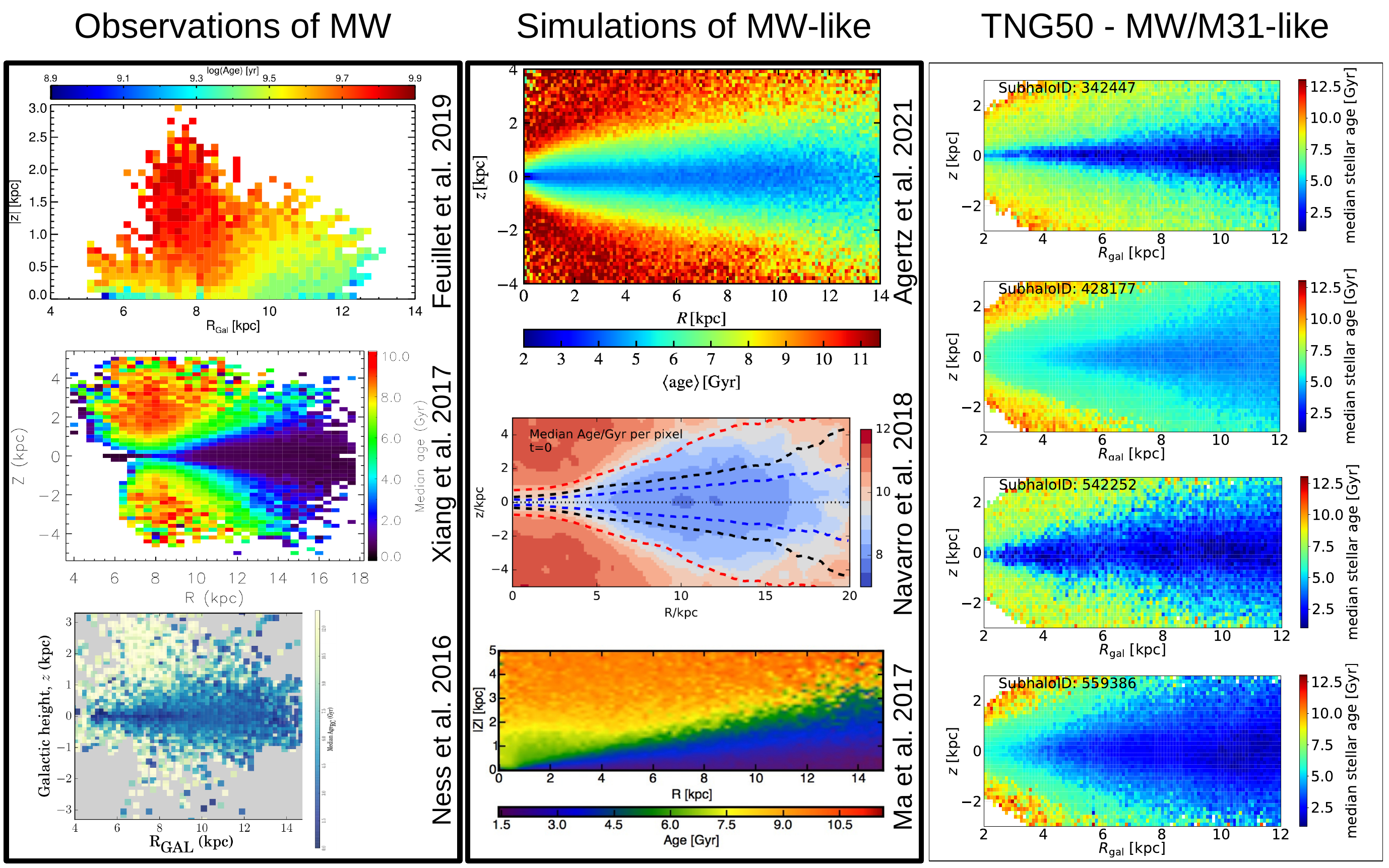}
\caption{\label{fig:flaring_obs} \textbf{Stellar age distributions in the MW/M31 midplane}: vertical distance from the midplane as a function of galactocentric distance with the color code representing the mean or median stellar age. Here we compare observations from \citet{2016Ness,2017Xiang,2019Feuillet} (left), simulations from \citet{2017Ma,2018Navarro,2021Agertz} (center), and a sample of five TNG50 MW/M31 analogs (right). Barring those in the right column, the plots are replicated from the respective research studies without modification.}
\end{figure*}

The common emergent picture is that of a \textit{funnel} shape, in which at each radius $R_{\rm gal}$, the mean age of disk stars increases with $|z|$, whereas at fixed $|z|$ the mean or median stellar age decreases with galactocentric radius. This can be appreciated in Fig.~\ref{fig:flaring_obs}, where we have compiled examples from the literature, for the Galaxy \citep[left column][]{2019Feuillet,2017Xiang,2016Ness} and from MW-like zoom-in cosmological simulations \citep[central column][]{2021Agertz,2017Ma,2018Navarro}, and where it can be seen that young stars at large galactocentric distances can indeed be found at large altitudes. 

For comparison, on the right column of Fig.~\ref{fig:flaring_obs}, we show five TNG50 MW/M31 analogs that are representative of the whole sample and that qualitatively reproduce previously-quantified observational and theoretical scenarios, for $R_{\rm gal}=3-14$ kpc and $|z|=0-4$ kpc. Here we divide all stars in $100\times 100$ bins on the $|z|-R_{\rm gal}$ plane, and the colors denote the mean stellar ages in each pixel. Also in TNG50 MW/M31-like galaxies, the mean stellar age increases with $Z$ at fixed $R_{\rm gal}$, and decreases with $R_{\rm gal}$, at fixed $Z$, but with different detailed patterns depending on the galaxy. 

However, it is important to realize that, even if these funnel features have been often connected to the flaring of the disk, they in fact do not imply it. Flaring is the increase of the stellar disk thickness at increasing radii. The change with radius and height of the typical ages of the disk stars \textit{do not} imply flaring: an un-flared inside-out model could still produce the age maps with a funnel shape. 

To demonstrate this we produce two MW disk mocks from the model described in \cite{2020Frankel} and inspect for both cases the stellar density and mean age as a function of radius and height (we do not show the plots here). Without flaring and inside-out growth, no funnel shape is present. However, when we activate inside-out growth, we can recover the same pictures as in Fig.~\ref{fig:flaring_obs}.

\begin{figure}
\centering
\includegraphics[width=0.48\textwidth]{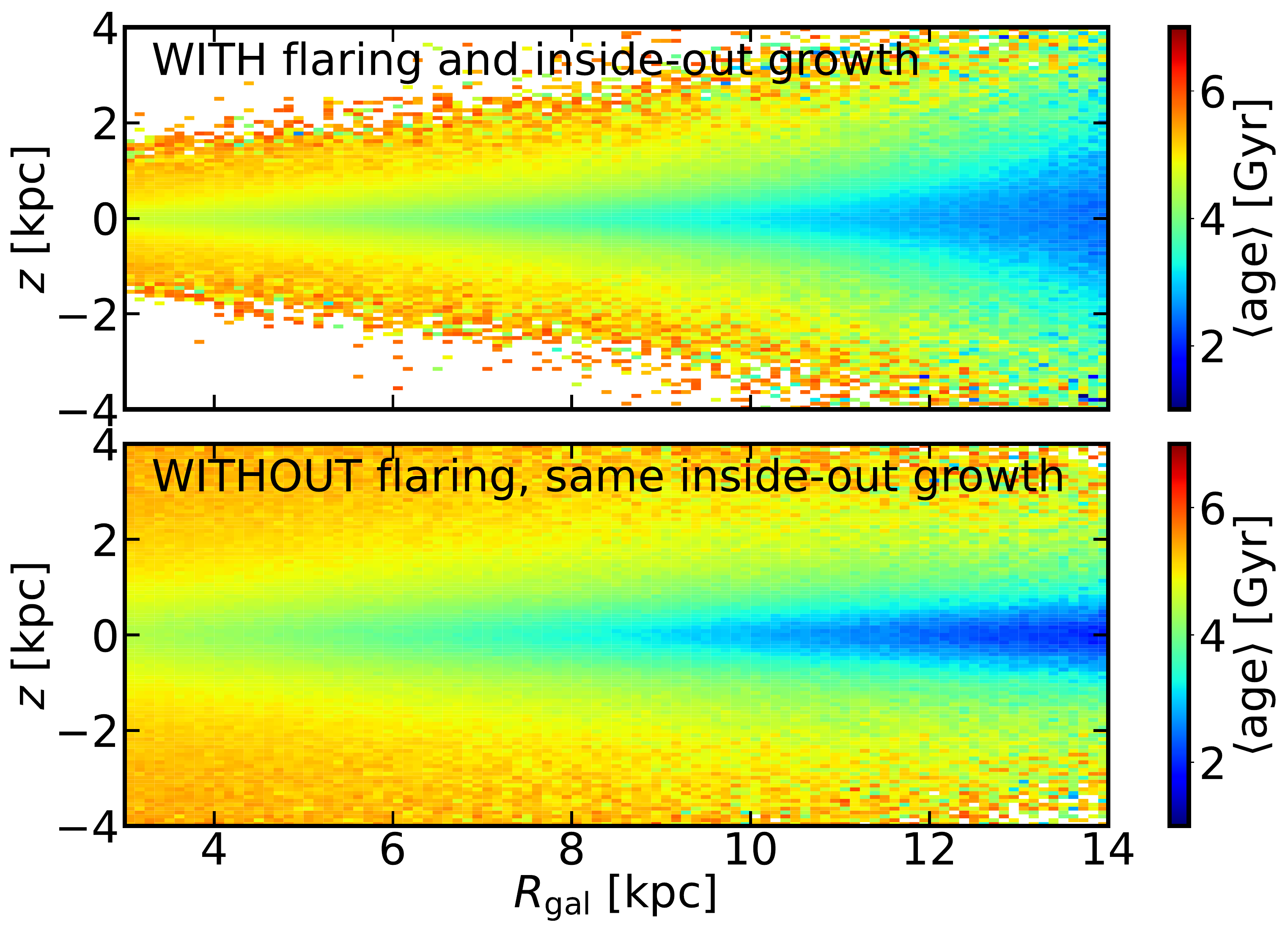}
\caption{\label{fig:demonstration_noflaring}{\bf Stellar mean age distributions for two MW-like disk mocks.} We illustrate the $R-z$ plane of two model galaxies color-coded by mean age (as in Fig.~\ref{fig:flaring_obs} from a toy disk model that includes flaring (top) and one that does not (bottom). The toy model is adapted from the best-fit of \citet{2020Frankel, 2019Ting}. In the top panel, the vertical distribution of stars is a $\mathrm{sech}^2$ function with an age- and radius-dependent scale-height as described in \citet{2020Frankel} that captures the flaring of the Galactic disk as described and fit by \citet{2019Ting}. In the bottom panel, the model is the same, but the vertical distribution of stars is replaced by a non-flaring toy distribution, where the scaleheight is essentially only a function of age. In particular, we took $h_z(t) = 0.15+0.1 (t/1 \mathrm{Gyr})$ kpc a (not physically motivated, but simple) example. The funnel shape of iso-age contours in the bottom panel arises naturally from the combination of inside-out formation of the disk and subsequent vertical heating, not from disk flaring.}
\end{figure}

To illustrate this further, we show in Fig.~\ref{fig:demonstration_noflaring} the $R-z$ plane color-coded by mean age that result from two toy models, one with flaring and one without flaring. Despite the absence of flaring, the model without flaring exhibits the typical funnel structure purely as a result from inside-out formation and vertical heating. The model with flaring is adapted from the model described in \cite{2020Frankel}, which builds on the best fit vertical distribution of \cite{2019Ting}. In summary, the model describes a star formation history of the disk from the inside-out, forming on an exponential radial profile, and the subsequent orbit evolution in the plane (radial migration and radial heating as diffusion in angular momentum and radial action) and out of the plane as a scale-height that increases with radius and time. The toy model without flaring has exactly the same formation scenario and parameters and the same in-plane orbit evolution, but the vertical distribution is now modelled as a scale-height that increases with time as $h_{\rm z}(t) = 0.15 + 0.1(t/1 \mathrm{Gyr})$ independent of radius. As it can be appreciated, a funnel-like distribution of stellar ages can be in place with or without actual flaring, i.e. with or without orbital changes of the stars.

\subsection{Comparison to previous simulations}
\label{sec:comparison}

As we have already decribed in Section \ref{sec:theoretical}, even if the flaring of the stellar disk has been studied and investigated in a number of previous theoretical works in the literature, a comparison among them is non-trivial. This is due to the different ways to quantify the flaring and the diverse ways in which the flaring is manifest in the various simulated galaxies and simulation models. In this section we attempt a closer comparison between the TNG50 results and those from selected zoom-in simulations of MW-mass disk galaxies, such as NIHAO-UHD \citep{2020Buck}, Latte \citep{2017Ma}, VINTERGATAN \citep{2021Agertz}, and Auriga \citep{2017Grand}.

\begin{figure*}
\centering
\includegraphics[width=0.49\textwidth]{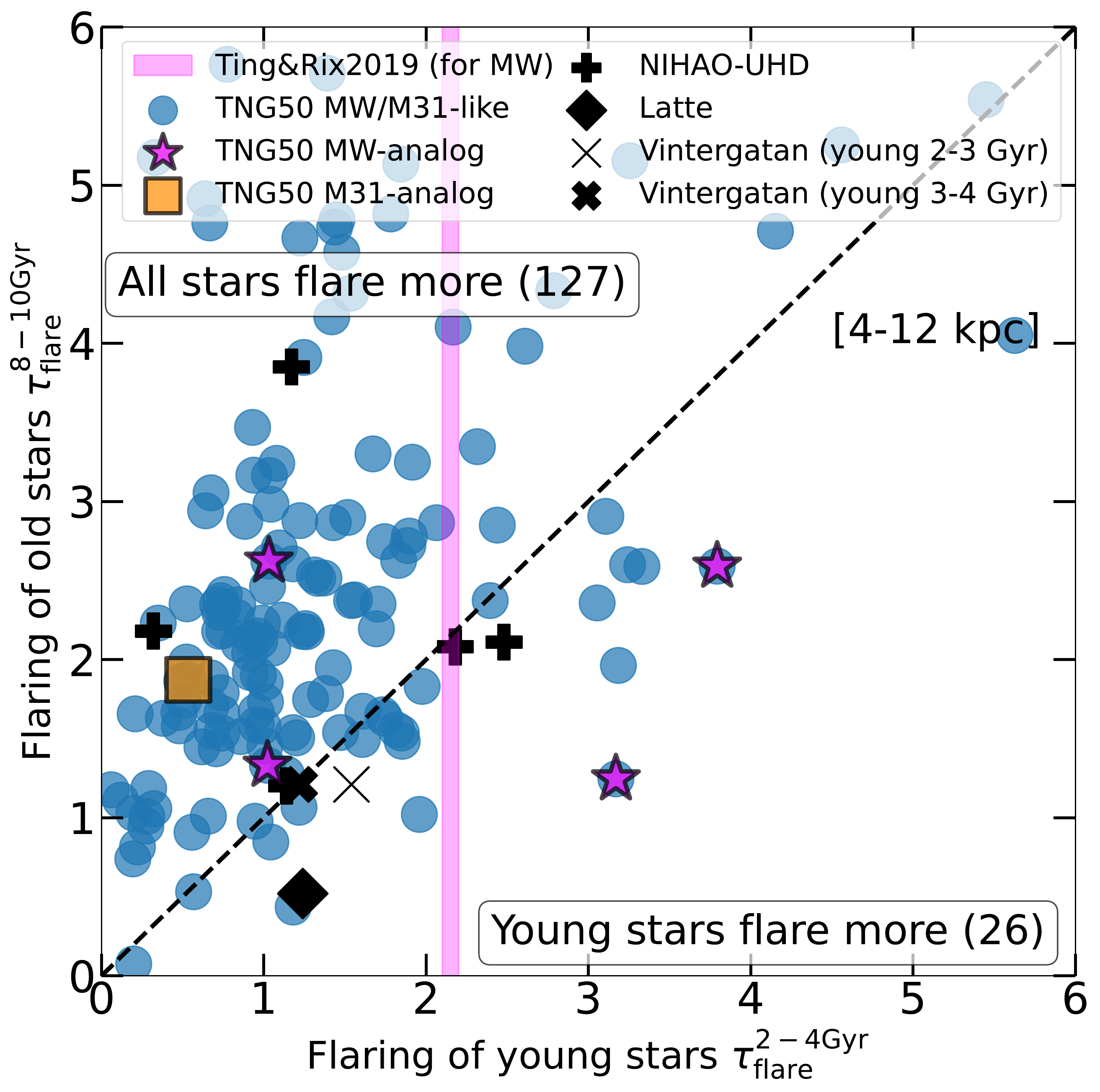}
\includegraphics[width=0.49\textwidth]{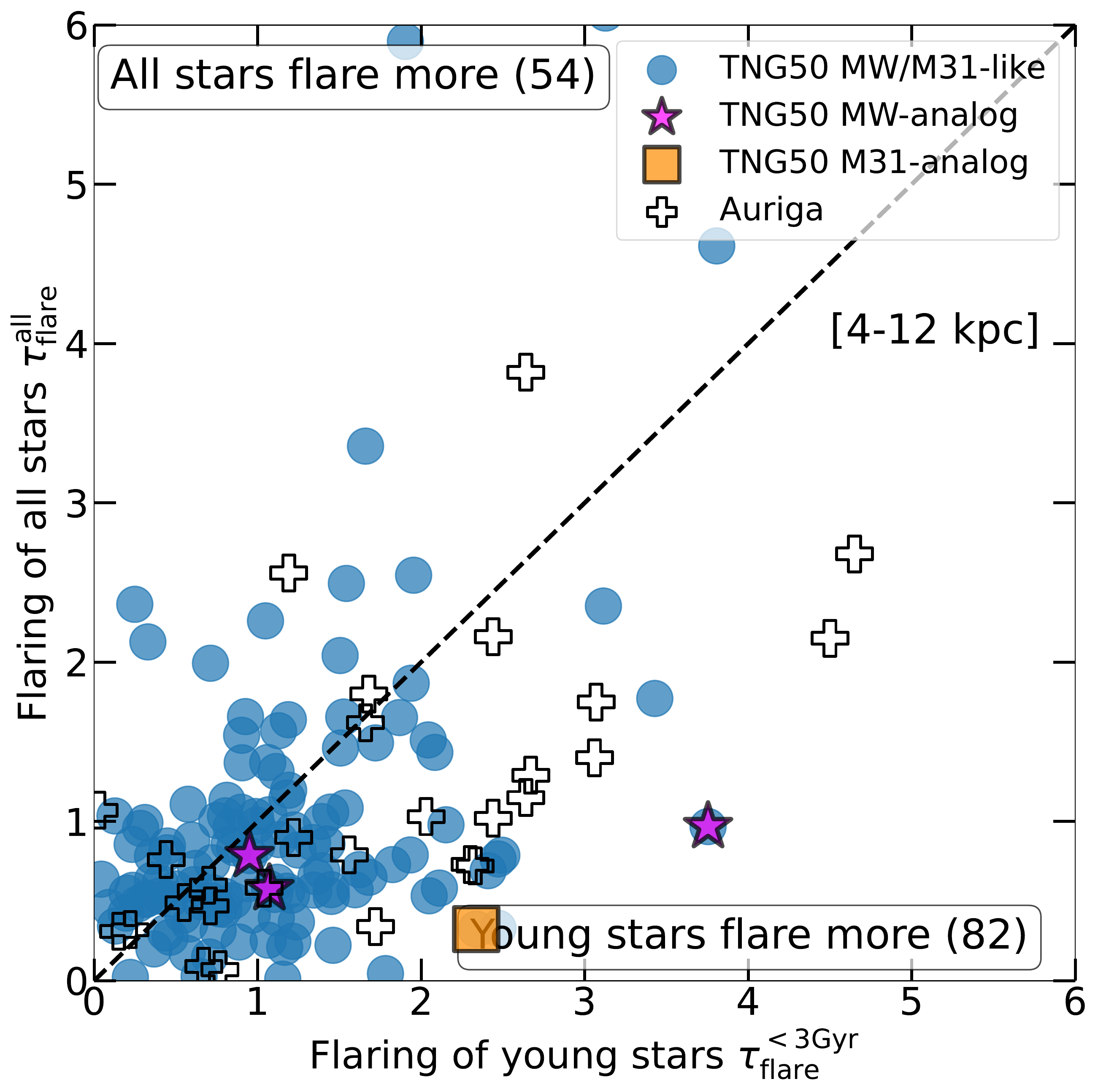}
\caption{\label{fig:tng50vsothers} \textbf{Flaring of TNG50 MW/M31 analogs in comparison to the results of other cosmological MW-like galaxy simulations}. We show again the flaring of old and young stars for TNG50 MW/M31 analogs in comparison to, on the left, simulations from \citet[][VINTERGATAN: thin and thick crosses, with young stellar populations of 2-3 Gyr and 3-4 Gyr, respectively]{2021Agertz}, \citet[][Latte: diamonds]{2017Ma}, and \citet[][NIHAO-UHD: plus symbols]{2020Buck}; on the right, simulations from the Auriga project \citep{2017Grand}. The magenta area denotes the stellar flaring of young stars of the Milky Way extrapolated from \citet{2019Ting}, as in Fig.~\ref{fig:young_old}. 
We note that in all the simulation models used for this comparison (except for Latte that uses ${\rm sech}^2$), the scaleheight of the mono-age stellar populations in the disk is evaluated from a single exponential fit. In these plots, the heights are hence measured using exponential profiles also for TNG50.
}
\end{figure*}

We note that in all the simulations used for this comparison, the scaleheight of the mono-age stellar populations that constitute the stellar disk is evaluated from a single exponential fit, which is not our fiducial choice -- this is the case for all but for Latte, where a single ${\rm sech}^2$ profile is used instead.
We hence proceed as follows. From the available literature, we extrapolate the scaleheight of young and old stellar population at 4 and 12 kpc from the galactic center (to be consistent with all the selected models). Unless otherwise stated we try to stick to our definition of \textit{young} (2-4 Gyr) and \textit{old} (8-10 Gyr) age bins. 
Then, using Eq.~\ref{eq:tau_flare} (but with the heights at the physical distances of 4 and 12 kpc instead of $1\times R_{\rm{d}}$ and $5\times R_{\rm{d}}$), we evaluate the amount of flaring $\tau_{\rm flare}$ of each model and we plot them in concert with the TNG50 results in Fig.~\ref{fig:tng50vsothers}. In order to make an apples-to-apples comparison, we measure the scaleheights of also the TNG50 sample by fitting single exponential profiles to the stellar vertical density distribution.

In the left panel of Fig.~\ref{fig:tng50vsothers}, TNG50 is compared with NIHAO-UHD (black plus symbols), Latte (black diamond) and VINTERGATAN (black crosses). We note that in Latte, the ``old'' stellar population is composed by all stars older than 8 Gyr while in
VINTERGATAN the bins of stellar ages are different from the one used in this work, being $\Delta$age = 1 Gyr. To be consistent, we then plot both of them, a thick cross denoting stellar populations of 3-4 Gyr and a thin cross representing stars of 2-3 Gyr. 

Because of the diverse treatment of the flaring of the mono-age stellar populations adopted for the Auriga galaxies, we choose to separate this comparison from the others, and show it in the right panel of Fig.~\ref{fig:tng50vsothers}.
In this case, we apply to our TNG50 sample the same choices as in the Auriga's paper. Indeed, here the young stars are defined to be all disk stars below 3 Gyr and their flaring is shown against all disk stars. 
For the 30 MW-like galaxies in the Auriga sample, the vertical density distribution of the disk is fitted at a series of different radii with a single exponential profile. However, as already mentioned in \S\ref{sec:mass_profile}, the vertical profiles of the TNG50 MW/M31-like galaxies are more often better described with a double functional than a single functional profile when all disk stars are considered. Nevertheless, also in this case we measure the heights for the TNG50 galaxies by fitting a single exponential profile. As in the left panel, the flaring is evaluated between 4 and 12 kpc.

As it is clear from Fig.~\ref{fig:tng50vsothers}, the TNG50 MW/M31-like sample returns and brackets all the other theoretical findings, including the most extreme cases: very small flaring of the old population, as in Latte, or cases where the old stars flare much more than the young ones, as in one of the NIHAO-UHD galaxies. We see also that in the TNG50 sample we have galaxies where the flaring values are larger than in previous simulations.
The consistency of the outcomes in Fig.~\ref{fig:tng50vsothers} is indeed a remarkable result. Until now it was not possible to say whether the diversity of flaring manifestations predicted by simulations was a genuine manifestation of galaxy-to-galaxy variation or was due to different numerical codes, galaxy-formation models, galaxy/halo selection or even different numerical resolution. The results with TNG50 demonstrate that MW/M31-like galaxies can exhibit very diverse levels and flavours of flaring simply due to galaxy-to-galaxy diversity.

\begin{figure*}
\centering
\includegraphics[width=0.45\textwidth]{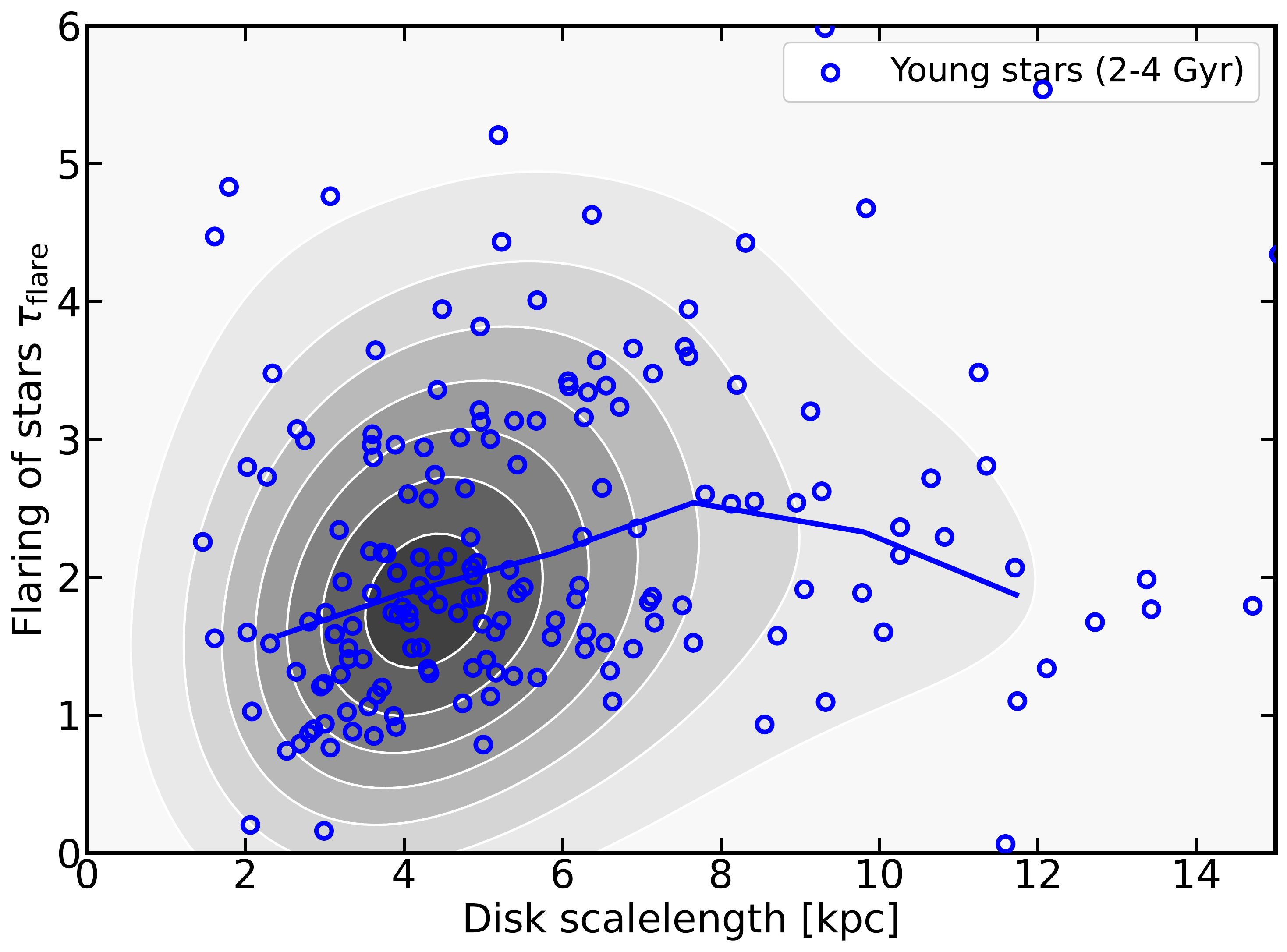}
\includegraphics[width=0.45\textwidth]{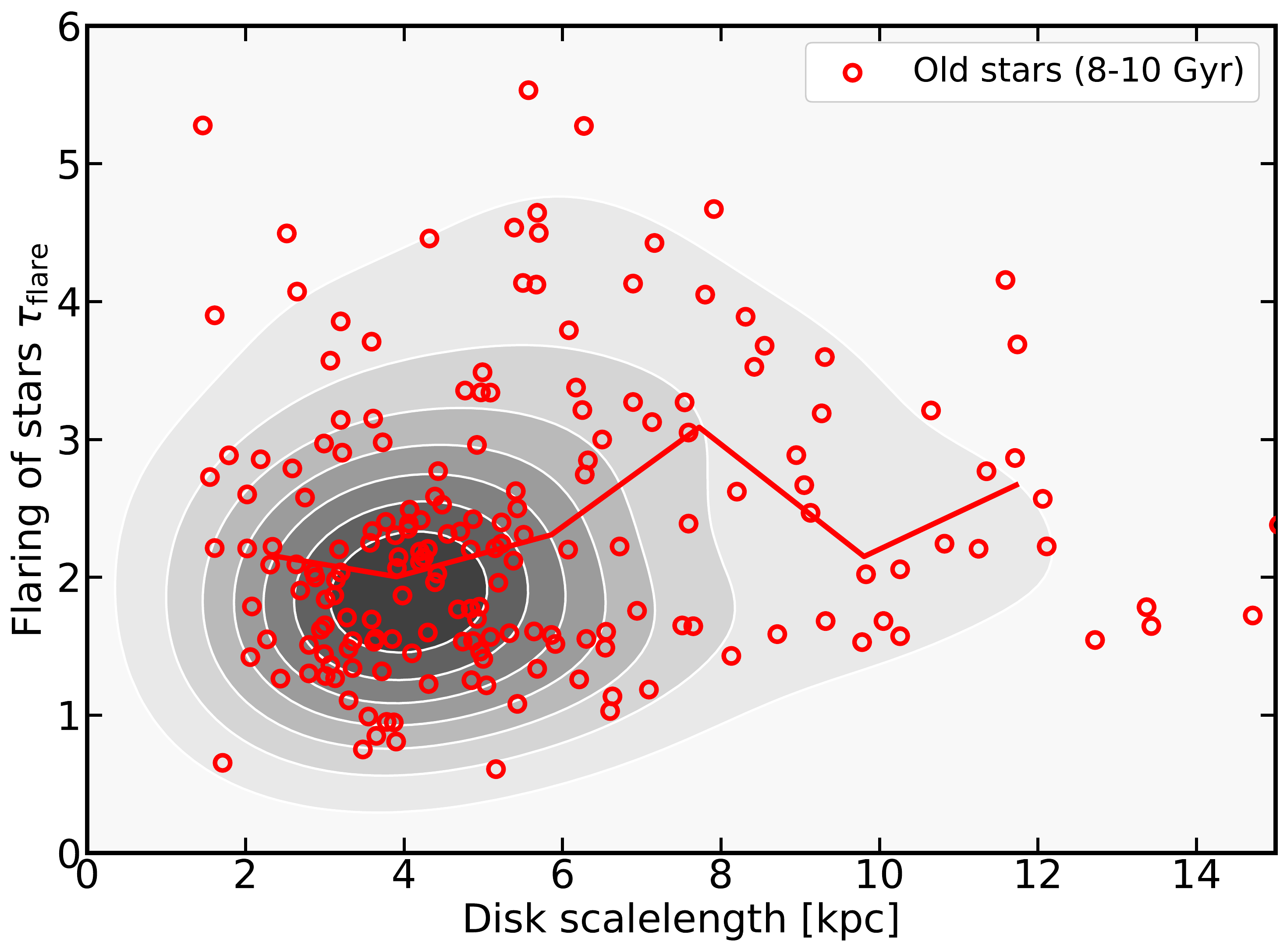}
\includegraphics[width=0.45\textwidth]{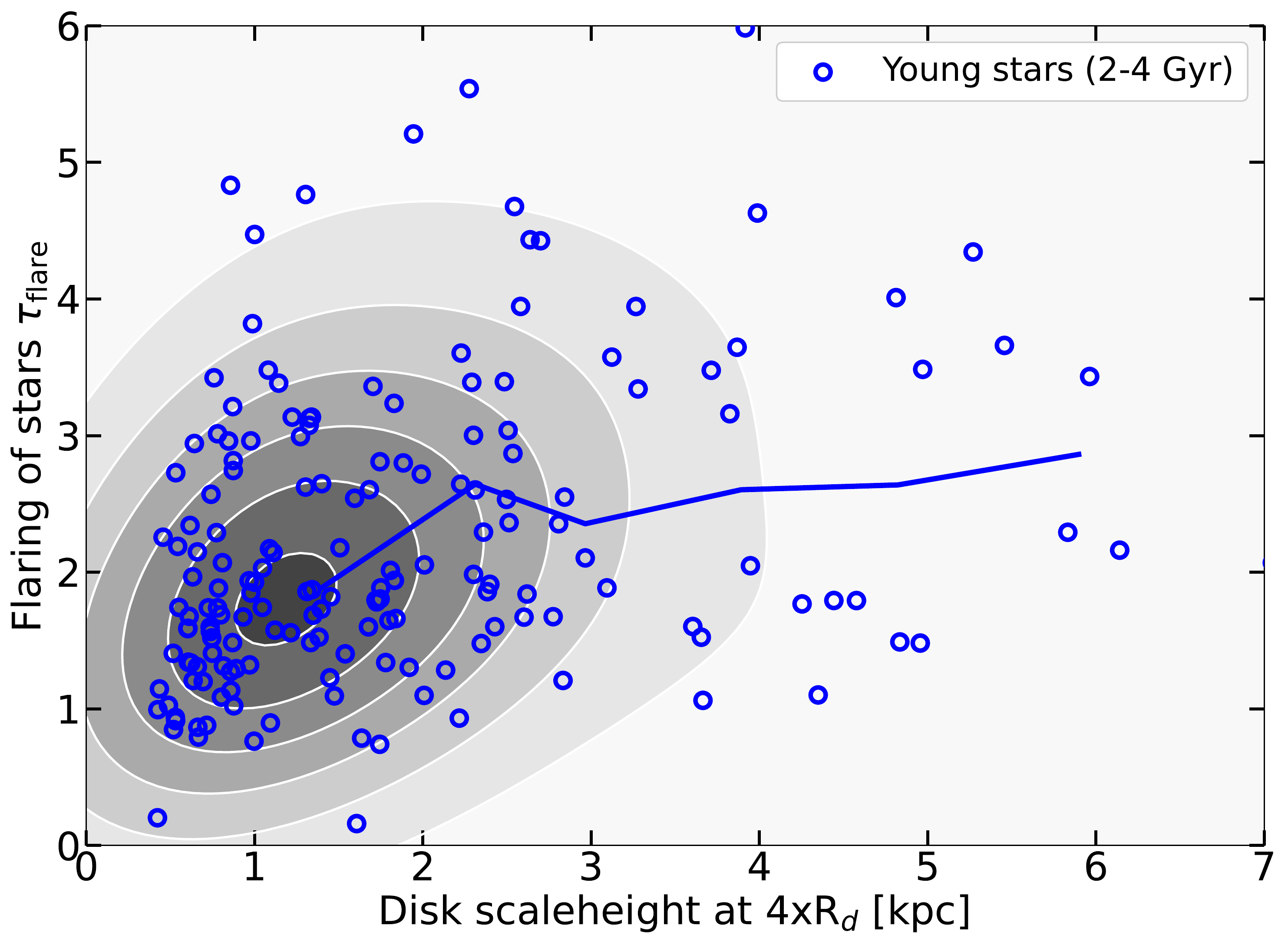}
\includegraphics[width=0.45\textwidth]{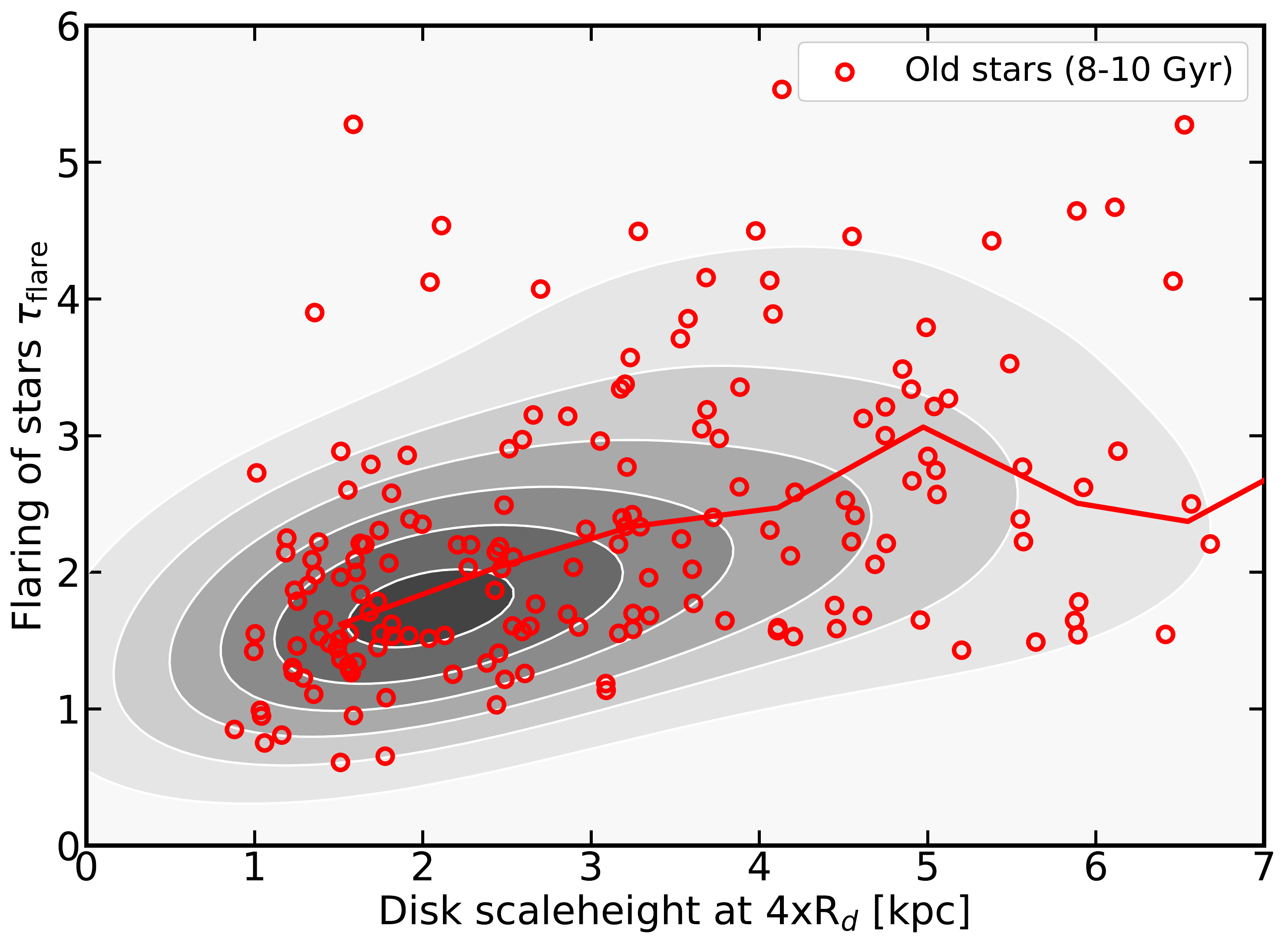}
\includegraphics[width=0.45\textwidth]{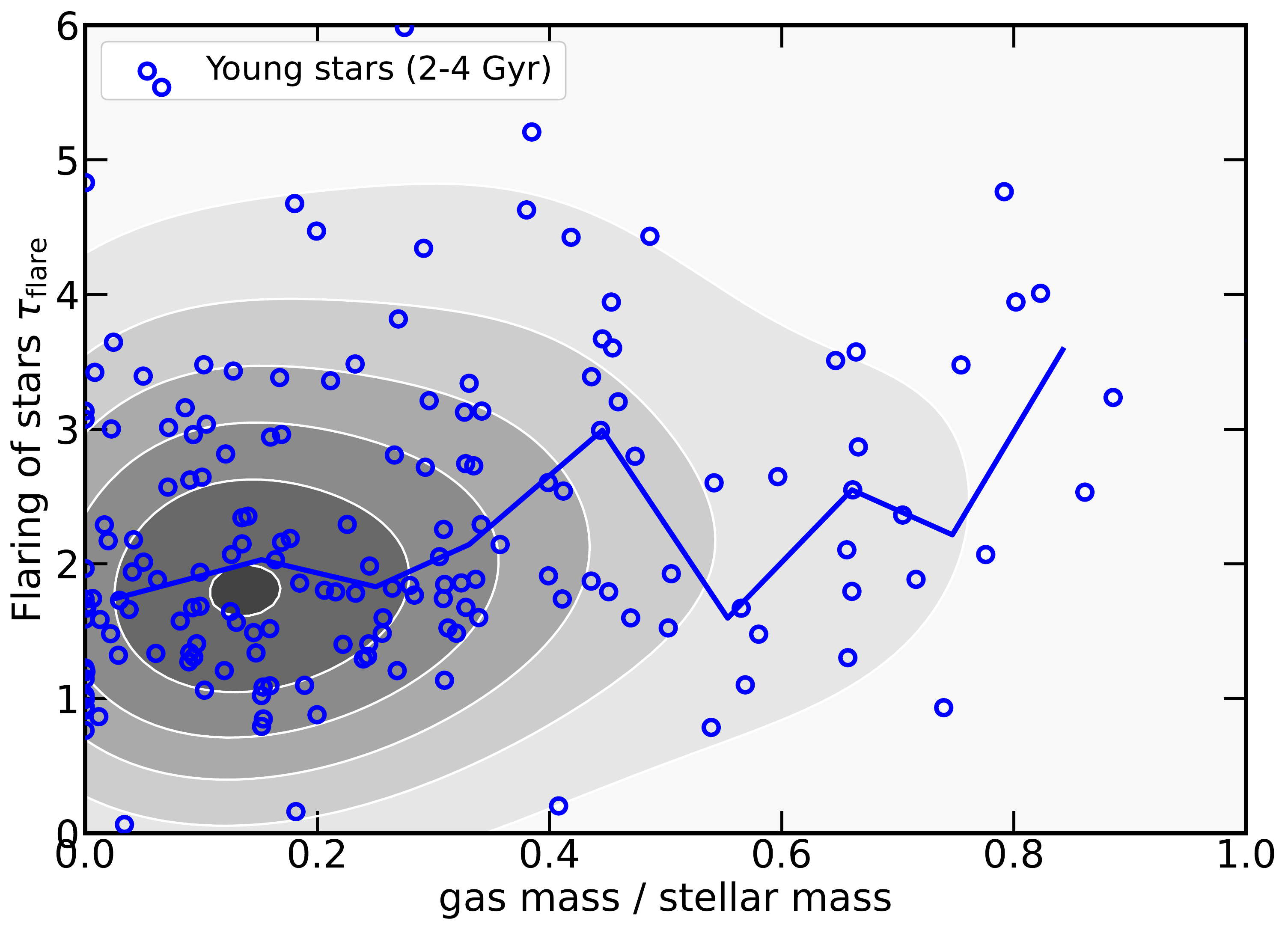}
\includegraphics[width=0.45\textwidth]{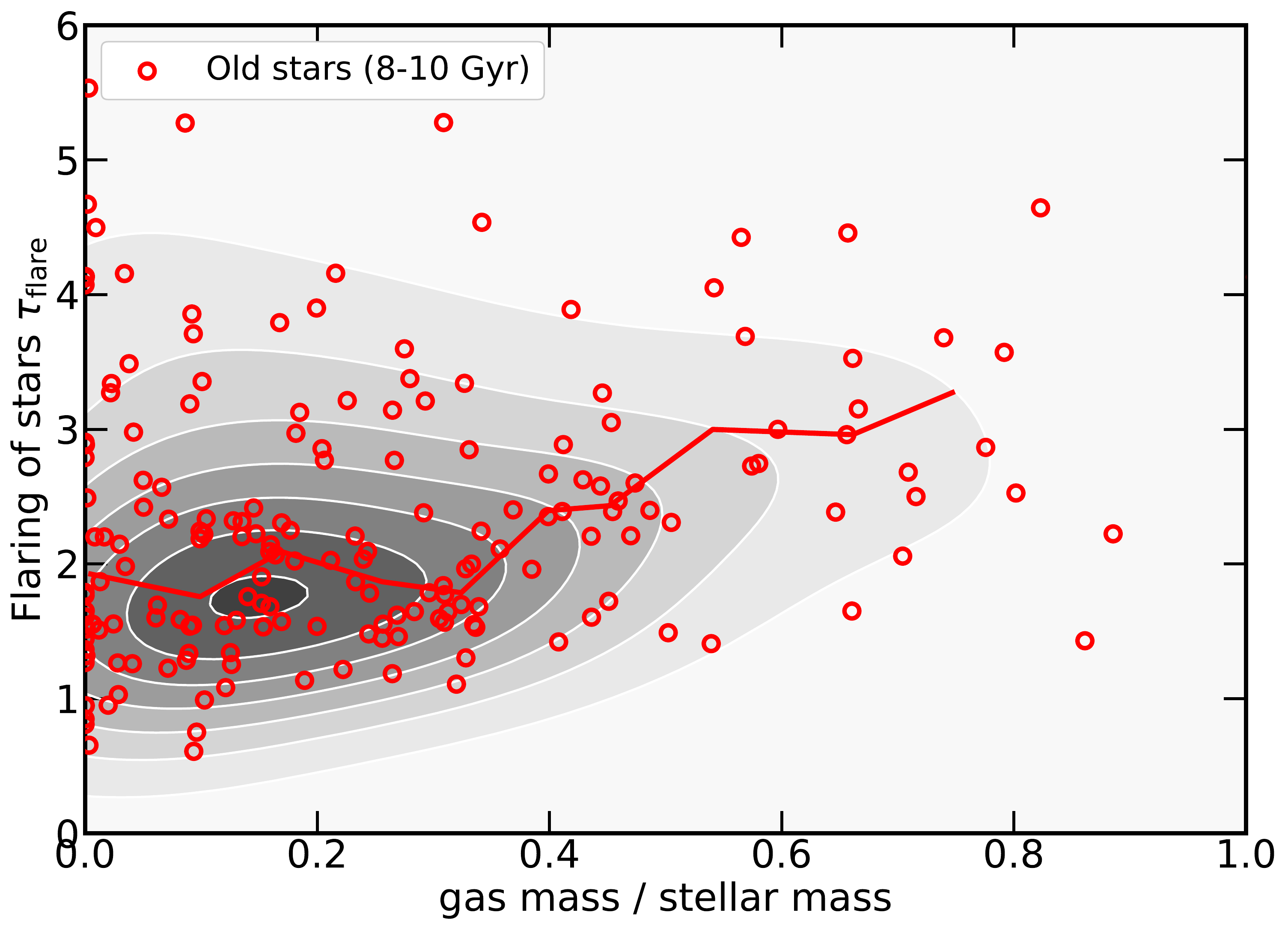}
\caption{\label{fig:young_old_diskprop} {\textbf{Disk flaring vs. disk properties for TNG50 MW/M31-like galaxies.}} We plot the amount of flaring of young (left column, in blue) and old (right column, in red) stellar populations as a function of stellar disk scalelength (top), disk scaleheight at 4$\times R_{\rm{d}}$ (middle) and gas fraction within the disk (bottom). In each panel, the galaxy number density is estimated with a Gaussian kernel and represented with the shaded contour areas. Solid curves are medians in bins of the quantity on the $x$-axes. Galaxies with thicker stellar disks and larger gas disk fractions seem to be characterized by somewhat larger degrees of disk flaring.}
\end{figure*}

\subsection{Disk flaring vs. $z=0$ structural and global properties of MW/M31-like galaxies}


In \S\ref{sec:kinematics}, we have shown that the flaring of the stellar disk thickness is a direct manifestation of larger stellar velocity dispersions and lower stellar surface densities in the outer disk regions. Are there other global and/or structural properties of galaxies that correlate with the disk flaring?
In this section we proceed to see whether or not the flaring of the mono-age stellar populations is connected with the global or structural $z=0$ properties of the TNG50 MW/M31-like galaxies. 

For the sake of clarity, we divide the $z=0$ properties in two subgroups: disk structural properties, including gas mass fraction in the disk region (Fig.~\ref{fig:young_old_diskprop}) and mass properties (Fig.~\ref{fig:young_old_massprop}). In each figure, the amount of flaring of young (left columns) and old stellar populations (right columns) are shown separately, with lines representing the running medians.

From Fig.~\ref{fig:young_old_diskprop}, we see that there is no clear trend or correlation between the degree of flaring and the scalelength of the disk. On the other hand, a clearer positive trend appears when we plot the dependence of flaring with the scaleheight of the stellar disk, both for the young and for the old stars, albeit with a significant scatter: namely, galaxies with thicker stellar disks appear to host a larger amount of disk flaring. As Fig.~\ref{fig:young_old_massprop} shows, the current galaxy stellar mass, stellar disk mass and disk-to-total mass ratio of a MW/M31-like galaxy are not predictors for disk flaring (even though this statement may differ across a larger range of stellar or halo masses). However, interestingly and related to the discussions of Section~\ref{sec:kinematics}, larger gas mass fractions in the disk of these galaxies imply larger degrees of flaring.

We confirm, although do not show, that the findings above hold irrespective of whether the flaring is evaluated based on stellar half-mass heights or scaleheights from parametric fits of the vertical stellar mass distribution.

Finally, we have compared the amount of flaring for young and old stellar populations with additional galactic and environmental properties (we omit the corresponding plots for brevity). The proportion of barred galaxies is not significantly different in the two subsets of systems where the young population flares more than the old stellar populations or viceversa. The presence itself of a bar does not seem to correlate either with the amount of flaring. We have examined whether the presence of other (satellite) galaxies in the proximity of our MW/M31 analogs at $z=0$ may be a predictor of, or may be associated to, larger degrees of flaring -- we cannot discern any statistically-robust trend.


\subsection{Disk flaring vs. merger histories of MW/M31-like galaxies} 
Is the amount of disk flaring determined by the past merger history of a galaxy? 

\begin{figure}
\includegraphics[width=0.48\textwidth]{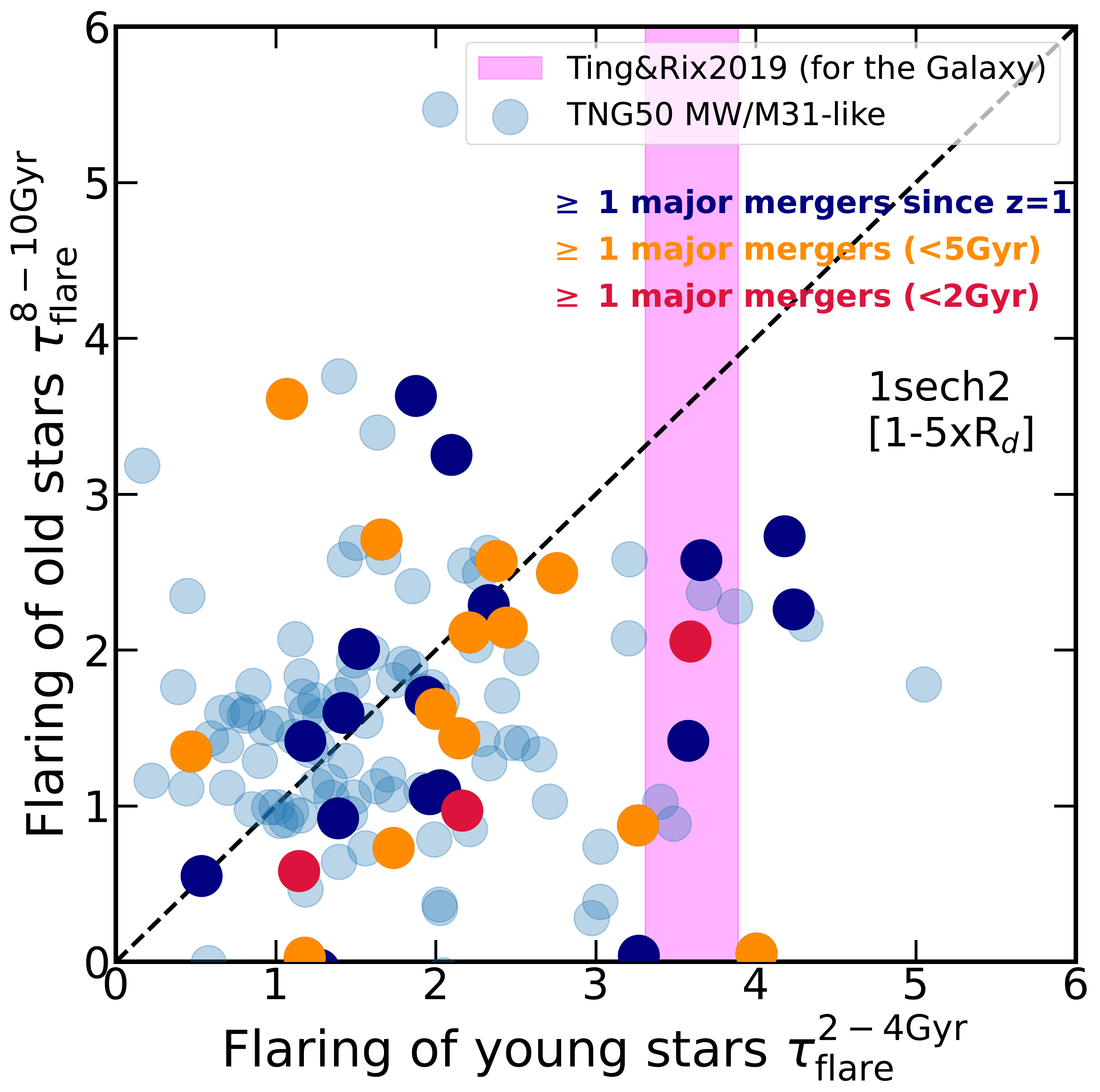}
\caption{\label{fig:young_old_mergers} {\bf Flaring of young vs. old stellar populations in TNG50 MW/M31-like galaxies in connection to their past merger history.} Here we show the same results, layout and method of measuring heights as in Fig.~\ref{fig:young_old}, but we additionally highlight the different merger histories: i.e. galaxies that have undergone at least one major merger in the last 2 Gyr (red), in the last 5 Gyr (orange) and since $z=1$ (navy). Whereas a recent or ancient major merger does not seem to systematically influence the degree of flaring, at the galaxy population level, it would appear that galaxies with recent major mergers are more likely to host young stellar populations that flare more than their older counterparts.}
\label{fig:mergers}
\end{figure}

As already mentioned in the Introduction, our TNG50 MW/M31-like sample is built with no constraint on past history, so that the $z=0$ MW/M31-like galaxies from TNG50 are the result of very diverse merger histories. In Fig.~\ref{fig:young_old_mergers}, we investigate whether the amount of flaring is correlated with aspects of the past merger history. To this aim, we highlight those galaxies that have undergone at least one major merger in the last 2 Gyr (red), 5 Gyr (orange) and since $z=1$ (navy). There seems to be a trend whereby MW/M31-like galaxies that underwent recent major mergers are more likely to have young stars flaring more with respect to the old population. To test the statistical significance of this statement (for the case of the major merger since $z=1$, where the counts are higher), we perform two sample tests of proportions, obtaining a p-value of 0.12 for both the Z-test and the chi-square test: this means that the trend that we see Fig.~\ref{fig:young_old_mergers} is only weak.

Now, we can speculate that even if stars younger than 2 Gyr were not born yet at the time of the major merger, they were born later in a ``perturbed'' stellar disk likely heated-up by the merger event. Alternatively, as discussed in \citet{2017Ma}, the young stars flare more because they inherit the flaring of the gas from which they formed, being this possibly perturbed by the recent mergers. However, the occurrence of recent or less recent major mergers does not seem to imply overall systematically stronger disk flaring. In fact, we have not identified trends between disk flaring and a number of merger-history statistics, including the number of all major and minor mergers, the ex situ total mass and the ex situ fraction of the galaxy.




\subsection{On possible resolution effects}
Before closing, we offer a few remarks on the possible effects of numerical resolution. In this paper we have characterized the stellar disk structures and flaring predicted for MW/M31-like galaxies by one simulation only, the highest-resolution run of the IllustrisTNG project: TNG50. As shown in \cite{2019Pillepich}, stellar disk heights are affected by the underlying resolution: TNG50 galaxies are on average thinner than those in the 8-times lower mass resolution counterpart TNG50-2: see their Figures B2 and B3. However, by how much galaxies are thinner in the higher-resolution simulation in comparison to lower-resolution runs depends on redshift and stellar mass range.  
With this in mind, we notice that all the results shown in this paper are based on the relative enhancement, $\tau_{\rm flare}$, of scaleheights across the simulated stellar disks. It is therefore plausible that any effect of numerical resolution on the values of the disk scaleheights should be mitigated by this definition. Furthermore, the levels of disk flaring predicted by the TNG50 simulation are consistent with those returned in MW analogs simulated at much better numerical resolution, such as Latte and Vintergatan. This gives further confidence on the quantitative soundness of the results provided herein.

\section{Summary and conclusions}
\label{sec:conclusions}

In this paper we have presented a comprehensive study of the stellar disk structure of a large, unbiased sample of Milky Way (MW) and Andromeda (M31)-like galaxies from TNG50, the highest-resolution cosmological magneto-hydrodynamical large-volume simulation of the IllustrisTNG project (\S\ref{sec:TNG50}).

Our TNG50 MW/M31-like sample includes objects with stellar disky morphology, with a stellar mass in the range of $M_* = 10^{10.5 - 11.2}~\Ms$, and within a MW-like Mpc-scale environment at $z=0$ (\S\ref{sec:selection}). 
We have focused on the vertical structure of the TNG50 MW/M31-like galaxies and, in particular, on the flaring of their stellar disks, distinguishing across mono-age stellar populations. Throughout this paper, by disk flaring we mean the quantification of by how much the scaleheight of the stellar disk changes (i.e. increases) with galactocentric distance, once stars are selected according to their ages (or other properties).

Our analysis and results rely on the resolution, sample size, and realism of the TNG50 simulated galaxies. In fact, thanks to its high numerical resolution approaching the one typical of ``zoom-in'' simulations, TNG50 returns at $z=0$ 198 different realizations of MW/M31-like galaxies. The stellar disk scalelength of TNG50 MW/M31-like galaxies ranges across $\sim1.5-17$ kpc, with a good qualitative and quantitative agreement when compared to other models and available observational findings of local disky and spiral galaxies (\textcolor{blue}{Pillepich et al. in prep.}). Moreover, the vertical stellar distribution in most TNG50 MW/M31-like galaxies can be well described with a double (squared hyperbolic secant) profile, allowing us to distinguish between a thin and a thick disk {\it geometric} component. For some galaxies, the stellar thin (thick) disk is found to be as thin (thick) as the observed one for the Milky Way, i.e. with a scaleheight of about $175-360 \, (900-1300)$ pc.\\

Our main results on disk flaring can be summarized as follows:

\begin{itemize}

    \item By fitting the vertical stellar density distribution of each simulated galaxy with a double or single functional profile, we have estimated the stellar disk scaleheights at a series of different galactocentric distances and for different mono-age stellar populations (Figs.~\ref{fig:vertical_profile} and \ref{fig:vertical_profile_ages}). TNG50 predicts, in general, systematically higher values of the stellar scaleheight moving outward from the galactic center, i.e. it predicts ``disk flaring''. In fact, we show that, with one unique set of physical ingredients, TNG50 is able to reproduce diverse levels and kinds of flared stellar disks (Fig.~\ref{fig:flare}). 
    \\
    \item Because, according to TNG50, the increase of stellar scaleheight with galactocentric distance can be linear, exponential or other, depending on the galaxy and stellar population, and because disk galaxies in a narrow range of stellar mass can exhibit disk scalelengths varying by up to factors of 8, with this paper we propose and advocate for an easy, non-parametric, fit-independent measurement of the degree of flaring. Namely, we propose to quantify flaring simply based on the relative enhancement of the scaleheight between 1 and 5 times the scalelength of each MW/M31 galaxy (Eq.~\ref{eq:tau_flare}). \\
    
    \item We have compared the amount of flaring displayed by the young (2-4 Gyr) and old stellar populations (8-10 Gyr) finding that which stars flare more, and by how much, changes from galaxy to galaxy, with both typically exhibiting $1.5-2$ thicker disk heights in the outskirts than towards the center (Fig.~\ref{fig:young_old}). The young stellar populations in about eleven MW/M31-like galaxies exhibit a similar degree of flaring of our Milky Way, for which we have extrapolated the scaleheight as a function of radius and ages from \cite{2019Ting}.    \\
    
    

    \item We have applied our method to the data available in the literature for selected zoom-in simulations of MW-like galaxies: namely, Latte, VINTERGATAN, NIHAO-UHD and Auriga. The amount of flaring of old and young stellar populations found in TNG50 encompasses qualitatively and quantitatively all the aforementioned simulations, implying that the stellar-disk flaring of the unbiased sample of TNG50 MW/M31 analogs, returned by a fixed physical model, covers all the previous findings of zoom-in simulations performed with different codes, in some cases with better resolution, different galaxy formation models and varying assumptions on the past assembly history of the simulated galaxies (Fig.~\ref{fig:tng50vsothers}).
    \\
    \item The scaleheights we measure in the TNG50 simulated galaxies are a manifestation of the underlying orbits and overall potential, with their values exhibiting a clear correlation with the local stellar vertical velocity dispersion and the local stellar surface density (Fig.~\ref{fig:young_old_sigmaVSurfDens}), as predicted by theoretical models. Namely, stellar disks are thicker if hotter and with lower surface densities.\\

    \item According to our analysis, galaxies with hotter or less dense outer stellar disks flare more strongly, in both young and old stellar populations, and the diversity in flaring seems to be mostly driven by a diversity in vertical stellar velocity dispersion in the outer disk regions (Fig.~\ref{fig:tau_kinematics}).\\
    
    \item On the other hand, the flaring of the mono-age populations does not manifestly depend on global $z=0$ structural properties of the MW/M31 sample. However, two key albeit mild trends can be highlighted: old stellar populations flare more in galaxies with larger disk scaleheight and larger gas mass fraction in the disks, the former  relationship being in place also for young stars (Figs.~\ref{fig:young_old_diskprop}, \ref{fig:young_old_massprop}).\\

    \item Finally, the stellar population properties in the mid plane of each TNG50 MW/M31-like galaxy exhibit a qualitative agreement with what typically observed: the average stellar age \textit{decreases} at increasing galactocentric distance (and at fixed vertical height), and \textit{increases} at increasing vertical height (and at fixed radius -- Fig.~\ref{fig:flaring_obs}, right panels compared to left and middle ones). However, we argue that these observed trends do not necessarily imply flaring, and shall not be used in fact as a proxy for flaring, as a change in average stellar age with galactocentric distance can be realized with a varying trend with radius of stellar age distributions and not necessarily by a change in stellar orbits (Fig.~\ref{fig:demonstration_noflaring}). 
    \\
    
\end{itemize}

Overall, with the analysis of this paper we have demonstrated that a sample of 198 $z=0$ MW/M31-like galaxies -- selected to be disky, isolated (but allowing Local Groups), and within a narrow range of stellar mass (covering the MW and M31 stellar masses) --, exhibit a great diversity in their vertical disk structures. 
The heterogeneity of the stellar disk properties and the diverse flavors and amounts of stellar disk flaring are a key result. Indeed, TNG50 is a cosmological magneto-hydrodynamical simulation that -- with a given set of physical-model ingredients and with a resolution that bridges the gap between the volume and zoom-in simulations -- is able to reproduce, replicate, and expand upon the features of the stellar disk flaring already investigated and shown in the literature, based on zoom-in cosmological simulations.

The fact that there is no trivial correlation between the amount of flaring and the $z=0$ structural {\it global} properties of galaxies is intriguing. We have also tentatively investigated whether flaring may be correlated with the number of the more or less recent major/minor/total mergers, finding no obvious dependence (Fig.~\ref{fig:mergers}). These considerations, together with the findings on kinematics, lead us to speculate that the increasing of the stellar disk scaleheight with the galactocentric distance could be ascribed to a possible heating of the stellar disk due to external perturbations such as flybys (without the need for coalescence).
In fact, keeping in mind that we have not imposed constraints on past history, the pathways leading to $z=0$ MW/M31-like galaxies could be very diverse across the whole sample \citep[see also][]{2022SotilloRamos}. Future and further analyses are required to connect the disk flaring to the past of MW/M31-like galaxies and to the dynamical evolution of their stars.

\section*{Data availability statement}
 The entire data of the IllustrisTNG simulations, including TNG50, are publicly available and accessible at \url{www.tng-project.org/data} \citep[][]{2019Nelson-a}. Additional and easier-to-use data products and particle data cutouts related to the 198 MNW/M31-like galaxies from TNG50 used in this paper are also publicly available (\textcolor{blue}{Pillepich et al. in prep.}). With this paper, we make public also a series of catalogs for the various measures of disk flaring and for the flags of warped and disturbed TNG50 MW/M31-like galaxies.

\section*{Acknowledgements}
DS, MD, and AP acknowledge support by the Deutsche Forschungsgemeinschaft (DFG, German Research Foundation) -- Project-ID 138713538 -- SFB 881 (``The Milky Way System'', subprojects A01 and A06). DN acknowledges funding from the Deutsche Forschungsgemeinschaft (DFG) through an Emmy Noether Research Group (grant number NE 2441/1-1).
The TNG50 simulation used in this work has been run on the HazelHen Cray
XC40-system at the High Performance Computing Center Stuttgart under the Gauss centers for Super-computing (GCS) Large-Scale Project GCS-DWAR (2016; PIs Nelson/Pillepich).

\footnotesize{
\bibliography{biblio}
}

\appendix
\label{appendix}

\section{Disk flaring vs. additional structural properties of MW/M31-like analogs}
\label{app:flarVSprop}
\begin{figure*}
\centering
\includegraphics[width=0.46\textwidth]{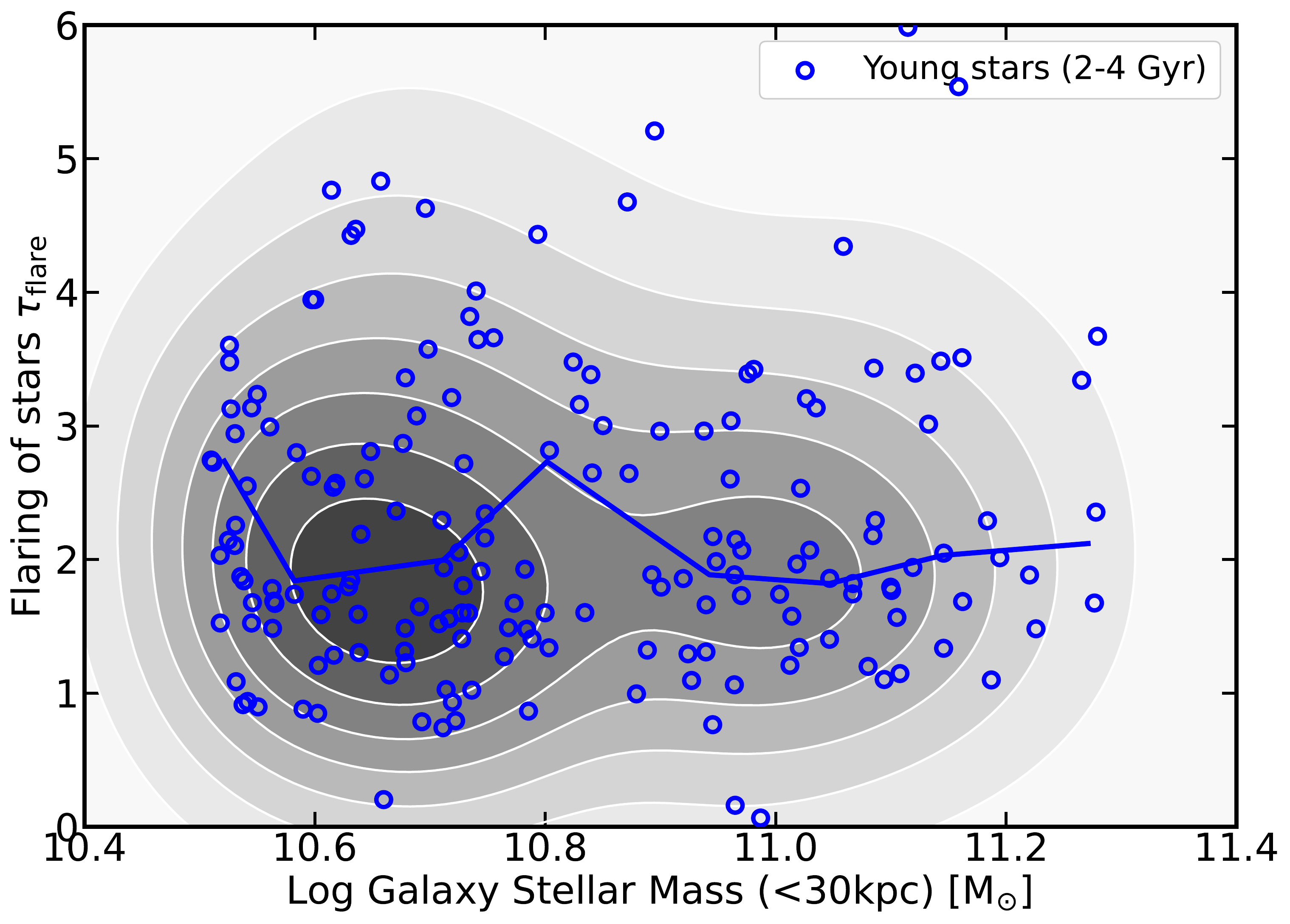}
\includegraphics[width=0.46\textwidth]{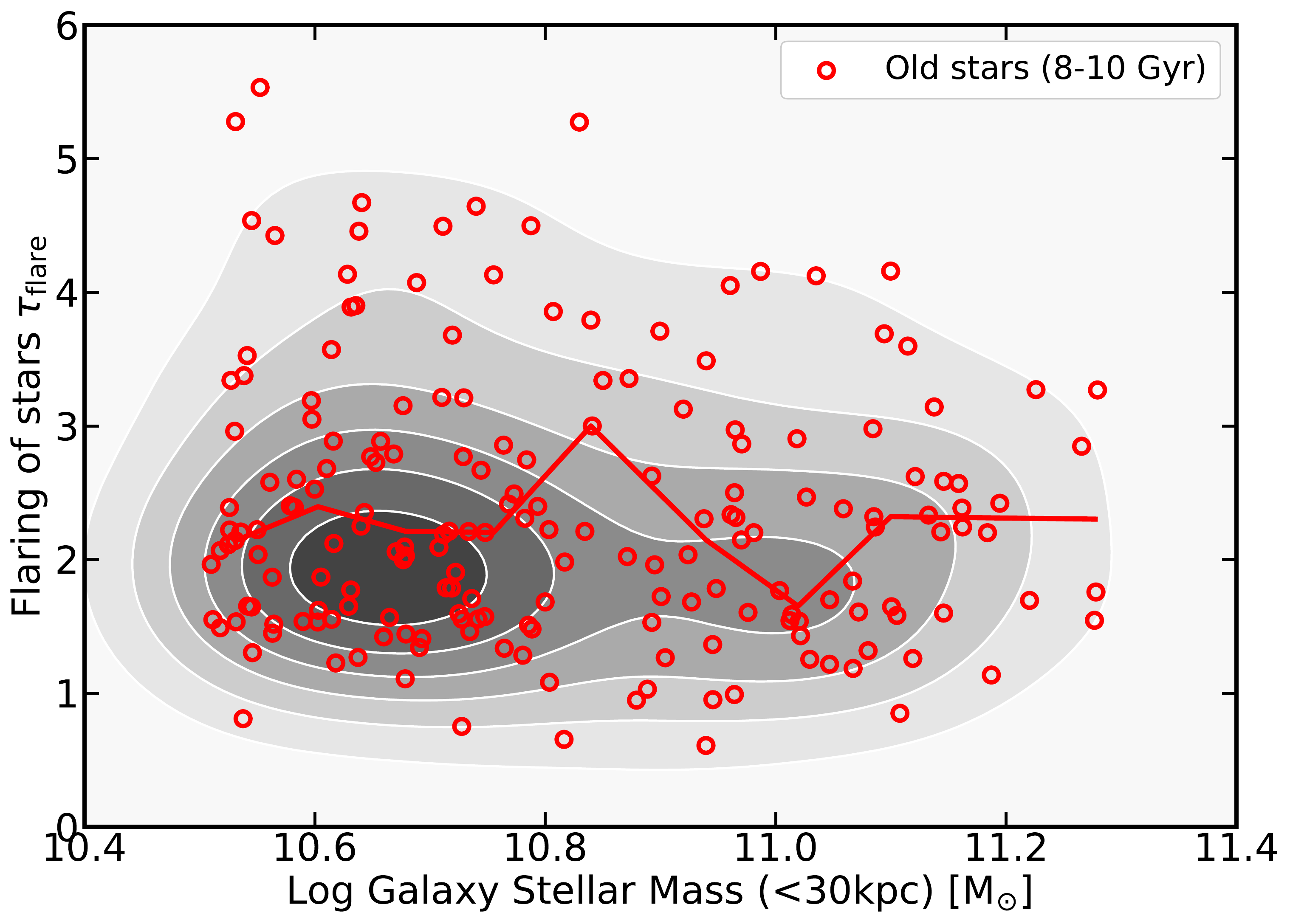}
\includegraphics[width=0.46\textwidth]{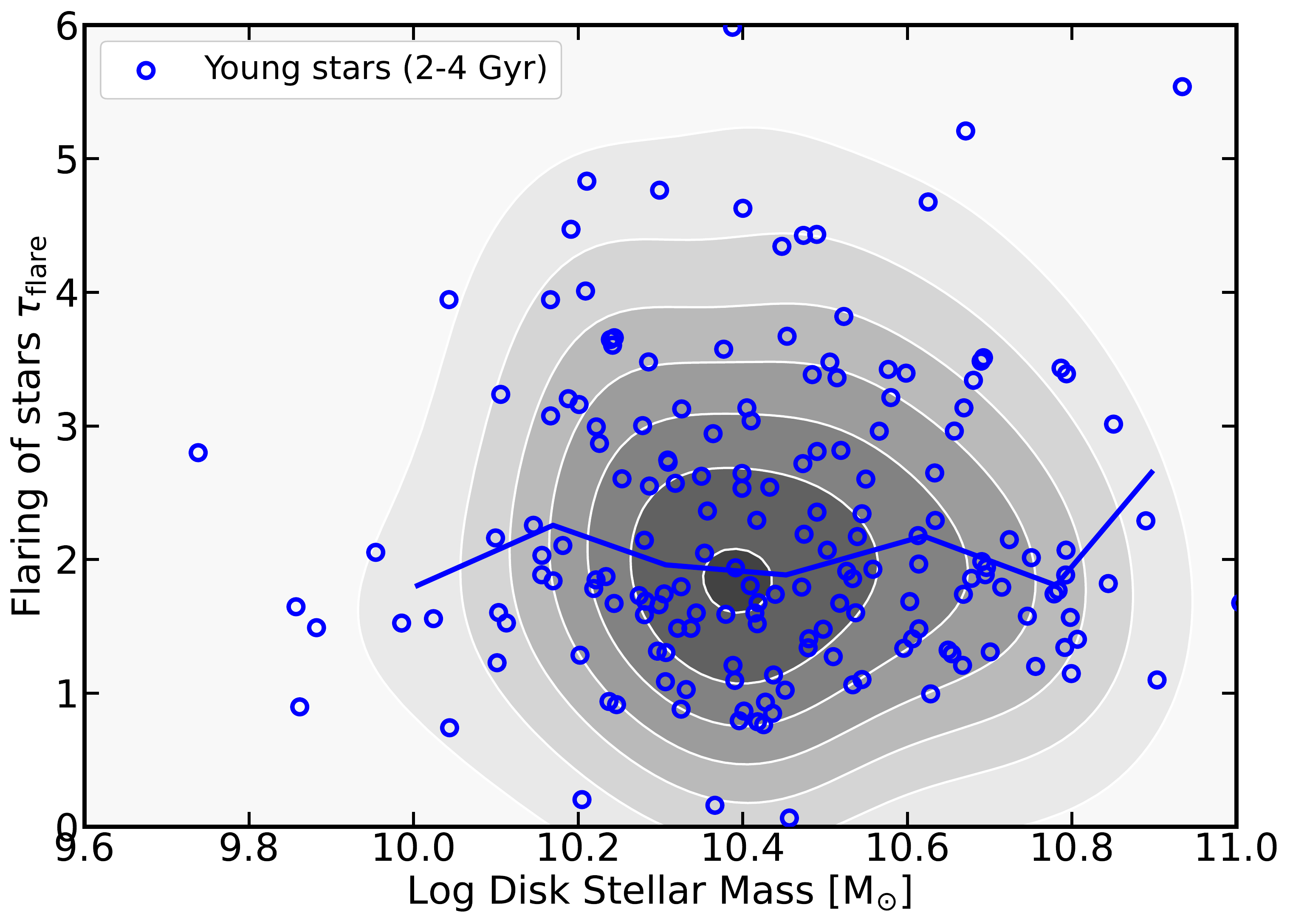}
\includegraphics[width=0.46\textwidth]{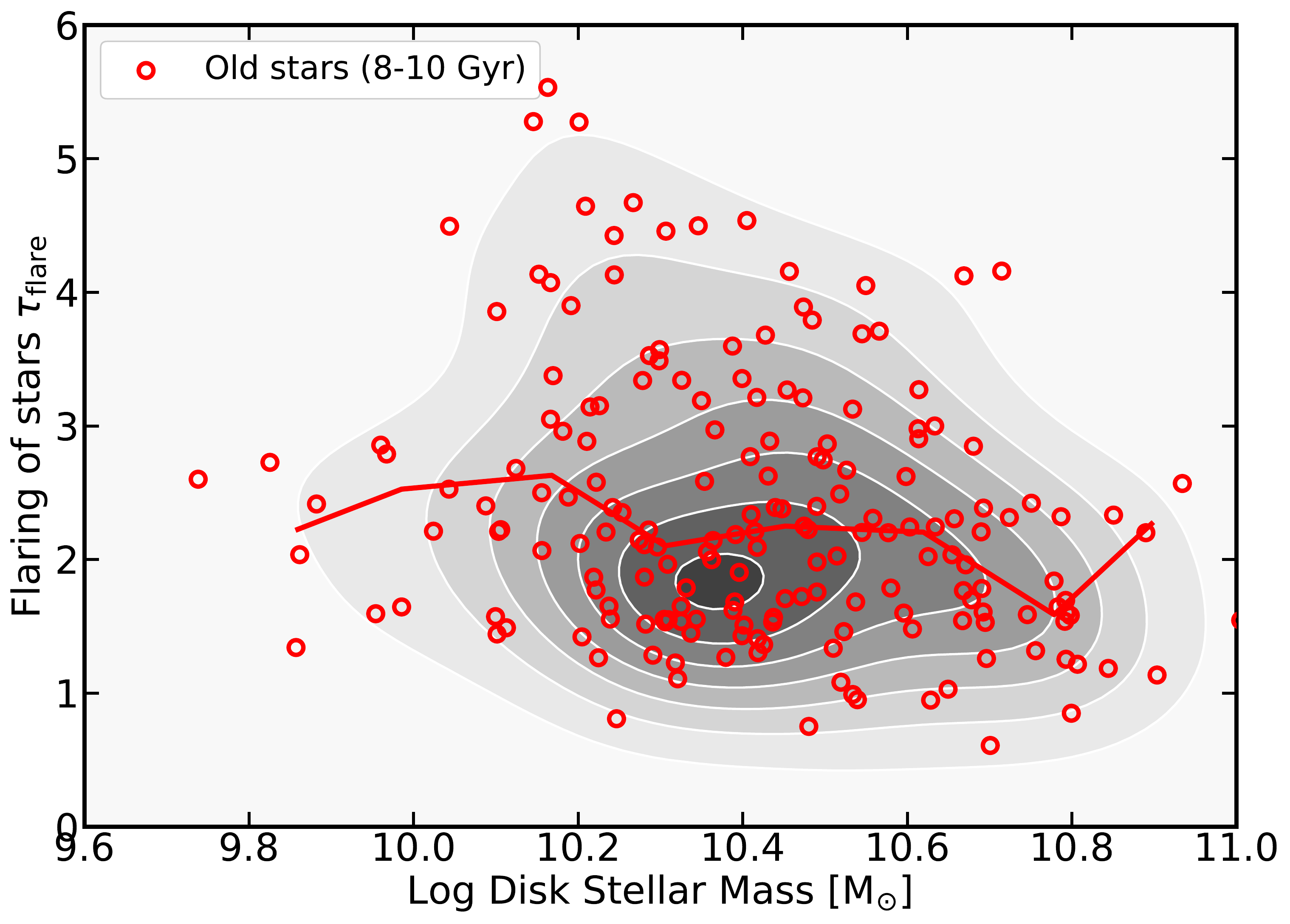}
\includegraphics[width=0.46\textwidth]{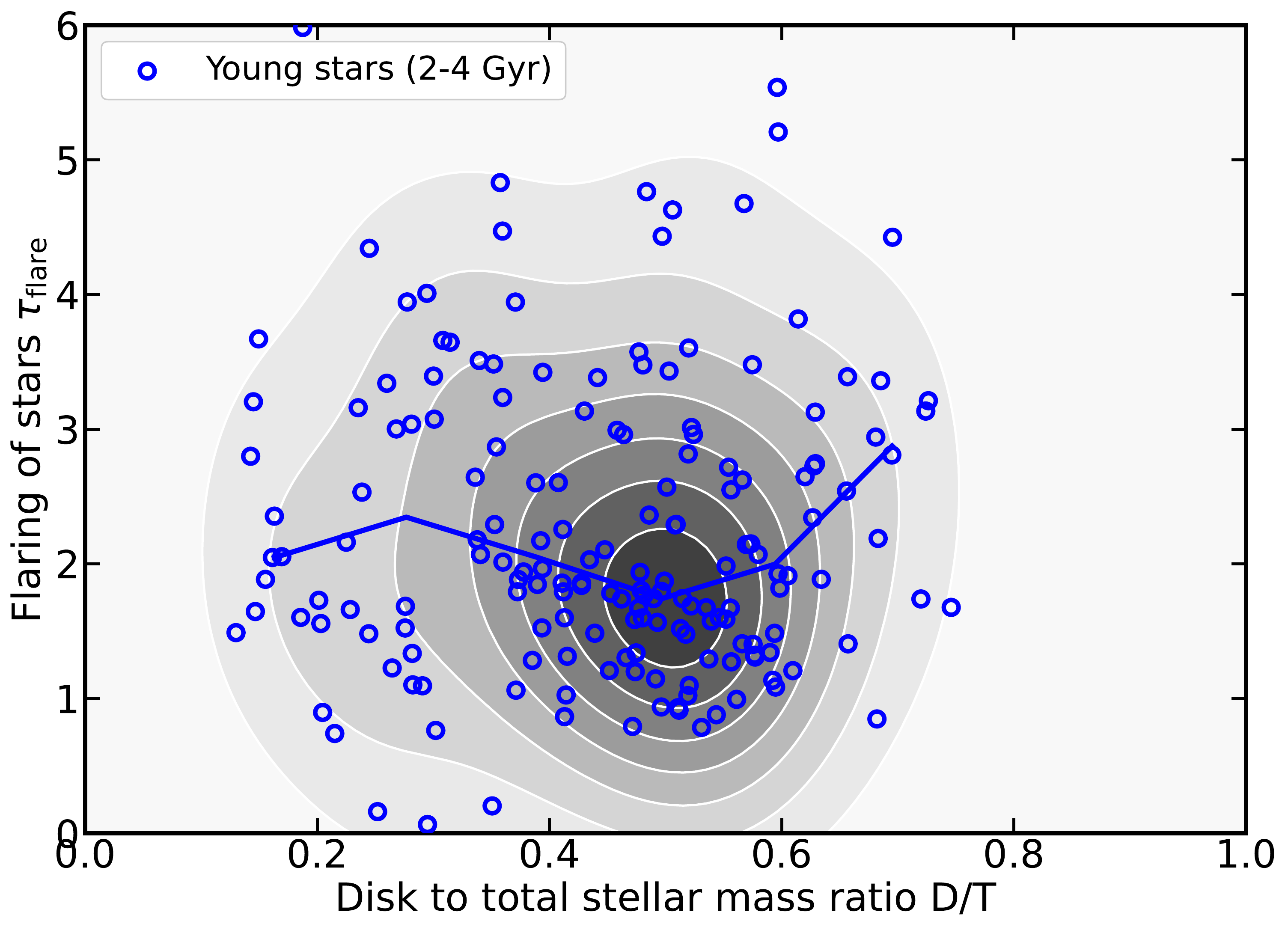}
\includegraphics[width=0.46\textwidth]{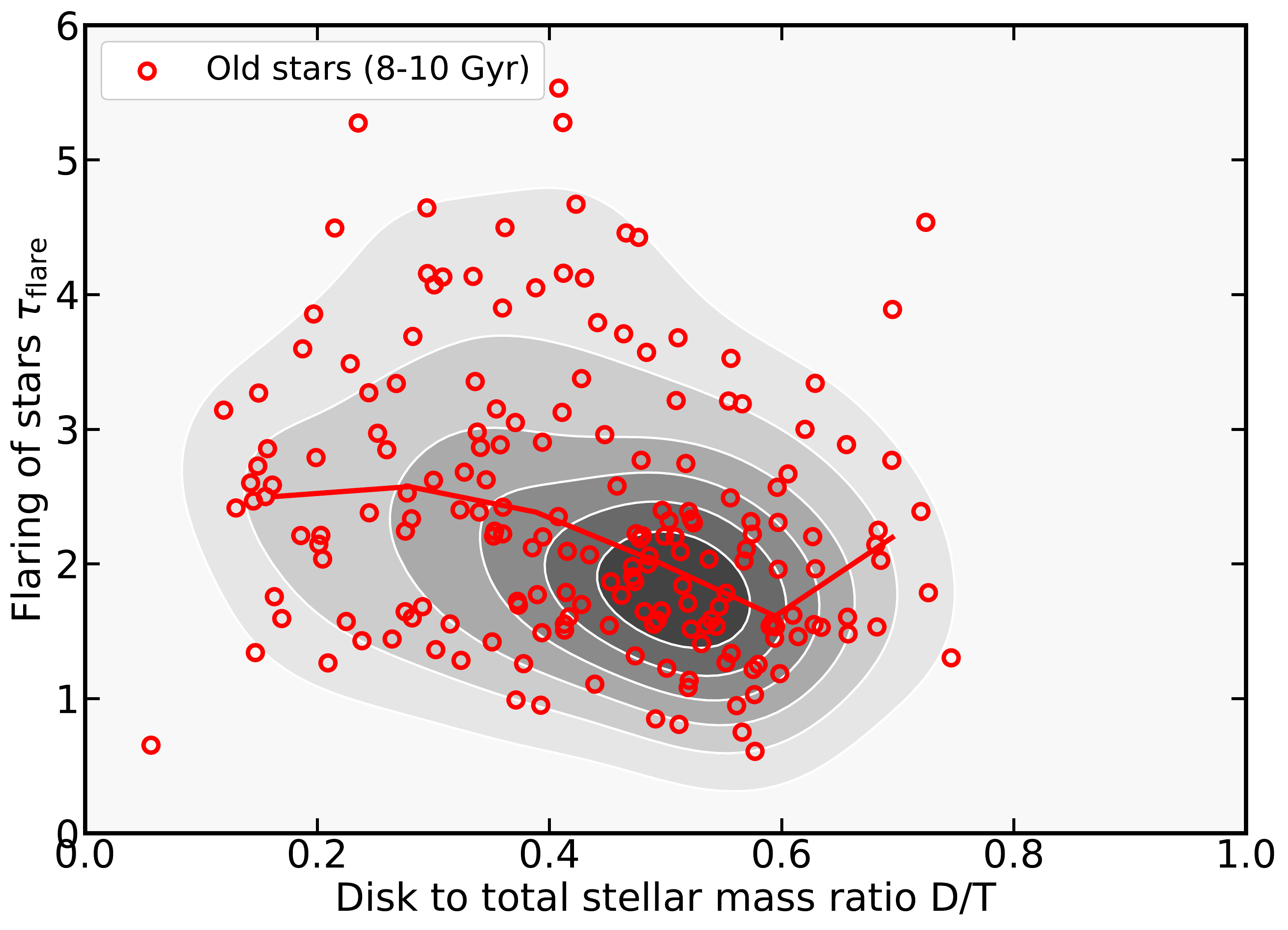}
\caption{\label{fig:young_old_massprop}\textbf{Disk flaring vs. galaxy masses for TNG50 MW/M31-like galaxies.} We show the amount of flaring of young (left column, in blue) and old (right column, in red) stellar populations as a function of three selected global galaxy properties: stellar mass (top), disk stellar mass (middle), and disk-to-total stellar mass ratio (bottom). In each panel, the galaxy number density is represented with the shaded contour areas. Solid curves are medians in bins of the quantity on the x-axes. No discernible trend seem to be in place.}
\end{figure*}

We also examine the level of flaring, observed in both young and old stars, as a function of selected galaxy mass properties: stellar mass, disk stellar mass and disk-to-total stellar mass ratio (Fig.~\ref{fig:young_old_massprop}).


\label{lastpage}
\end{document}